\documentclass[referee]{aa}

\usepackage{hyperref}

\usepackage{silence}
\usepackage{soul}  %% Package for strikethrough

\usepackage{subfigure}

\usepackage{graphicx}
\usepackage{txfonts}

\usepackage{xcolor}

\usepackage{lineno}
\def\Msol {$\hbox{M}_\odot$\xspace}
\def\Msolperyr {$\hbox{M}_\odot$\,yr$^{-1}$\xspace}

\usepackage[T1]{fontenc}
\graphicspath{{./}{figures/}}
\begin{document}

\titlerunning{Formation and evolution of a protoplanetary disk}
\authorrunning{The Treilles group}

\title{Formation and evolution of a protoplanetary disk: combining observations, simulations and cosmochemical constraints}
%\subtitle{I. Overviewing the $\kappa$-mechanism}

\author{
Alessandro Morbidelli\inst{1,2}, 
Yves Marrocchi\inst{3},
Adnan Ali Ahmad\inst{4},
Asmita Bhandare\inst{5,6},
Sébastien Charnoz\inst{7},
Beno\^{i}t~Commer\c{c}on\inst{6},
Cornellis P. Dullemond\inst{8},
Tristan Guillot\inst{2},
Patrick Hennebelle\inst{4},
Yueh-Ning Lee\inst{9,10,11},
Francesco Lovascio\inst{6},
Raphael Marschall\inst{2},
Bernard Marty\inst{3},
Anaëlle Maury\inst{4,13,14},
and
Okamoto Tamami\inst{12,2}
}

\institute{
%1 
Coll\`ege de France, Centre National de la Recherche Scientifique, Universit\'e Paris Sciences et Lettres, Sorbonne Universit\'e, 75014 Paris, France
\\
email: Alessandro.Morbidelli@oca.eu
\and
%2
Laboratoire Lagrange, Centre National de la Recherche Scientifique, Observatoire de la C\^ote d’Azur, Universit\'e C\^ote d’Azur, 06304 Nice, France
\and
%3
Universit\'e de Lorraine, CNRS, CRPG, UMR 7358, 54000 Nancy, France
\and
%4
Universit\'e Paris Cit\'e, Universit\'e Paris-Saclay, CEA, CNRS, AIM, F-91191, Gif-sur-Yvette, France
\and
%5
Universit\"ats-Sternwarte, Fakult\"at f\"ur Physik, Ludwig-Maximilians-Universit\"at M\"unchen, Scheinerstr.~1, 81679 M\"unchen, Germany 
\and
%6
ENS de Lyon, CRAL UMR5574, Universit\'e Claude Bernard Lyon 1, CNRS, Lyon, 69007, France
\and
%7
Universit\'e Paris Cit\'e, Institut de physique du globe de Paris, CNRS, 1, rue Jussieu, Paris, F-75005, France
\and
%8
Institute for Theoretical Astrophysics, Center for Astronomy, Heidelberg University, Albert-Ueberle-Str. 2, 69120 Heidelberg, Germany
\and
%9
Department of Earth Sciences, National Taiwan Normal University, Taipei 116, Taiwan
\and
%10
Center of Astronomy and Gravitation, National Taiwan Normal University, Taipei 116, Taiwan
\and
%11
Physics Division, National Center for Theoretical Sciences, Taipei 106, Taiwan
\and
%12
Tokyo Institute of Technology, Ookayama, Meguro-ku, Tokyo 152-8551, Japan
\and
%13
Institute of Space Sciences (ICE), CSIC, Campus UAB, Carrer de Can Magrans s/n, E-08193, Barcelona, Spain
\and
%14
ICREA, Pg. Lluís Companys 23, Barcelona, Spain
}

\date{Received xxxx; accepted xxxx}

% \abstract{}{}{}{}{}
% 5 {} token are mandatory

\abstract
{The formation and evolution of protoplanetary disks remains elusive. We have numerous astronomical observations of young stellar objects of different ages with their envelopes and/or disks; moreover {in the last decade} the numerical simulations of star and disk formation have made tremendous progress, with realistic equations of state for the gas and treating the interaction of matter and the magnetic field, using the full set on non-ideal magneto-hydrodynamics equations. Yet, it is not fully clear how a disk forms: from inside out or outside in, where the material accreted onto the disk falls and comes from, the evolution of dust in the disks, the appearance of structures. These unknowns limit our understanding of how planetesimals and planets form and evolve. }
{We attempt to reconstruct the evolutionary history of the protosolar disk, guided by the large amount of cosmochemical constraints derived from the study of meteorites, while using astronomical observations and numerical simulations as a guide of which scenarios may be plausible.}
{Our approach is highly interdisciplinary. We do not present new observations or simulations, but combine in an original manner a large number of published results concerning young stellar objects observations, numerical simulations, and the chemical, isotopic and petrological nature of meteorites to reconstruct the history of the protoplanetary disk at the origin of our Solar system.}
{{We achieve a plausible and coherent view of the evolution of the protosolar disk that is consistent with the cosmochemical constraints and compatible with observations of other protoplanetary disks and sophisticated numerical simulations. The evidence that high-temperature condensates, CAIs and AOAs, formed near the protosun before being transported to the outer disk can be explained by either an early phase of vigorous radial spreading of the disk, or fast transport of these condensates from the vicinity of the protosun towards large disk radii via the protostellar outflow. The assumption that the material accreted towards the end of the infall phase was isotopically distinct allows us to explain the observed dichotomy in nucleosynthetic isotopic anomalies of meteorites and leads to intriguing predictions on the isotopic composition of refractory elements in comets. When  the infall of material waned, the disk started to evolve as an accretion disk. Initially, dust drifted inwards, shrinking the radius of the dust component to $\sim 45$~au, probably about 1/2 of the width of the gas component. Then structures must have emerged, producing a series of pressure maxima in the disk which trapped the dust on My timescales. This allowed planetesimals to form at radically distinct times without changing significantly of isotopic properties. There was no late accretion of material onto the disk via streamers. The disk disappeared in 5~Myr, as indicated by paleomagnetic data in meteorites.}}
{The evolution of the protosolar disk seems to have been quite typical in terms of size, lifetime, and dust behavior, suggesting that the peculiarities of the Solar system with respect to extrasolar planetary system probably originate from the chaotic nature of planet formation and not at the level of the parental disk.}
\keywords{meteorites, chondrules, CAI, protoplanetary disks, young stellar objects, dust particles}

\maketitle

\section{Introduction}\label{sec:introduction} 

The stars and the possibility of existence of planetary bodies in orbit around them have fascinated all human societies since the oldest ages. Modern astronomical observation techniques have revealed the extreme diversity of extra-solar systems that bear little resemblance to our well-ordered Solar system. The formidable complexity of processes at play during star and planet formation is the focus of many scientific disciplinary fields. Astronomical observations, theoretical modeling, and meteorite characterizations all provide important information that are not often combined for establishing comprehensive scenarios. Thanks to the "Fondation des Treilles" located in the south-east of France, a group of 15 astronomers, astrophysicists, and cosmochemists spent several intense days exchanging scientific results and discussing possible interpretations of data and constraints provided by their respective scientific fields. The present article aims at presenting a scenario, and its possible variants when still unclear, that can integrate most of the current constraints regarding the evolution of pre-stellar cores, star formation, gas and dust dynamics in protoplanetary disks, and the petrographic and isotopic properties of meteorites. The authors emphasize that the scenario proposed here should not be regarded as set in stone, but needs to be tested and challenged by new observations, models, and geochemical measurements. Our ambition is thus to generate a scientific debate and lay out paths for future multi-disciplinary researches. 

This manuscript starts in Sect. 2 with a presentation of the minimal amount of prerequisites that the reader should master about stellar formation stages, geochemical concepts, and meteorites properties, in order to follow the rest of the work. Then the manuscript discusses the formation and evolution of a protoplanetary disk in a chronological sequence. Section 3 discusses the Class-0 and Class-I stages, during which the star-disk system, still embedded in its envelope, evolves after having formed from the collapse of a protostellar core. It focuses in particular on how the material from the cloud feeds the forming disk and on the transport of grains formed in the vicinity of the star to large star-centric distances. Section 4 discusses the evolution of later disks of Class II, the evolution of dust, the emergence of structures, and the possible late delivery of fresh material via streamers. Throughout the presentation we will highlight the uncertainties, identify future observations or model results that can discriminate scenarios and discuss how common or peculiar the protosolar disk, constrained by the cosmochemical properties of meteorites, appears to be relative to the collection of observed extrasolar disks. Section 5 will close the manuscript with our conclusions. In order to improve readability, we decided to mention the essence of each relevant argument, without entering into details, for which we refer the reader to some key references.

%%%%%%%%%%%%%%%%%%%%%%%%%%%%%%%%%%%%%%%%%%%%%%%%%%%%%%%%%%%%%%%%%%%%%%%%%%%%%%
%%%%%%%%%%%%%%%%%%%%%%%%%%%%%%%%%%%%%%%%%%%%%%%%%%%%%%%%%%%%%%%%%%%%%%%%%%%%%%
%%%%%%%%%%%%%%%%%%%%%%%%%%%%%%%%%%%%%%%%%%%%%%%%%%%%%%%%%%%%%%%%%%%%%%%%%%%%%%
%%%%%%%%%%%%%%%%%%%%%%%%%%%%%%%%%%%%%%%%%%%%%%%%%%%%%%%%%%%%%%%%%%%%%%%%%%%%%%
%%%%%%%%%%%%%%%%%%%%%%%%%%%%%%%%%%%%%%%%%%%%%%%%%%%%%%%%%%%%%%%%%%%%%%%%%%%%%%
%%%%%%%%%%%%%%%%%%%%%%%%%%%%%%%%%%%%%%%%%%%%%%%%%%%%%%%%%%%%%%%%%%%%%%%%%%%%%%
\section{An introduction to star formation and cosmochemistry}
\label{sec:Glossary}

\subsection{Star formation}
\label{intro-star}
%The formation of stars takes place in filamentary molecular clouds (primarly composed of molecular hydrogen and helium), when the high-density interstellar medium (ISM) partly collapses and fragments into bound prestellar cores. 
The formation of stars and their surrounding disks is triggered by the gravity-dominated collapse of bound fragments known as prestellar cores, within filamentary molecular clouds of the interstellar medium (ISM) primarly composed of molecular hydrogen, helium, and dust grains \citep{Rosen2020, Zhao2020}. 

\begin{figure*}[t!]
\begin{center}
\includegraphics[width=0.6\textwidth]{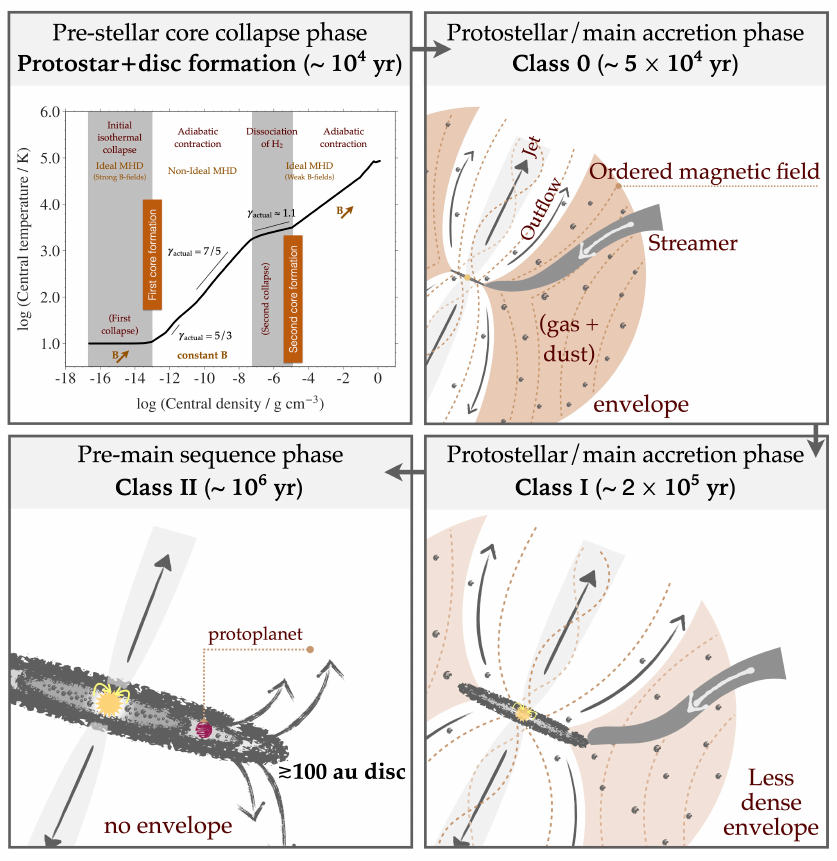}
\end{center}
\caption{
Schematic of the evolutionary sequence for the formation of a solar-type star, starting from the collapse of a molecular cloud core towards its subsequent evolution into a young stellar object. The top left panel shows the evolution of central gas temperature as a function of density during protostellar formation (reproduced from \citealt{Bhandare2020}). The collapse leading to the formation of the first and second cores is characterized by a weak increase of temperature during the increase in density. 
The right top and bottom panels show the main accretion phase, divided into the Class 0 and I stages, during which the star gains its mass and the disk is assembled, while both are still embedded in the parent envelope (also called protostellar core, after \citealt{Andre00}). At these stages, outflows and jets are also observed. {Additionally, asymmetric accretion, sometimes funneled via streamers to the young star-disk system can also be seen.} The bottom left panel shows the Class II stage where the system evolves into a young star surrounded by a protoplanetary disk, further moving onto the main sequence.}
\label{fig:seqevol}
\end{figure*} 

In his pioneering work, \cite{larson_1969} identified the two-step evolutionary sequence that results in the birth of a low-mass protostar (Fig.~\ref{fig:seqevol}, top left panel). Initially, the collapse occurs isothermally as any compressive heating is radiated away at infrared wavelengths by dust grains thermal emission. Once the optical depth of the dust and gas mixture reaches unity as a result of the increase in density, radiative cooling becomes inefficient and a first Larson core in hydrostatic equilibrium forms. This core, of radius $\sim 5$ au, continues to contract adiabatically, eventually reaching the temperatures ($\sim$ 2000 K) required to dissociate molecular hydrogen, an endothermic process that consumes the compressive energy that was previously heating the gas. This causes a second, more violent gravitational collapse whose result is a second Larson core in hydrostatic equilibrium: the protostar, of radius $\sim 10^{-2}$~au. 

%\st{During each collapse event, leading to the formation of the first and second Larson cores, angular momentum conservation implies that part of the collapsing gas has to accelerate its rotation and start to orbit the center of mass, thus forming a disk. However, molecular clouds and star-forming cores are embedded in large-scale magnetic fields }\citep{Pattle23}, \st{which are observed to remain at least partially organized inside the protostellar cores and down to the disk scales} \citep{Galametz18, Huang24}. \st{In magnetized models of disk formation, angular momentum conservation is violated by the interaction of the partially ionized collapsing gas with the magnetic field, as discussed below. Depending on the strength of this coupling, an inner disk around the second core and an outer disk around the first Larson core can form} \citep{Machida2011}. \st{These disks can have different masses and radial extents. How these two disks interact and eventually merge to form a unique protoplanetary disk is not clear and is the main focus of the discussion in Section}~\ref{MHD}. \st{For sake of notations, we consider synonymous the terms circumstellar disk, protostellar disk and protoplanetary disks, all representing the disk of material formed around a young stellar object, in which planets may form.}
From the observational point of view, Class~0 protostars are the youngest (proto)stellar objects, observed only $t < 10^5$~yr after their formation, while most of the mass is still in the form of a dense cloud/envelope collapsing onto the central protostellar embryo \citep{Andre93, Evans09, Maury11} (see Fig.~\ref{fig:seqevol}, top right panel). This accretion process is associated with the ejection of material under the form of a fast, highly-collimated protostellar jet, accompanied by a lower-velocity, wider outflow. The Class I phase is characterized by a much thinner envelope which allows the first stellar photons to escape the system, leading to larger infrared luminosities (see Fig.~\ref{fig:seqevol}, bottom right panel). More than $50\%$ of the final stellar mass is acquired by the protostar from the envelope during the Class~0 phase alone, while $99\%$ of the final stellar mass is accreted by the end of the protostellar phase (Class 0+I), which is the stage of relatively large accretion rates ($> 10^{-6}$ \Msolperyr). When the only material remaining around the star is in the form of a circumstellar disk, the star enters the so-called Class~II phase (Young Stellar Objects - YSOs - see the bottom left panel of Fig.~\ref{fig:seqevol}). Starting from  the Class~II stage the accretion on the central object is residual (typically $<10^{-7}$ \Msolperyr), and originates directly from the disk (magnetospheric accretion): the young star is now revealed and mostly formed.

During the formation of the protostar, angular momentum conservation implies that part of the collapsing gas has to accelerate its rotation and start to orbit the center of mass, thus forming a disk\footnote{For sake of notations, we consider synonymous the terms circumstellar disk, protostellar disk, and protoplanetary disks, all representing the disk of material formed around a young stellar object, in which planets may form.}. The final angular momentum of the central star is only $\sim 1\%$ of the angular momentum of the gas initially contained in the parent protostellar core ($\sim10.000$ au in size). If angular momentum were preserved during the collapse this inevitably would lead to the formation of protostellar disks $>$~500~au in radius \citep{Hueso05, Walch09}. However, a vast majority of observations now show that most disks are compact, with radius $r<60$ au (see \citealt{Maury19, Sheehan22}, and \citealt{Tsukamoto2023} for a recent review). The small disk sizes, the detection of magnetic fields at all scales in protostellar cores \citep{Galametz18, Huang24}, and the comparison of disk observations to predictions from magnetized models \citep{Maury18, Lebreuilly24, Yen24}, suggest a disk formation scenario where magnetic fields in protostellar cores regulate the early evolution of the disk sizes. Observations also find that at least a fraction of the gas reservoir in the parent protostellar core experiences an efficient reduction of its angular momentum when transported to disk scales, as expected from magnetic braking (\citealt{Galametz20, cabedo_magnetically_2023, Gupta22}, see also \citealt{Maury2022} for a recent review).

Because of their compact nature, the young disks in Classes 0 and I are still quite poorly characterized: for example, dust masses measured for the youngest disks span a large range of values and bear large uncertainties. The current limitations are mostly due to the hypothesis made on dust emissivity to compute the dust  masses, and the lack of observations at wavelengths where the dust emission remains sufficiently optically thin to probe the full column density of dust. Despite these large uncertainties, the typical dust masses inferred from mm-observations at scales of a few 10s of au are $\sim 3-9{\rm\,M_\oplus}$ for the young (Class 0 and I) disks in Orion  \citep{Sheehan22}.

The recent advent of large surveys of Class II disks has allowed to reveal a large population of compact dusty disks \citep[see, e.g.][for recent reviews]{Miotello23, Manara23} with sizes similar to those of the youngest protostellar progenitors. These surveys have confirmed that the dust populating the disks around $\sim 1-2$ Myrs old Class II Young Stellar Objects (YSOs) is different from the one in the diffuse ISM in terms of spectral index (the slope of the intensity of radiation as a function of wavelength) of their emitted light. {This} is interpreted as a signature {that  dust particles in disks are significantly larger and} suggests that a substantial fraction of the {original} dust reservoir may have already been converted {to pebbles-size (mm to cm) and even planetesimal-size (tens of km) objects}  \citep{Natta2007, Birnstiel2010}, during the early disk phases.  The total mass of the dust, estimated from the emission of mm-size grains observed in the mid-plane and extrapolated over the inferred grain size distribution is  $\sim 3-6 {\rm\,M_\oplus}$ for the Class II disks in Ophiuchus and in Orion  \citep{Williams2019, Encalada2021}. The largest of these evolved disks, when they are spatially resolved, often show structures in both their dust and gas spatial distribution, such as rings, gaps, spirals and asymetries \citep{Andrews2018, Pinte2023, Miotello23}.

\subsection{Cosmochemistry}

Asteroids are relevant to the study planet formation because they are the leftovers of the original population of so-called 'planetesimals', out of which the planets are thought to have formed. In addition, they have preserved, to a certain extent, the original dusty material that made up the protosolar nebula.

\begin{figure*}[t!]
    \centering
    \includegraphics[width=0.95\linewidth]{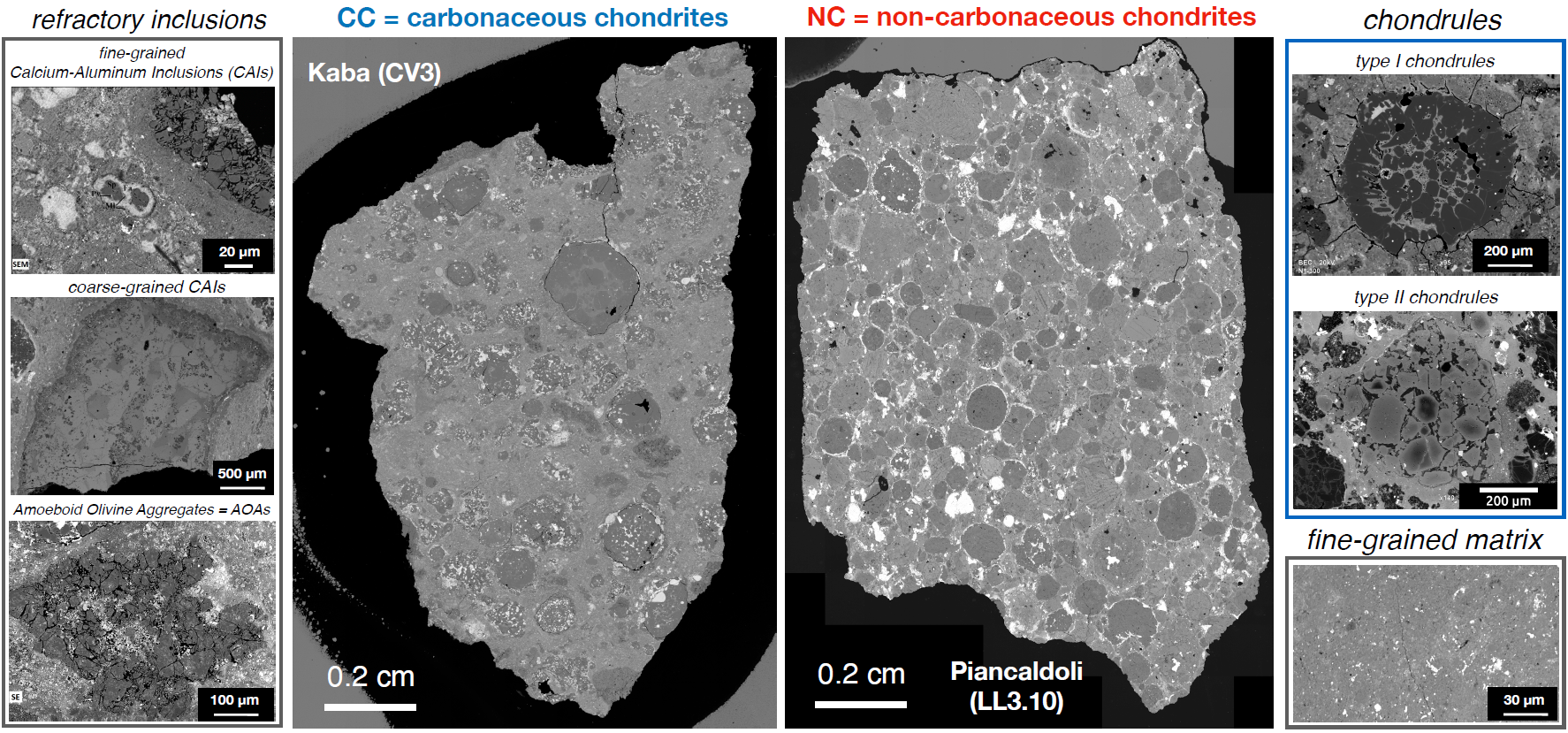}
    \caption{Back-scattered electron images of refractory inclusions (fine-grained CAIs, coarse-grained CAIs, AOAs), two representative examples of carbonaceous chondrite and non-carbonaceous chondrites, type I and type II chondrules and fine-grained matrix}
    \label{fig:chondrites}
\end{figure*}

Meteorites are asteroidal fragments recovered on Earth's surface promptly after their fall or during field trips to cold/hot deserts. Although rare meteorites could correspond to fragments of Trojan asteroids (the objects sharing the orbit of Jupiter; \citealt{marrocchi_tarda_2021}), most meteorites originate from the main asteroid belt \citep{colas2020}. At first order, meteorites are divided into two overall categories:

\begin{itemize}
\item Undifferentiated meteorites, referred to as chondrites, which result from the mechanical agglomeration of small constituents formed at different times and locations during the evolution of the protoplanetary disk \citep{scott_chondrites_2014,piralla_unified_2023}. Chondrites are composed of $\mu m$- to cm-size refractory inclusions, sub-millimeter to mm-scale chondrules (enigmatic grains issued from the crystallization of molten or partially molten silicate-rich droplets in the disk) and a  matrix of fine-grained, thermally unprocessed material, rich in carbon and volatile elements (Fig.~\ref{fig:chondrites}).
These three components can be considered representative of the grains existing in the disk when these meteorites formed.
In turn, the refractory inclusions in chondrites are in the form of (i) calcium-aluminum-rich inclusions (CAIs) and (ii) amoeboid olivine aggregates (AOAs; Fig.~\ref{fig:chondrites}). Given the small sizes, all these components will be generically referred to as "dust" hereafter, even though only the matrix is representative of the porous dust aggregates often envisioned to exist in a protoplanetary disk; instead, refractory inclusions are condensates of a hot gas, whereas chondrules are most likely the product of the repetitive magmatic events that have affected chondrule precursors \citep{jones_petrographic_2012, marrocchi_formation_2019}.  Chondrites never experienced melting after their agglomeration and thus represent key witnesses of the {primordial} dust existing in the solar protoplanetary disk {throughout} its formation and evolution. Chondrites are themselves divided in different groups, depending on their specific petrological and chemical properties, and into types, depending on their thermal or hydrous alteration. At a lower level of resolution, the different groups can be broadly {categorized} into the carbonaceous (CC) and non-carbonaceous (NC) classes, the former having a more abundant carbon-rich matrix (Fig.~\ref{fig:chondrites}).  
\item {Differentiated meteorites, or achondrites, which originate from planetesimals (i.e., bodies larger than $\sim 10$ km) that melted internally.} Due to the different densities, the metal and silicate separated, with the metal sinking to the center of the object, forming a metallic core, and the silicate remaining in the mantle and crust. This melting and metal-silicate differentiation (i.e. segregation) erased the primitive nature of the dust that the differentiated planetesimals formed from. These processes result in the formation of stony and iron meteorites thought to respectively sample the crust and the core of differentiated bodies. An additional type of differentiated meteorites composed of nearly equal part of iron and silicates also exists, called stony-iron meteorites. Their formation is thought to be related to collisional impacts between bodies rather than the incomplete differentation of an original planetesimal due to partial internal melting \citep{windmill_isotopic_2022}.
\end{itemize}

When a planetesimal forms by accretion of dust, its {internal temperature is somewhat higher than the local temperature in the disk due to release of gravitational energy during accretion. The temperature difference is not expected to be very significant (a few degrees), even in the case of instantaneous accretion. However}, if the planetesimal contains radioactive elements, the decay of these elements releases energy. This heats up the planetesimal {interior because it takes time for the released energy to diffuse to the planetesimal surface where it is radiated away. The maximal internal temperature is obtained when the energy irradiated away by the surface equals that released by the radioactive decay}. A number of short-lived and long-lived radioactive elements existed in the material comprising the planetesimals. Among these, the most important in terms of energy budget was $^{26}$Al, which decays to ${}^{26}$Mg with a half-life of 0.717 Myr. Of course, all $^{26}$Al has disappeared by now, but its original existence is proven by the observation of a correlation between $^{26}$Mg/$^{24}$Mg and $^{27}$Al/$^{24}$Mg in the various minerals. In fact, $^{27}$Al and $^{24}$Mg are stable isotopes of Al and Mg, so that $^{27}$Al/$^{24}$Mg is representative of the Al/Mg ratio at the time when a given dust formed; the correlation of $^{26}$Mg/$^{24}$Mg with the Al/Mg ratio reveals that a fraction of Al was in $^{26}$Al and the original $^{26}$Al/$^{27}$Al ratio sets the slope of the correlation \citep{lee_demonstration_1976}. 

The maximal temperature that a planetesimal reaches in its interior depends weakly on its size (unless the body is {smaller than $\sim 10$~km}) but is very sensitive to the initial content of $^{26}$Al. Thus, a planetesimal that formed early, when a large amount of $^{26}$Al was still present in the disk, reached a temperature large enough to melt; instead, a planetesimal that formed late, after most of $^{26}$Al had already decayed in the disk, did not heat up considerably. Defining "parent body" the planetesimal from which a given meteorite comes from, we can conclude that the parent bodies of differentiated meteorites originated earlier than the parent bodies of chondrites. 

This conclusion is confirmed by radioactive chronometers that date the differentiation of the parent bodies of achondrites or the formation of chondrules, the main constituents of chondrites \citep{barrat_4565-my-old_2021, kleine_chronology_2012, kruijer_protracted_2014, kruijer_age_2017, fukuda_temporal_2022, piralla_unified_2023}.  A radioactive chronometer is provided by a pair of elements of which one, the parent, is radioactive with a known decay constant and the other is the daughter product. These two elements are separated when some event occurs: for instance metal-silicate differentiation separates Hf from W, evaporation/condensation separates U from Pb and melting and crystallization mobilizes Al relative to Mg. Each of these pairs of elements can therefore be used to date a different event. The U-Pb system is particularly important because there are two radioactive isotopes of U, ${}^{238}$U (half-life = 4.468 Gyr) and ${}^{235}$U (half-life = 0.7038 Gyr), which decay respectively to ${}^{206}$Pb and ${}^{207}$Pb. This dual system allows to determine the absolute age of fractionation of Pb from U, without knowing the initial abundances of the Pb isotopes (but still requiring the initial uranium isotope ratio to be determined; \citealt{brennecka_238u/235u_2010}). {The use of $^{26}$Al-$^{26}$Mg as a chronometer instead requires that the homogeneous distribution of $^{26}$Al in the circumsolar disk and to know the initial $^{26}$Al/$^{27}$Al (assumed to be the value measured in the oldest CAIs), while the use of the $^{182}$Hf-$^{182}$W  to date metal-silicate differentiation of a planetesimal requires to know the initial $^{182}$Hf/$^{183}$Hf and $^{182}$W/$^{184}$W (again, assumed to be those measured in CAIs) and the Hf/W ratio (assumed to be the one measured in chondrites).  }

The U-Pb chronometer shows that CAIs are the first minerals that formed in the Solar system, 4.5672 \citep{connelly_absolute_2012} or 4.5687 Gyr ago \citep{bouvier_age_2010, desch_statistical_2023, piralla_unified_2023}. This time is defined as time 0 of the whole cosmochemical chronological sequence. Chondrules instead formed in a time period ranging from $\sim 0.7$ to $\sim 4$ Myr after the formation of the CAIs \citep{bollard_early_2017, budde_hf-w_2018, fukuda_temporal_2022, piralla_unified_2023}. 

The time of differentiation of the parent bodies of achondrites can be dated using the Hf-W system;  the formation age of these parent bodies, which has to predate their differentiation, is then deduced using thermal evolution models \citep[e.g.,][]{kruijer_protracted_2014}. The results show that these objects all formed within one million years from time 0 \citep{spitzer_isotopic_2020,spitzer_nucleosynthetic_2021}.

The actual formation of the parent bodies of chondrites cannot be dated by any chronometer because the dust accretion is a low-temperature mechanical process that does not alter isotopic ratios. However, because chondrules formed in the disk and chrondrites are mostly made of chondrules, it is obvious that their parent bodies had to form after the chondrules that they contain, i.e. a few millions of years after CAIs (ages ranging from $\sim 2$ to $\sim 4$  {Myr } depending on the chondrite group).  This confirms that achondrites and chondrites are samples of two generations of planetesimals that assembled at distinct epochs in the protoplanetary disk ("early" and "late", respectively), as expected from their radically different thermal evolutions.

\begin{figure}[h!]
    \centering
    \includegraphics[width=0.9\linewidth]{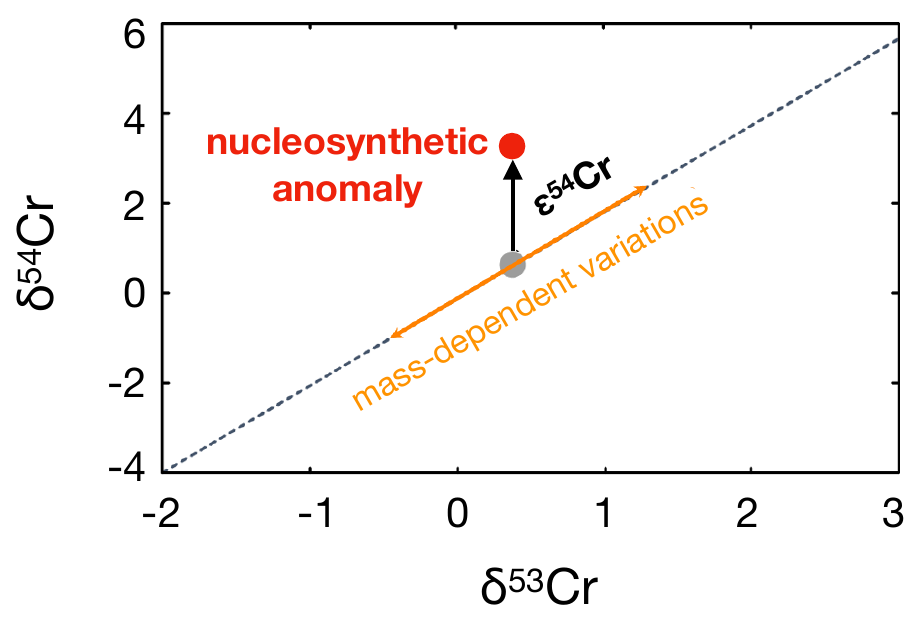}\\
    \includegraphics[width=0.9\linewidth]{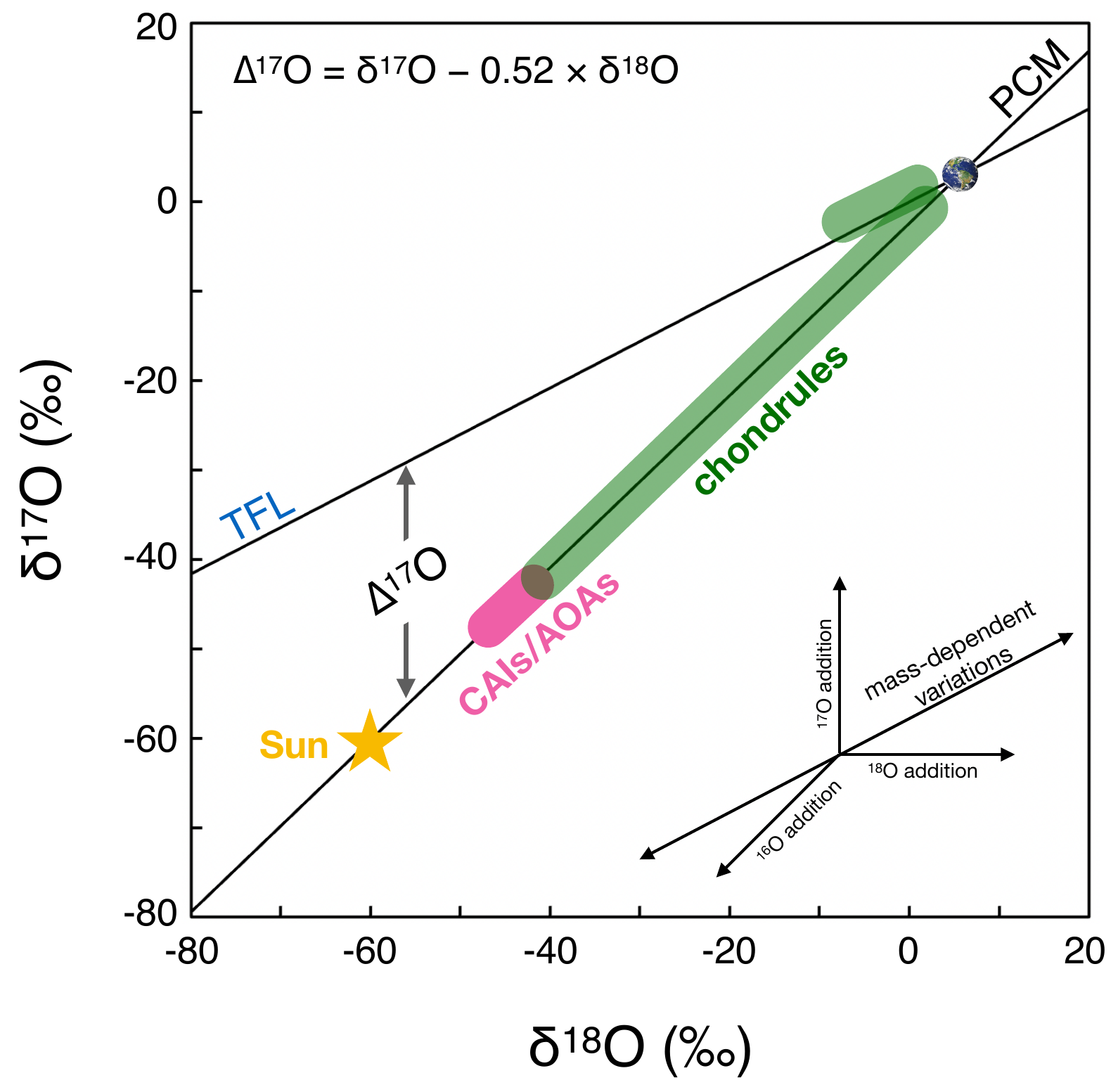}
    \caption{Top: Chromium three-isotope plot showing the line defined by mass-dependent variations. $^{54}$Cr-excesses relative to this line cannot be explained by any physico-chemical processes and thus correspond a mass-independent anomaly. Bottom: the same but for the oxygen three-isotope plot. Terrestrial rocks all plot along a 1/2 slope line passing through (0,0), called terrestrial fractionation line (TFL). Again, any deviation from this line corresponds to an mass-independent anomaly. The plot also shows the composition of the Sun, of refractory inclusions (CAIs and AOAs) and chondrules, which are all anomalous with respect to Earth. Notice that meteorites and chondritic constituents plot along a line of slope $\sim 1$, dubbed Primitive Chondrule Minerals (PCM) line, \cite{ushikubo_primordial_2012}. }
    \label{3-isotopes}
\end{figure}

\subsubsection{Isotopic anomalies}\label{sec-anomaly}
Meteorites and their components also contain a variety of isotopes of a given element that are stable (i.e. they do not decay). The ratio of these isotopes provide numerous pieces of information. 

Some evolutionary (physico-chemical) processes separate isotopes according to their mass difference (mass-dependent variations). For instance, O has three isotopes: $^{16}$O, $^{17}$O, $^{18}$O, (see Fig.~\ref{3-isotopes}, bottom). Any physico-chemical process (e.g. evaporation, crystallization, chemical reactions etc.) would evolve the isotope ratios $^{17}$O/$^{16}$O vs. $^{18}$O/$^{16}$O along a line of slope 1/2 (as the difference in mass between  $^{17}$O and $^{16}$O is 1/2 - precisely 0.52- of that between $^{18}$O and $^{16}$O; see Fig.~\ref{3-isotopes}, bottom). The position of different minerals along such a line is a measure of how affected these minerals have been by these mass-dependent fractionation processes. Due to the small isotopic variations commonly observed within samples, the isotopic compositions are commonly expressed in {$\delta$} notation, which corresponds to the relative difference between the isotopic ratio in a sample and that in a reference standard (for oxygen isotopes the standard is mean ocean water, dubbed SMOW; for chromium it is NIST SRM3112a). For instance, the oxygen isotopic composition is expressed as follows: 
%\newline
\begin{equation}\label{eq:delta-notation}
    \delta^{17,18}{\rm O} = \left[ \frac{\left(\frac{^{17,18}{\rm O}}{^{16}{\rm O}}\right)_{\texttt{Sample}}}{\left(\frac{^{17,18}{\rm O}}{^{16}{\rm O}}\right)_{\texttt{SMOW}}}-1 \right] \; \times \;10^3.
\end{equation}

The deviation from the mass-dependent line reveals that the object in consideration has anomalous isotopic ratios, in the sense that there are no evolutionary processes that can link them to those of the reference standard, given that the latter can only change isotopic ratios along a mass-dependent slope line (1/2 slope for O). In other words, the object and the standard must have accreted materials with distinct isotopic signatures. These anomalies are called mass-independent, to distinguish them from the more trivial ones along the mass-dependent slope line. These anomalies therefore reveal information about the original material from which the planetesimals form.

To quantify the mass-independent anomalies, it is customary to measure on the three-isotope plot the distance of the considered object along the $y$-axis from the mass-dependent slope line passing through the standard, as indicated in Fig.~\ref{3-isotopes}, top panel. For oxygen, this is denoted by 
\begin{equation}\label{eq:delta-notation}
    \Delta^{17}{\rm O} = \delta^{17}{\rm O} -0.52 \times \delta^{18}{\rm O}.
\end{equation}

For elements other than oxygen the situation is analog. Chromium, for instance, has three isotopes $^{52}$Cr, $^{53}$Cr, $^{54}$Cr. For historical reasons \citep{trinquier_widespread_2007,trinquier_origin_2009} the y-axis reports the ratio $^{54}$Cr/$^{52}$Cr and the x-axis the $^{53}$Cr/$^{52}$Cr ratio, so the mass-dependent fractionation line has slope 2 (instead of 1/2 as for oxygen) and the anomaly is defined as
\begin{equation}\label{eq:eps-notation}
    \epsilon^{54}{\rm Cr} =  (\delta^{54}{\rm Cr} -2 \times \delta^{53}{\rm Cr})\times 10 \ .
\end{equation}
Note the multiplicative factor given that the mass-independent anomalies are much smaller than for oxygen. 

Mass-independent anomalies can be of nucleosynthetic origin or due to the selective photo-dissociation of parent molecules before their accretion into planetesimals. Nucleosynthetic anomalies correspond to the heterogeneous distribution of grains inherited from the ISM characterized by either enrichment or depletion in nuclides derived from multiple nucleosynthetic sources, such as irradiation in the envelopes of evolved stars, supernova, and kilonova explosions, reactions in plasmas. A priori, such heterogeneous distribution can be the consequence of accretion of materials with different interstellar origins in different parts of the disk, or the evaporation of some presolar grains in the hotter parts of the disk. The mass independent anomalies of Cr, Ti, Ca, Fe, Mo, Zn, Nd, Zr, Ru are all of nucleosynthetic origin. Mass-independent anomalies for oxygen may instead be due to the preferential photo-dissociation of the C$^{17}$O and C$^{18}$O molecules with respect to C$^{16}$O (the latter being more abundant and therefore optically thicker; \citealt{lyons_photochemical_2005}). The free $^{17}$O and $^{18}$O atoms are then incorporated (e.g. in water or silicates) in different proportions in different parts of the disk, distributing different materials along the Primitive Chondrule Minerals (PCM) line (Fig.~\ref{3-isotopes}, bottom).  Similarly, $^{15}$N enrichment over $^{14}$N in NH$_3$ or HCN can come from the preferential dissociation of $^{15}$N$^{14}$N over $^{14}$N$_2$.

\begin{figure}[t]
    \centering
    \includegraphics[width=\linewidth]{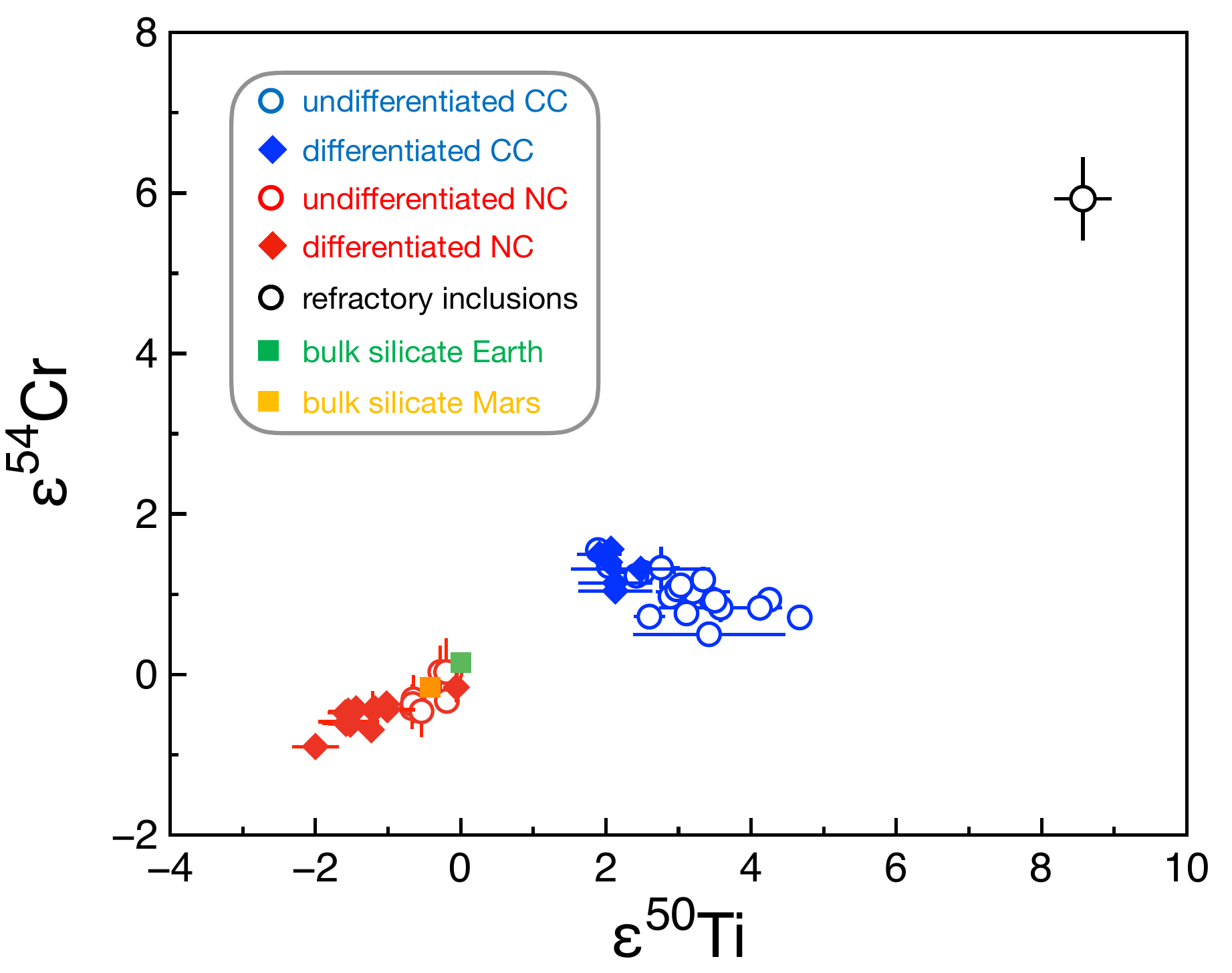}
    \caption{Nucleosynthetic isotopic anomalies in ${}^{50}$Ti and ${}^{54}$Cr of refractory inclusions (CAIs and AOAs), CC and NC meteorites (both undifferentiated and differentiated) and the silicate parts of Earth (BSE) and Mars (BSM). Data from \cite{burkhardt_terrestrial_2021} and references therein}
    \label{fig:anomalies}
\end{figure}

The improvements in the precision and accuracy of cosmochemical analyses have recently revealed the existence of an overarching dichotomy of nucleosynthetic isotopic anomalies between carbonaceous and non-carbonaceous chondrites (CC and NC respectively, see Fig. \ref{fig:anomalies}. Initially inferred from the Ti and Cr isotopic compositions of NC and C chondrites \citep{trinquier_widespread_2007, trinquier_origin_2009, warren_stable-isotopic_2011}, this dichotomy now extends to other isotopic systems (e.g. Ni, Mo, Zn isotopes) \citep{kruijer_age_2017, nanne_origin_2019, brennecka_astronomical_2020, steller_nucleosynthetic_2022, savage_zinc_2022}. Importantly, this dichotomy holds also for differentiated meteorites \citep{kruijer_age_2017}, which has led to redefine the CC and NC classes on the basis of this dichotomy, so as to include both differentiated and undifferentiated meteorites. The dichotomy is interpreted as an indication that meteorite parent bodies derive from two genetically distinct reservoirs. Because CC chondrites are rich in water and other volatile elements, it is generally accepted that they formed farther from the Sun than NC meteorites.

\begin{figure*}[h!]
\includegraphics[width=0.95\textwidth]{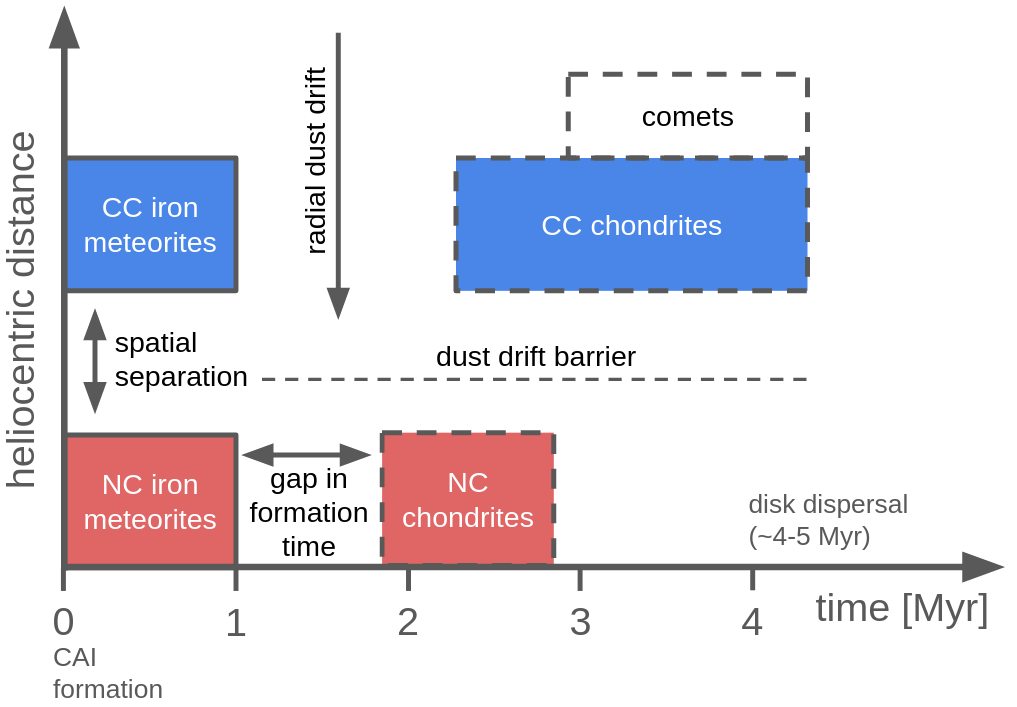}
\caption{{{{
{Schematic of times and distances of formation inferred for the major meteorite classes (time zero is defined as the time of CAI formation). Both NC and CC iron meteorite (achondrites) parent bodies formed contemporaneously at different heliocentric distances. To prevent the mixing of the two isotopic reservoirs some barrier against the radial drift of dust must have formed (see Sect.~\ref{trapping}). The chondrites formed later, after an apparent temporal gap relative to iron meteorite parent bodies, and up to the time of disk dispersal at ~4-5~Myr. Comets formed beyond the CC chondrites but it is unknown what their refractory isotopic composition is and therefore if they belong to the NC or CC group (see Sect.~\ref{comets}).}
}}}
}
\label{fig:nc-cc-scheme}
\end{figure*}

Combining this isotopic classification with the chronological constraints discussed before, Fig.~\ref{fig:nc-cc-scheme} provides a schematic synthesis, which highlights the very existence of distinct isotopic reservoirs and their preservation over millions of years. They provide terrific constraints on the formation and evolution of the protosolar disk, as will be discussed in the forthcoming sections of this paper.  
%, with the NC and CC reservoirs respectively representing the inner and outer regions of the solar system \cite{kleine_non-carbonaceouscarbonaceous_2020}. In combination with the accretion timescale of meteorite parent bodies (\cite{kruijer_age_2017}), this allows integrating meteorite constraints into large-scale disk models with the barrier separating the disk into two reservoirs being attributed to the early formation of Jupiter’s core (\cite{kruijer_age_2017}), a long-lived pressure maximum (\cite{brasser_partitioning_2020}), or evolving ice and silicate lines in the protoplanetary disk (\cite{lichtenberg_bifurcation_2021}).  

It is important to note that the isotopic compositional differences between NC and CC meteorites are of the order of one part per 10,000 or less. Apart from these tiny, but significant differences, the Solar system appears extremely homogeneous, which supports the idea that the abundance of short-lived radioactive nuclei, such as $^{26}$Al, was about the same everywhere in the disk \citep{desch_statistical_2023}, leading to the same heating regardless of where the planetesimals formed \citep{spitzer_nucleosynthetic_2021}. This is confirmed by more specific analyses on the coherence of various radioactive chronometers \citep{nyquist_distribution_2009, desch_statistical_2023, desch_statistical_2023-1, piralla_unified_2023} that would not be possible if the radioactive elements were initially distributed in a grossly inhomogeneous way.

\section{The disk-building phase (Class 0 \& Class I protostars)}

In this section we discuss the early stages of protoplanetary disks, and the protosolar disk in particular, as they form by accretion of material from the parent envelope and start to evolve. These disk-building processes occur as the star-disk system evolves through the Class 0 and Class I phases. We first address cosmochemical evidence for the transport of dust from the inner to the outer part of the protosolar disk. We then discuss whether this transport could have occurred during the radial spreading of the disk or {via} the protostellar outflow. Finally we discuss how the isotopic dichotomy of the meteorites (see Sect.~\ref{sec-anomaly}), as well as the strong differences in isotopic properties of the volatile elements between comets, asteroids, and the Sun, constrain where material was continuously accreted from the envelope into the disk. We compare this evidence with the results of modern numerical simulations.  

\subsection{Evidence of dust transport from the inner to the outer disk}\label{sec:dust transport}

CAIs are iconic objects representing the oldest solids formed in the Solar system, with estimated radiometric formation ages ranging from 4,567.2 to 4,568.7 Myr {ago}  \citep{bouvier_age_2010, connelly_absolute_2012, desch_statistical_2023, piralla_unified_2023}. They are composed of refractory minerals that are interpreted to be high-temperature equilibrium condensates from a gas of solar composition \citep{larimer_chemical_1967, grossman_condensation_1972}. The fact that the value of $\Delta^{17}$O of CAIs is very close to that of the Sun (Fig.~\ref{3-isotopes}) confirms this interpretation. The discovery of the original presence of the short-lived isotope ${}^{10}$Be (half-life of 1.387 Myr) in CAIs has been interpreted as evidence of strong irradiation caused by solar flares \citep{gounelle_irradiation_2006}. The nature of the flares (gradual vs. impulsive), the target (gas vs. dust), the duration, and the location (molecular cloud vs. protoplanetary disk) remain largely unconstrained. Nevertheless, the detection of ${}^{7}$Be (half-life of 53 days, \citealt{chaussidon_li_2006}) and the correlated excesses of ${}^{10}$Be and ${}^{50}$V \citep{sossi_early_2017} were interpreted as evidence of in-situ irradiation of solid refractory dust at $\sim 0.1$ au from the protosun. More recent geochemical measurements and analytical calculations suggest that CAI condensation could have occurred at larger heliocentric distance up to 1 au \citep{jacquet_beryllium-10_2019, bekaert_fossil_2021, fukuda_irradiation_2021, dunham_uniform_2022}. In any case, a consensus exists that CAIs formed in the inner disk and early in disk history.

This observation has profound implications on our understanding of the protosolar disk evolution, because CAIs are mostly observed in carbonaceous chondrites which, due to the presence of pervasive water alteration, are considered to have accreted in the outer Solar system (i.e. beyond the so-called "snowline" and possibly beyond Jupiter's orbit, \citealt{kruijer_age_2017}). This implies that CAIs must somehow have been transported from the inner protosolar disk to the outer and colder regions, where they remained stored for several Myr before having been accreted in planetesimals \citep{desch_effect_2018, jongejan_effect_2023}. Other pieces of evidence of outward transport of dust arise from the presence of (i) CAIs and chondrules in cometary samples \citep{zolensky_mineralogy_2006, nakamura_chondrulelike_2008, ogliore_incorporation_2012} and (ii) AOA-like materials in the CI chondrites, a peculiar chondrule-free carbonaceous chondrite also thought to have formed in the outer part of the disk (i.e. >~15~au, \citealt{desch_effect_2018, morin_16o-rich_2022}).

Although such observations clearly attest that outward dust transport occurred during the early evolution of the protosolar disk, they are mostly based on the characterization of cm-sized coarse-grained CAIs that experienced subsequent melting after their formation \citep{krot_refractory_2019}. This could be fundamental, because lifting and/or transporting cm-sized dust would set a strong constraint on the disk dynamics. However, this overlooks the wide variety of much smaller refractory inclusions present in chondrites. In fact, in addition to cm-sized CAIs, unmelted refractory inclusions also exist in the forms of:

\begin{enumerate}
	\item Small CAI-like monomers of 10-30 $\muup$m-sized dispersed throughout the chondritic matrices and composed of layered refractory minerals \citep[e.g.][Fig.~\ref{fig:chondrites}]{macpherson_fluffy_1984}. They are also found in NC chondrites \citep{Haugbolle2019}.
	\item Fine-grained "fluffy" CAIs resulting from mechanical agglomeration of individual CAI-like monomers (e.g. \cite{kawasaki_variations_2019}). 
	\item Fine-grained aggregates of olivine grains associated with variable proportions of CAI-like nodules referred to as AOAs (Fig.~\ref{fig:chondrites}). Notably, AOAs represent the most abundant type of refractory inclusions \citep{krot_amoeboid_2004} and are considered to have condensed at lower temperature than CAIs \citep{ruzicka_amoeboid_2012} in a turbulent disk with strong thermal heterogeneities \citep{marrocchi_rapid_2019}. 
\end{enumerate}

All refractory inclusions together represent up to  $10~\%$ of the mass in carbonaceous chondrites while only up to $0.1~\%$ in non-carbonaceous chondrites.
The large majority of refractory inclusions are smaller than the commonly considered CAIs and most of them escaped the later melting event(s) recorded by large igneous CAIs.

A fundamental constraint stands in the similar ${}^{26}$Al formation ages of both unmelted and igneous refractory inclusions \citep{macpherson_well-resolved_2012, krot_refractory_2019, kawasaki_variations_2019}. This implies that condensation, transport, and melting processes occurred almost contemporaneously, within $\sim 0.2$ Myr during the initial disk history \citep{kawasaki_variations_2019, marrocchi_rapid_2019}, in a region where silicates had not condensed (otherwise CAIs would be polluted and appear as AOAs).
	
\subsection{Rapid radial spreading of the disk as a mechanism of transport}
\label{spreading}

This section discusses our theoretical understanding of how the disk could rapidly expand in the radial direction, promoting the transport of dust from the vicinity of the Sun to large distances. First, we present the results of one-dimensional (1D) idealized models, which show {the necessary} conditions {that} need to be satisfied for the radial transport of dust to be effective. Then, we present the results of more comprehensive two-(2D) and three-dimensional (3D) magneto-hydrodynamic (MHD) models, which however have a poorer spatial resolution and are limited to shorter timescales; these models give important indications to discuss the realism of the conditions unveiled in 1D models.  

%\ab{Relevant literature to be checked: https://arxiv.org/abs/2404.15715}\\
%\am{Also (but for Class II disks): https://ui.adsabs.harvard.edu/abs/2023ApJ...954...41T/abstract}

\subsubsection{1D models of disks spreading}\label{1D}

Although {simplistic}, 1D disk models are instructive because they can easily explore different scenarios and the effects of various parameters. Despite the differences from model to model, most {have in common to} simulate the radial evolution of the gas under the effect of its own viscosity. The latter is usually parameterized with the $\alpha$-prescription (\citealt{Shakura1973}) which, from dimensional arguments, assumes that the gas viscosity $\nu$ is proportional to $H^2\Omega$, where $H$ is the pressure scale-height of the disk (depending on local temperature in the vertically-isothermal approximation) and $\Omega$ is the local orbital frequency. The coefficient of proportionality is called $\alpha$ and its value sets the evolution of the disk. The temperature in the disk is usually computed by balancing viscous heating on the midplane, vertical transport of energy from the midplane to the disk's surface (dependent on the assumed disk's opacity) and black body emission from the surface to space. When the computed temperature drops below the value predicted by a simple model of disk heating by stellar irradiation \citep{ChiangGoldreich}, it is reset to that value. The dust is initially introduced together with the gas as microscopic particles, with a given gas-to-dust mass ratio (typically 100, using an averaged estimate found in the ISM). The growth rate of dust is computed at every distance from a simple statistical collisional model, assuming perfect coagulation until a fragmentation velocity threshold is reached. Most 1D codes consider a single dust size at each distance, evolving over time. In turn, the size of the particles sets the Stokes number, which governs the coupling between gas and particles due to gas drag. Turbulent diffusion is taken into account to compute the scale height of the particle layer relative to the pressure scale height of the gas and to compute the collisional velocity among dust particles (see for instance \cite{Drazkowska2016} for a detailed description of one of such codes).  

The first 1D model of formation and evolution of a protoplanetary disk using the $\alpha$-prescription and accounting for the infall of gas from the molecular cloud, was presented in \citet{Hueso05}. The infall of gas from the protostellar core into the disk was treated as a source term, following the model developed in the seminal work of Frank Shu \citep{Shu77}. That model assumed a cloud in rigid rotation and the conservation of angular momentum during the infall (following \citealt{Ulrich1976ApJ}). It shows that different shells of the cloud infall sequentially in order of increasing angular momentum. Thus, the earliest material that reaches the disk has the lowest angular momentum (coming from the molecular cloud's deepest regions) and falls in the disk near the star while, at a later time, high angular momentum material reaches the disk at a larger distance. The distance in the disk where the material falls is called \emph{centrifugal radius} $R_c$ and corresponds to the place where the rotationally supported disk has the same angular momentum of the infalling material. Different recipes for $R_c(t)$ can be introduced, from different assumptions on the angular momentum distribution in the parent protostellar core, and can be adapted to mimic the possibility that part of the core's angular momentum is removed by magnetic braking during the collapse \citep{Galli06}. 

\citet{Hueso05} used Shu's recipe for the growth of the centrifugal radius and showed that the disk spreads very rapidly in the outward direction because of its viscosity and of the very steep surface density gradient due to the strong concentration of material close to the star \citep{1974MNRAS.168..603L}. Thus, even if $R_c(t)$ grows over time, it is always smaller than the actual disk's outer radius. 
 
\cite{Dullemond_Apai_2006} and \cite{YangCiesla2012} expanded the Hueso and Guillot approach by introducing the coupling between dust and gas. They showed that dust is initially transported outward very efficiently, because of its initial small size and strong coupling with the gas and the vigorous radial expansion of the early gas disk. These results suggested the first explanation for the higher abundance of CAIs in outer disk planetesimals (namely in carbonaceous chondrites) relative to inner disk planetesimals (i.e. in non-carbonaceous chondrites).

\begin{figure}[t]
    \centering
    \includegraphics[width=\linewidth]{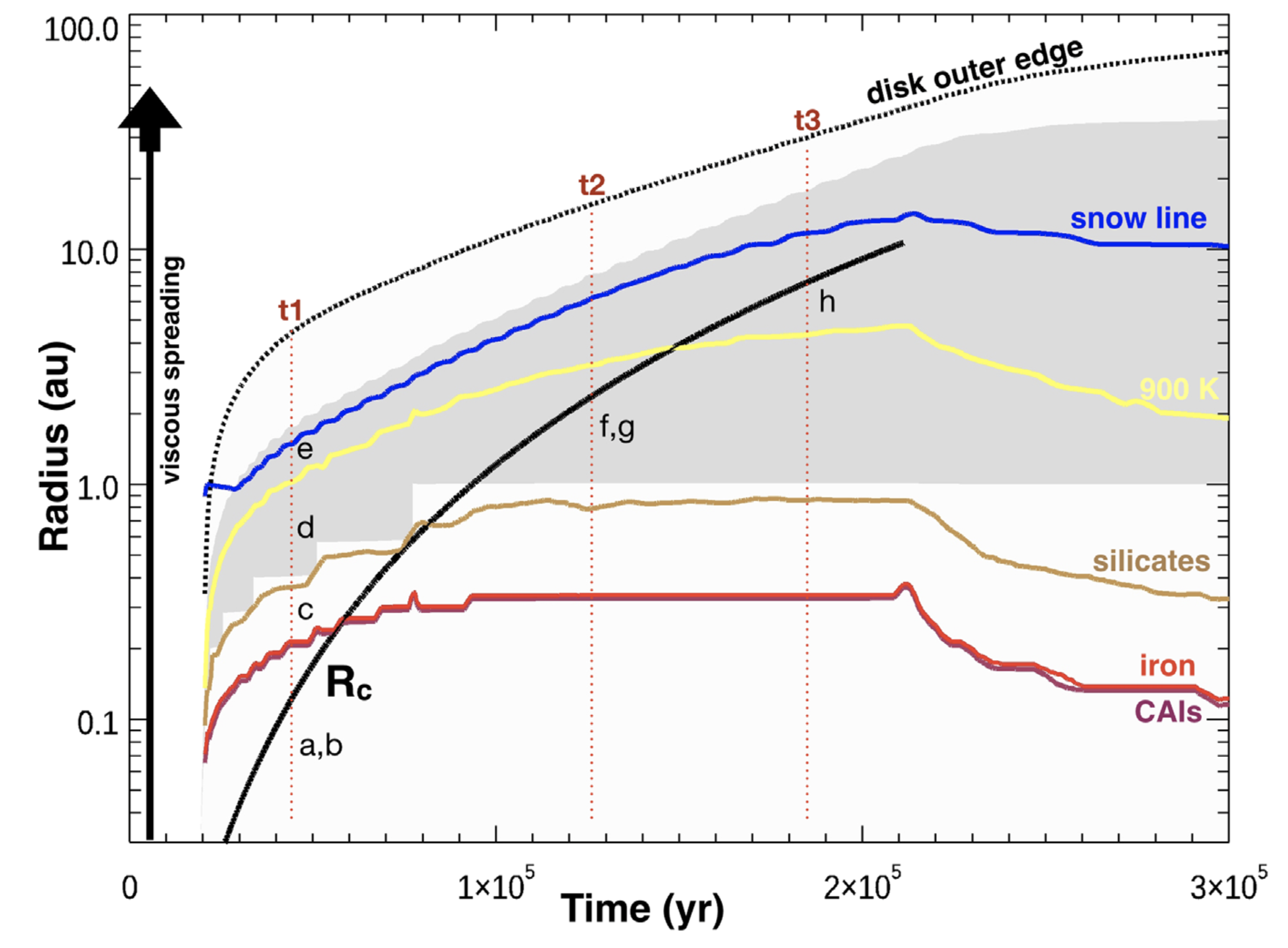}
    \caption{ Time evolution of the centrifugal radius (assuming the canonical model of \citealt{Shu77}), the locations of the disk's outer edge and of different condensation fronts, as labeled. The grey zone denotes a region where $\alpha$ is assumed to be lower than in the other parts of the disk (a.k.a. dead-zone). Processes taking place at three snapshots are indicated. At t=50 kyr (t1), $R_c(t)$ is within the condensation fronts of the most refractory species (Ca and Al, Fe) so that all the ISM material injected in the disk is vaporized (a,b). At the same time, farther out in the disk, condensation of CAIs and iron (c) and silicates (d) takes place from a gas which is spreading outward. Iron condensed and transported with the spreading gas can be processed farther out (e). At 120 kyr (t2), $R_c(t)$ has reached a region in the disk where the temperature is below 1500 K. Presolar refractory material and silicates can be injected into the disk without being vaporized (f) but they can experience thermal processing (g). At 180 kyr (t3), the material from the parent cloud is injected into the disk's colder regions, likely retaining its primordial composition, because the temperature is too low for thermal processing (h). Figure adapted from \cite{pignatale_making_2018}. }
    \label{fig:condensations_fronts}
\end{figure}

\cite{pignatale_making_2018} and \cite{pignatale_fingerprints_2019} then introduced chemical considerations, and considered several different dust and gas species that condense at different temperatures. They showed that, as the gas spreads outwards it cools, so it can condense in sequence first the most refractory elements (Al,Si,Ca, the constituents of CAIs, condensing when the temperature drops below 1650~K) which then, if still at equilibium with the gas, transform into refractory silicates akin to AOAs when the temperature drops below 1500~K. Similarly, iron in gas condenses in solid iron below 1500~K; silicate rich gas condenses in minerals below 1000~K; H$_2$O condenses when the temperature drops below 150~K. At the same time, because the disk grows in mass and becomes optically thicker and viscous dissipation is enhanced, the temperature at every given radius increases. Consequently, the various condensation fronts discussed above move outward (Fig.~\ref{fig:condensations_fronts}). Pignatale et al. showed that only in the first 50 Kyr CAIs can condense from the disk because at later time most of the disk is fed beyond the Ca-Al condensation front. However olivine-rich and iron-rich minerals can continue to condense until a later time. Moreover, as in the model by Yang and Ciesla, the CAIs are efficiently transported outwards in the disk's radial expansion. When the infall of fresh material on the disk ends, \cite{pignatale_making_2018}  showed that the disk becomes an accretion disk (i.e. the radial motion of the gas is directed inwards) and dust also flows inwards (following the gas radial motion but also because their azimuthal velocity is decelerated by the drag exerted by the gas in sub-Keplerian rotation). 

\cite{Marschall2023} revisited the Pignatale et al. model with several different assumptions, but reaching overall qualitatively similar results. They emphasized the importance that $R_c$ remains small (i.e. within 1~au) during the entire infall period -- which could be the case if magnetic breaking reduces the angular momentum of the infalling material \citep{hennebelle_magnetically_2016} -- to sustain the radial expansion of the disk for longer and transport {\it all} CAIs into the outer disk.

{We stress that all the viscous 1D models discussed here predict a smooth and slow evolution of the disk's temperature (e.g. Fig.~\ref{fig:condensations_fronts}) and of the stellar accretion rate. In reality, stars are observed to undergo episodic massive accretion events, known as FU Ori events, and the thermal histories of CAIs, AOAs and chondrules show rapid cooling, which implies that the disk was sometime, at least locally, out of thermal equilibrium. Such rapid changes in disk's conditions are completely missed in 1D models. Rapid cooling may have profound effects on the mineralogy of the condensed dust \citep{Charnoz-condensates}. The effects on the global distribution of dust and planetesimal formation are yet unclear.}

\subsubsection{Is the vigorous radial expansion of the disk realistic? What MHD simulations say}
\label{MHD}
Numerical simulations investigating the inner and outer disk formation around the two Larson cores have shown a large variety of outcomes, which heavily depend on the physics and the numerics used in these experiments. We refer the readers to \citet{Teyssier2019} for a review on the  methods used in disk formation numerical experiments, and to \citet{Machida2014} and \citet{Hennebelle2020} for a discussion on the effect of the initial conditions and of the sink particle/cell models.

As discussed in Sect.~\ref{intro-star}, stellar birth occurs via the formation of a hydrostatic core (named first Larson core), which then collapses when H$_2$ dissociates, thus forming the protostar (a.k.a. second Larson core). It is now commonly admitted that disk formation around the first Larson core is regulated by non-ideal MHD processes \citep{Tsukamoto2023}. In most models, the minimum resolution is of the order of the au, which allows to integrate the disk evolution over a few 10s kyr. In particular, \citet{Machida2010}, \citet{Hennebelle2020}, and \citet{Mauxion2024} show that the disk radii increase from a few au up to a variety of sizes peaking around $\sim 30$ au (in some cases reaching 100~au) through the Class~0 phase over 100 kyrs. Note that the disk expansion is not due to radial spreading but rather by accreting material from the envelope progressively farther from the star, as magnetic braking weakens. This behavior can be well-described when the early disk evolution is regulated by ambipolar diffusion. In this case  \cite{hennebelle_magnetically_2016} predict that the disk radius increases as $M_\star^{1/3}$, where $M_\star$ is the stellar mass. The disk radii measured in these numerical experiments are consistent with those observed around Class~0 and Class~I protostars, which suggests that there is no need of disk's viscous radial expansion to explain disk radii up to 100~au. Of course, without disk's radial expansion, the transport of CAIs from the stellar vicinity to large distances described in 1D models would not occur.  However, the poor resolution in these works did not allow to address what happens near the protostar, where CAIs are expected to form, so these results need to be taken with precaution. 

In recent years, a numerical effort has been made to achieve sub-au resolution. %In all models which include non-ideal MHD, a disk is forming, but the scale at which this disk forms depends on the initial set-up. 
In all high-resolution models that include non-ideal MHD, the scale at which a disk forms depends largely on the initial set-up. For instance, \citet{Vaytet2018} show that a disk of size 0.2~au forms quickly around the second Larson core while no disk forms around the first core. However, they manage to follow the disk evolution only over a physical timescale of a month after the formation of the second core.  %They show that the disk forming around the second core  
On the contrary, \citet{Tomida2015} found that a 5~au disk forms first around the first core. Both numerical experiments included Ohmic dissipation and ambipolar diffusion,  but started from different initial conditions, that is, uniform versus Bonnor-Ebert density profile, with a much longer free-fall time for the latter configuration. \citet{Wurster2021} investigated disk formation including the Hall effect and different initial configuration between the rotation axis and the magnetic field orientation (parallel or antiparallel). They show that, in the antiparallel case, the Hall effect accelerates the gas rotation at the first core scale, which increases the lifetime of the first core and thus allows for a disk to form around it. In all these models, only small disks ($\sim$au) form.  

In summary, these works suggest that the longer the free-fall time and/or the first core lifetime, the more time is available to establish a disk around the first core prior to the one around the second core. The question of the (co?)evolution of these two inner and outer disks, respectively around the second and first cores, remains open: do they coexist for some time and eventually merge? or does the inner disk around the second core form first and then expand to the $>5$~au first core scales?

\begin{figure*}[t!]
    \includegraphics[width=\linewidth]{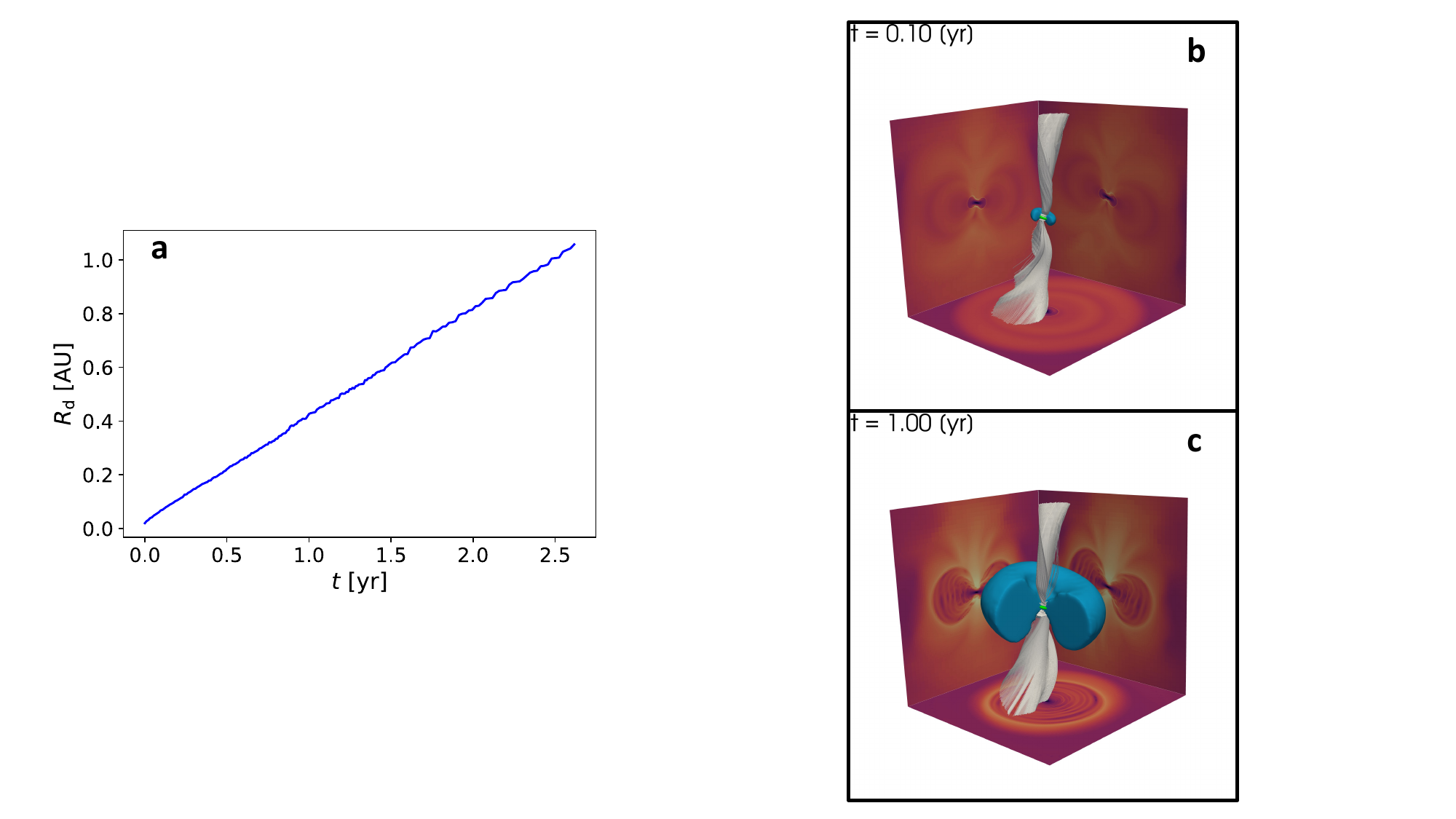}
    \caption{Illustration of the rapid expansion of the {inner} circumstellar disk. Panel (a) displays the disk radius as a function of time, where $t=0$ corresponds to the moment of birth of the disk. Panels (b) and (c) are 3D renderings at $t=0.1$ yr and $t=1$ yr, respectively, where the blue surface is the circumstellar disk and the green surface is the protostar. The white curves are velocity vector field streamlines, illustrating polar accretion. The images are cross-sectional slices displaying the radiative flux, which illustrates the location of accretion shocks. Note: no outer disk forms in this simulation. Figure adapted from \citet{Ahmad2024}.}
    \label{fig:diskexpansion}
\end{figure*}

The simplification of the MHD physics allows high-resolution simulations to cover a longer timespan. \citet{Machida2019} treated the effects of magnetic fields via  Ohmic dissipation, used a maximum resolution of $0.005$~au and did not include a sink particle to model the protostar. They followed the inner disk evolution for 2,000 years after the protostar formation, showing that the inner disk quickly expands within 500 years from sub-au scale to 5~au. \citet{Ahmad2024} conducted similar experiments also without a sink particle but without magnetic fields and using a maximum resolution of $2\times10^{-4}$~au. They explored different initial conditions (initial turbulence and solid body rotation). In their experiments, the inner disk forms independently of the outer disk, very quickly expands to au-size within a few years (see Fig.~\ref{fig:diskexpansion}) and eventually can merge with the outer disk, if it exists. Radial expansion at the early stages is thus a consequence of the collapse leading to the formation of the second core (i.e. the protostar). More work needs to be done in order to explore the universality of the inner disk expansion when the influence of magnetic fields is treated. \citet{Vaytet2018} show that the inner disk is weakly magnetized because of the MHD decoupling at the first core scale. Thus, a priori we expect similar time evolution with and without magnetic field and that magnetic braking does not impede the spreading of the inner disk. 

In this context, temperatures above that of sublimation of the most refractory minerals (2000~K) are achieved in the central region of the first Larson core, where temperature increases with the increasing density. At this temperature, hydrogen dissociation triggers the collapse of the second core and the formation of the inner disk around it, as described above. Thus, both this disk initially and the central part of the first core do not contain any solids: all elements are in vapor form, that is, the gas has the solar chemical composition. CAI condensation occurs when the gas cools. This can happen during the radial expansion of the inner disk, as in the 1D models described in the previous section, but also during turbulent mixing within the high-temperature regions of the first core, which can transport the gas from the hottest to cooler and less dense regions. In both cases (condensation in the disk or in the first Larson core), the CAIs would reach multi-au distances during the radial expansion of the inner disk, until it merges with the outer disk. The shock front associated with the expansion of the inner disk into the outer disk also acts as a temporary barrier to the inward radial drift of dust from the outer disk \citep{Bhandare2024, Ahmad2024}. %[ADD MORE REFS:  Bate 1998; Saigo et al. 2008; Tscharnuter et al. 2009; Machida et al. 2010; Bate 2010; Machida \& Matsumoto 2011; Bate 2011]}. 

In summary, although the picture is certainly more complicated than that described in 1D models as \cite{pignatale_making_2018}  or \cite{Marschall2023}, the idea of a vigorous radial expansion of the inner disk appears robust, associated with the radial transport of dust from the vicinity of the {proto}star outwards. The open question is where this radial expansion stops. This is related to the inner radius of the outer disk and its density: if the  outer disk has not formed yet or is still a flattened envelope, the radial expansion of the inner disk should continue to larger distances. As we have seen above, the study of meteorites reveals that CAIs have been transported up to the region of formation of carbonaceous chondrites, but we do not know exactly where the latter was, although it is guessed to be about 5-10~au. On the other hand there may be an alternative mechanism of transport of CAIs into the outer disk, as discussed next.

%\ab{Can also be included: Using a 1D accretion disk model along different disk layers, \citet{Zhou2022} show that dust can be transported radially outwards due to turbulent diffusion in a gravitationally unstable protostellar disk. }

\subsection{Dust transport in disk winds and outflows}
\label{ballistic}
%***Anaelle, Asmita***

Thermal pressure or magnetically driven outflows, jets, and disk winds are found to be common occurrences that help remove the excess angular momentum and energy as the disk builds up \citep{Pudritz2019, Pascucci2023}. 
%Recently, jet, outflow, and wind models have received renewed attention in their potential ability for an outward transport of some of the dust particles processed in the disk \citep[2019A&A...623A.147A} for an overview see Fig.~1 in][]{Haugbolle2019}. This section draws attention to an alternative mechanism for transport of the gas and dust mixture via outflows during the disk formation and evolution stages. Outflows launched at smaller radii ($<$~a few au) can provide an efficient pathway to transport and redistribute material across different disk radii. This has two implications, first, for the CAIs and AOAs, wherein outflows can replenish the outer disk with material from regions in the protostellar vicinity that has experienced extremely high temperature (> 1400 K) and pressure (> 10$^{-4}$~bar). And second, for the presence of dust larger than the typical ISM sizes of a few microns recently observed in the inner protostellar envelopes.
\begin{figure*}[t]
    \centering
    \includegraphics[width=\linewidth]{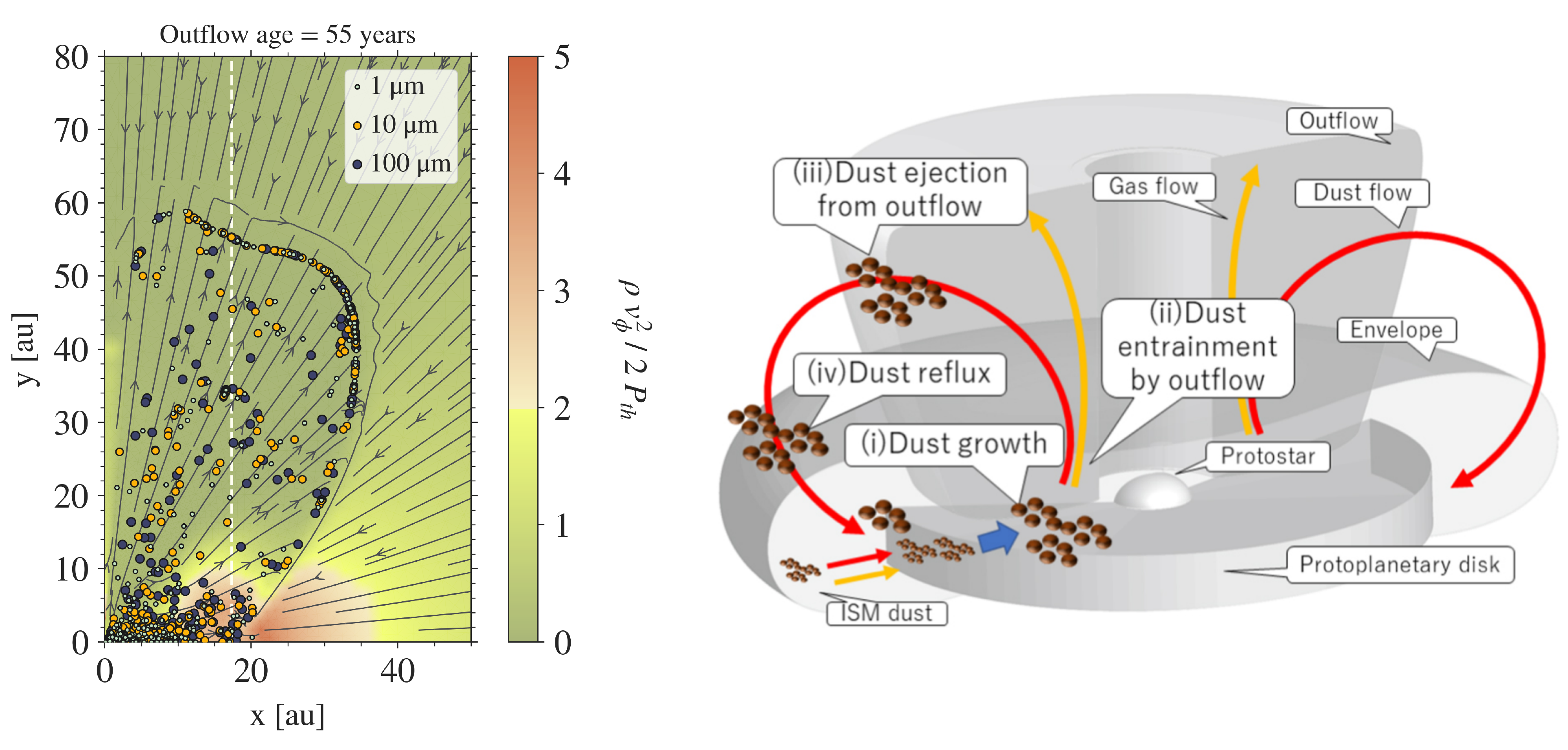}
    \caption{Left: A thermal pressure driven outflow launched during low-mass star and disk formation. The transient outflow seen in a 2D radiative hydrodynamic simulation lifts well-coupled dust up to 100~$\muup$m. The color shows the rotational and thermal pressure support within the collapsing protostellar core and the dashed white line indicates the protostellar disk radius. The figure is reproduced from \citet{Bhandare2024}. Right: Schematic of a protostar--disk--outflow system from a 3D MHD simulation indicating the recycling of dust via outflows. Figure is reproduced from \citet{Tsukamoto21}.}
    \label{fig:DustinOutflow}
\end{figure*}

X-winds \citep{Shu1996, Shu2001}, stellar, and disk winds have been modeled since long ago as processes that are potentially able to extract CAIs from the inner disk where they are condensed, and to transport them to the disk outskirts where they are incorporated in CC planetesimals \citep{Shang2000, Liffman2016, Scott2018} \citep[for an overview see Fig.~1 in][]{Haugbolle2019}. Arguments against the X-wind model launching CAIs selectively from within 0.1~au can be found in \citet{Desch2010}. For instance these authors questioned the very existence of solids within this distance from the star, their survival despite the large relative velocities, the possibility to avoid stellar engulfment, etc.   
%A strong criticism against the outflow transport model is made in \citet{Desch2010} but most of the arguments are tailored against the X-wind model \citep{Shu1996, Shu2001}, which predicted that CAIs formed within 0.1~au from the Sun and were launched from this region. 
However, modern protostellar outflow models \citep{Tsukamoto21, Koga2023, Basu2024, Machida2024} supported by observations \citep{Bjerkeli2016, Lee2022} predict a much broader base of the outflow spanning a wider temperature and density range. This potentially allows at least some dust particles to stably exist in the disk and be eventually launched by the outflow into lower density regions, thus possibly removing the problems highlighted in \citet{Desch2010}. % deeming a wide-angle outflow or a disk wind as a probable transport mechanism. If the outflow can eject large particles from the inner protostellar disk into the envelope, it is natural to expect that some of these particles would then fall into the outer disk, especially if the outflows are short lived. 

\citet{Tsukamoto21} and \citet{Koga2023} did not focus specifically on the origin of CAIs, but on the transport of generic "dust" in the outflow, depending on size, to confront with the observation of dust in stellar envelopes. The collapse simulations by \citet{Tsukamoto21} treating 3D non-ideal MHD effects and covering the first $\sim$10$^4$~years after protostar formation
%\citep{Tsukamoto21,Koga2023} %. Labeling this the ``ash-fall'' mechanism,\citet{Tsukamoto21} 
show that protostellar outflows are able to transport large dust particles, grown up to a centimeter in the inner disk, to the outer disk and envelope. In this ``ash-fall'' mechanism, the entrainment of such large dust in the outflow is possible due to the high gas densities in the launch region (n$_\mathrm{H} > 3\times 10^{11}$ cm$^{-3}$), implying a low Stokes number and hence a strong gas--dust coupling. The dust then decouples from the outflow and remains in the envelope and potentially is supplied to the outer disk. A schematic representation of the ash-fall mechanism for the redistribution of dust both in the outer disk and the envelope is shown in the right panel in Fig.~\ref{fig:DustinOutflow}. %: this is due both to the decrease in gas density along the outflow, and the centrifugal force applied to the dust particles, as the outflow takes away some of the disk's angular momentum and thus includes rotational motions \citep{Lee2018}.
\begin{figure*}[t!]
    \centering
    \includegraphics[trim=1cm 5cm 1cm 4cm, clip, width=\linewidth]{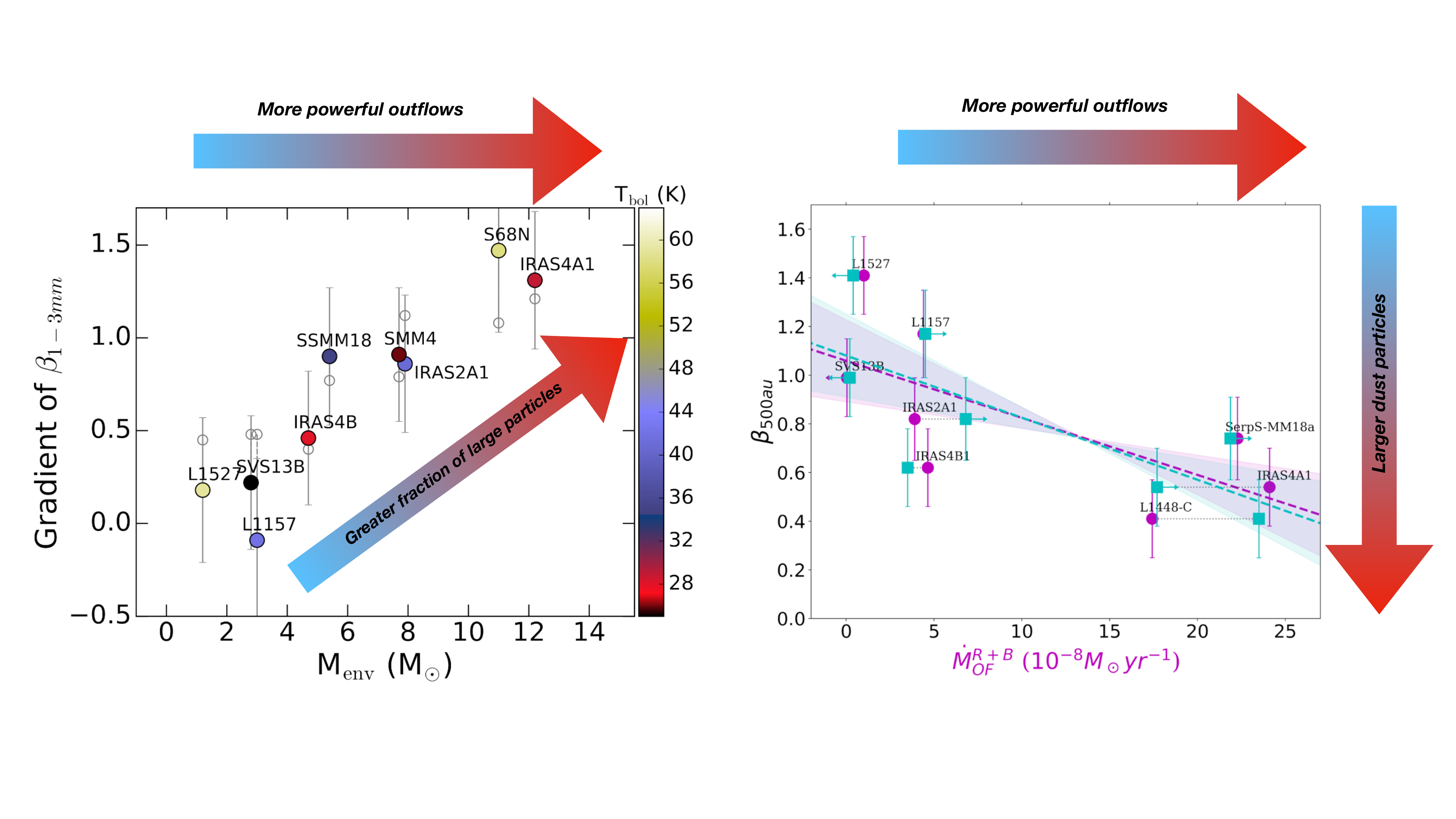}
    \caption{Observations of dust in the inner 500 au of extremely young protostars (aged $<0.1$ Myr) suggest the amount of large dust particles in the vicinity of the disk could be correlated to the flux of matter transported by the protostellar outflow. {Left:} Figure from \citet{Galametz2019} showing that all protostellar envelopes are characterized by low dust emissivities $\beta$, with the most massive envelopes associated to the most pronounced dust emissivity radial evolution. The radial gradient of $\beta$ being interpreted as tracing a greater fraction of large dust grains at small envelope radii, this correlation suggests that protostars embedded in the most massive envelopes (i.e. associated to more powerful outflows) show a greater proportion of large grains in the vicinity of the disk. {Right:} Figure from \citet{Cacciapuoti2024a} showing a tentative correlation between the presence of large grains and the outflow ability to recycle material, as dust emissivities are lower in sources where the outflow strength (momentum flux expressed as \Msolperyr) is greater in some of these young protostars.}
    \label{fig:ObsDust}
\end{figure*}

\citet{Bhandare2024} also investigated this scenario with a 2D $(r,z)$ simulation accounting only for pure radiation hydrodynamic effects but no dust growth. In their study a thermal pressure-driven outflow, launched from sub-au scales shortly ($>$~4 years) after the inner disk formation, lifts dust up to 100~$\muup$m in size during the early stages. Some of the dust is seen to experience temperatures above 1400~K (and pressure above 10$^{-4}$~bar) before being dragged to the relatively cooler outflow region, which suggests the possibility that CAIs condensed at the base of the outflow, not (solely) in the disk. The left panel of Fig.~\ref{fig:DustinOutflow} shows the well-coupled dust dragged in this model by the gas within the high-pressure, supersonic outflow, spread over a few tens of au. Since the velocity of this outflow decreases over time, the outflow should eventually be quenched (within $<$ 100 years) and could deposit this processed dust in the inner envelope and different parts of the disk. However, the possibility of dust growth {from ISM sizes up to and} beyond 100~$\muup$m, {reaching} sizes characteristic of some CAIs, remains to be investigated in this model.  

%Given the short-lived nature of the outflow, seen as a decrease in the outflow velocity over time, this dust will be resupplied back to the disk at different radii where temperature lowers down to 150~K \am{not clear to me why the outflow needs to be short-lived for this to occur - can you explain?}. 
%Magnetocentrifugal disk winds as dust carriers from evolved protoplanetary disks have also been investigated by \citet{Giacalone2019} and \citet{Rodenkirch2022}.

Recent observations suggesting the existence of "large" dust grains in the protostellar envelopes surrounding young disks support the idea that protostellar outflows are effective in transporting disk's dust particles outwards. Infrared observations of excess scattered light suggest dust grains with sizes up to a couple of microns could be present in the dense ISM where stars form \citep{Lefevre14,Dartois24}, an order or magnitude larger than the typical sizes of dust grains estimated by extinction measurements in the diffuse interstellar medium \citep{Mathis77}. In the dense protostellar cores and disks where the dust emission can only be probed at longer wavelengths, the exponent $\beta$ relating dust emissivity $\epsilon(\lambda)$ to wavelength $\lambda$ (i.e. $\epsilon(\lambda)\propto(\lambda/\lambda_0)^\beta$) is usually considered to be an indicator of dust size distribution at millimeter wavelengths. Numerous observational studies show that significant variations of $\beta$ in the millimeter-wavelength range are found in the disks and envelopes of young protostars, with $\beta$ values sometimes $<1$ \citep{Li17, Galametz2019, Bouvier2021, Xu2023}, significantly smaller than the ISM value of $\sim 1.7$ \citep{Agurto-Gangas2019}. These observational findings are shown in Fig.~\ref{fig:ObsDust}. Although the thermal emission of dust is sensitive to its composition, porosity, and size, studies of various interstellar dust analogues suggest that $\beta<$~1 as observed around protostars can only be produced by grains larger than 100~$\muup$m \citep{Ysard2019}. Moreover, polarimetric observations of dust \citep{LeGouellec19} and their comparison to synthetic observations of the polarized dust fractions predicted from protostellar evolution models \citep{Valdivia19, Giang2024} also suggest that significantly large ($>20~\muup$m) grains should be present in the inner protostellar envelopes to explain the large polarization fractions that are observed.

Such large dust particles are not expected to form in protostellar envelopes. Indeed, dust evolution models predict that the formation of sub-millimeter sized particles from sub-micrometer ISM-like grains would require coagulation times of 10$^7$ years at the typical densities of protostellar envelopes \citep{Ormel09,Silsbee22, Guillet2020}. \citet{Wong16} computed that large grains could be built only in the protostellar disk, because gas number densities n$_\mathrm{H}\sim 10^{10}$ cm$^{-3}$ (equivalent to $10^{-3}$~bar at 1500~K) are required to grow grains above 300~$\muup$m in less than a few free-fall times. Thus, the presence of large dust in envelopes should be explained by mechanisms which allow to transport large dust grains from a high-density formation site (i.e. the disk), to the inner envelope where they are tentatively observed. \citet{Lee2022} and \citet{Galametz2019} on the basis of these observations suggested that outflows could be such a mechanism. 

This scenario implies that the stronger the outflow, the more efficient should be the transport of processed dust grains from the inner disk to the intermediate envelope scales. This is observed \citep[][see Fig.~\ref{fig:ObsDust}]{Galametz2019} as a correlation between the low dust emissivities in the inner envelope and the mass of the protostellar envelope, and hence with the outflow momentum flux \citep{Bontemps1996}. Moreover, tentative evidence of a correlation between the outflow momentum and large variations of observed dust emissivity has been suggested \citep{Cacciapuoti2024a}, which could also support this scenario. 

However, even if confirmed, the presence of large grains in the inner envelopes of Class~0 protostars cannot be considered to be an unambiguous signpost of dust transport from the disk to the envelope. Indeed, current grain growth models are very simplistic, and it cannot be excluded that some micro-physical processes allow grains to grow in-situ to macrometer sizes faster than expected, despite the limited gas densities in envelopes. This would solve the main difficulty of the outflow transport model for explaining the observed low dust emissivities: the mass budget of dust which can be recycled in the outflow seems insufficient to significantly alter the observed spectral index of dust emissivity when mixed to pristine envelope material. On the other hand, dust growth in the envelope would not explain the observed correlation that the lowest emissivities are observed around protostars exhibiting the largest mass outflow rates, shown in Fig.~\ref{fig:ObsDust}.

%If confirmed, the presence of such large dust particles in the inner envelopes is puzzling, because of the long timescales and/or high densities that would be required to build them according to the paradigm established by most dust evolution models. 

%This timescale shrinks **to how much?** when considering hydrogen number densities $>$ 10$^6$ cm$^{-3}$ (densities that are reached in the center of protostellar envelopes), or if the coagulation processes start from already grown grains ** to which size?**, as expected to explain the coreshine effect observed in the near-infrared towards cold prestellar cores, for example.
%\citet{Wong16}  

In summary, in the outflow scenario, CAIs -- which are condensates formed at high temperature -- could thus be built either in the inner disk \citep{Tsukamoto21} or at the base of the outflow \citep{Bhandare2024}, where the temperatures and pressures are high enough, then transported outwards by the outflow. Observations of protostellar outflows suggest the mass budget of transported dust, integrated over the $10^4$ yr of the Class 0 phase, is up to $10^{-5}$~\Msol: although this budget may be insufficient to inject sizeable fractions of large dust grains in envelopes, it could be more than enough to explain the overall mass of CAIs/AOAs in the pre-solar disk, inferred from CC meteorites in the Solar system.

%In summary, the existence of large dust in protostellar envelopes can be considered to be empirical evidence that dust grown in the disk could be transported by the outflow into the disk surroundings: being able to promote mixing of dust populations at disk scales, such process could be responsible for some transport of CAIs from their formation site to larger disk radii. Moreover, since outflows transport large amount of material (up to $10^{-3}$~\Msol during the protostellar phase alone, \citealt{Cacciapuoti2024a,Machida2024}), such processes are expected to be able to inject \ab{substantial} amount of CAIs at large disk radii.
%It is then natural to expect that some of these particles would then fall into the outer disk, especially if the outflows are short lived. 

\subsection{Radial spreading or transport in outflows?}\label{comparing}

The two models described in the Sects. ~\ref{spreading} and~\ref{ballistic} potentially explain the transport of refractory material from the inner to the outer disk constrained by the meteoritic record discussed in Sect. ~\ref{sec:dust transport}. The scenario of ballistic transport in outflows is supported by the indication of existence of grown dust in the envelope, which seems to require dust ejection from the disk or the condensation within the outflow, as well as by modern models of outflows and disk winds in the embedded phase, which do reproduce the ejection of dust in a natural way (see Fig.~\ref{fig:DustinOutflow}). The radial transport in the disk, instead, can occur only if the radial expansion of the inner disk, formed around the protostar, is not impeded by the presence of an outer disk previously formed around the first Larson core, as discussed in Sect.~\ref{MHD}. 

The meteoritic evidence, however, adds another piece to the puzzle: no significant amount of CAIs should have remained in the inner disk, meaning that the transport of grains from the inner to the outer part of the disk should have been very efficient. The evidence lays in the isotopic dichotomy. The NC isotopic group comprises both achondrites (e.g. iron meteorites) and chondrites with no systematic difference between the two. Remember that achondrites and chondrites formed at two different epochs, the parent bodies of the NC iron meteorites being the first planetesimals to form in the disk \citep{kruijer_age_2017}, while chondrites formed at least 1 million of years later. NC chondrites do not contain CAIs (with the possible exception of tiny ones, of negligible total mass in the meteorite, \citealt{dunham_calciumaluminumrich_2023, Haugbolle2019}). CAIs carry strong isotopic anomalies relative to the NC group (Fig.~\ref{fig:anomalies}), so the fact that NC iron meteorites and NC chondrites share the same isotopic properties implies that CAIs were absent in the parent bodies of NC iron meteorites as well. If the transport process of dust towards the outward disk had left behind in the inner disk a significant fraction of the CAIs, this would not be possible. The absence of CAIs in NC chondrites could be explained by the drift into the Sun of the CAIs from the inner disk while the formation of a barrier prevented the replenishment of the inner disk of CAIs drifting from the outer disk \citep{desch_effect_2018}. However, this explanation is unlikely to hold for the NC iron meteorites because their parent bodies started to form at $\sim 10^5$~yr, that is, before the CAIs of the inner disk could have drifted into the Sun \citep[see][Figs. 8 \& 9]{desch_effect_2018}. The same argument applies to AOAs. To empty the inner Solar system of CAIs before the formation of iron meteorite parent bodies, the barrier against their drift into the inner disk should have established near time~0 and most inner-disk CAIs should have been large (probably cm-sized) so to drift rapidly into the Sun. Both possibilities are unlikely in an early massive, hot, and turbulent disk. Thus, more likely the lack of systematic isotopic difference between NC iron meteorites and NC chondrites, suggests that the transport of CAIs and AOAs from the inner disk to the outer disk had an efficiency close to unity. 

It is unclear whether the transport of CAIs produced in the disk by the outflow could have had such a high efficiency. Outflows process a large fraction of the initial mass reservoir available and the efficiency of disk wind increases at small disk radii. On the other hand, large CAIs close to the disk's midplane may be entrained with more difficulty in the flow out of the disk. The case of AOAs would be even more difficult to explain, because they presumably condensed farther from the star, where the outflow is less powerful and, therefore, supposedly less efficient in lifting dust.

However, this issue could be solved if CAIs condensed directly in the outflow. Then no condensation in the disk would be required and the condensates in the outflow would have, by construction, a very high efficiency (close to unity) of being outflown, reaching the envelope and potentially the outer part of the disk. Moreover, condensation in an outflow would provide very fast cooling rates, as deduced from the large mass-dependent isotopic fractionation of silicon isotopes in AOAs \citep{marrocchi_rapid_2019}. 

On the other hand, if the disk suffered a phase of vigorous radial expansion, the radial advection of CAIs and AOAs to the outer part of the disk would likely be efficient and could have continued well beyond the phase of condensation, leaving basically no CAIs and AOAs in the inner disk \citep[see Fig. 9 in][]{Marschall2023}. However, we stress that the radial transport of CAIs and AOAs during disk's expansion has so far been studied in 1D models only \citep{pignatale_making_2018, Marschall2023}, while models of higher dimension as those discussed in Sect.~\ref{MHD} might reveal that transport is less efficient\footnote{\cite{Desch2018} proposed that CAIs were transported to the outer disk by turbulent diffusion in the disk, rather than advection during the gas-disk radial spreading. But in this case only a tail of the CAI radial distribution would have reached the outer disk, leading again to the problem of understanding the lack of CAIs in the inner disk when NC iron meteorite parent bodies formed.}. 

%Although more work is clearly needed, our current conclusion is that, even though both processes collaborated in the transport of inner disk material to the outer disk, the disk radial spreading is more likely to be the dominant process to redistribute dust particles at disk scales. In particular the cosmochemical constraints seem difficult to satisfy without a phase of radial disk expansion. 

\subsection{Predominant continuous infall in the inner part of the disk as revealed by the fundamental isotopic dichotomy of the Solar system}\label{sec:dichtomy}

As discussed above, the isotopic composition of meteorites reveals a fundamental dichotomy, partitioning non-carbonaceous from carbonaceous meteorites (Fig.~\ref{fig:anomalies}). Because chondrites still display the same isotopic dichotomy as iron meteorites, despite they derive from parent bodies which accreted 1-4 Myr later, both NC and CC reservoirs must have co-existed for several million years in the disk, without significant mixing that would have homogenized such subtle isotopic differences. The nature of the physical barrier between the two reservoirs will be discussed in Sect.~\ref{trapping}. Here we focus on the early separation of the disk into two distinct isotopic reservoirs, which can reveal how the disk accreted material while it was embedded in the envelope. 

\begin{figure*}[t!]
    \centering
    \includegraphics[width=0.45\linewidth]{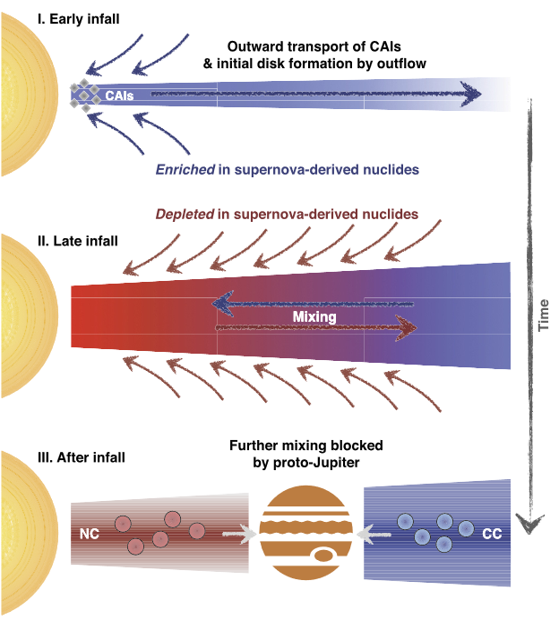}\includegraphics[width=0.45\linewidth]{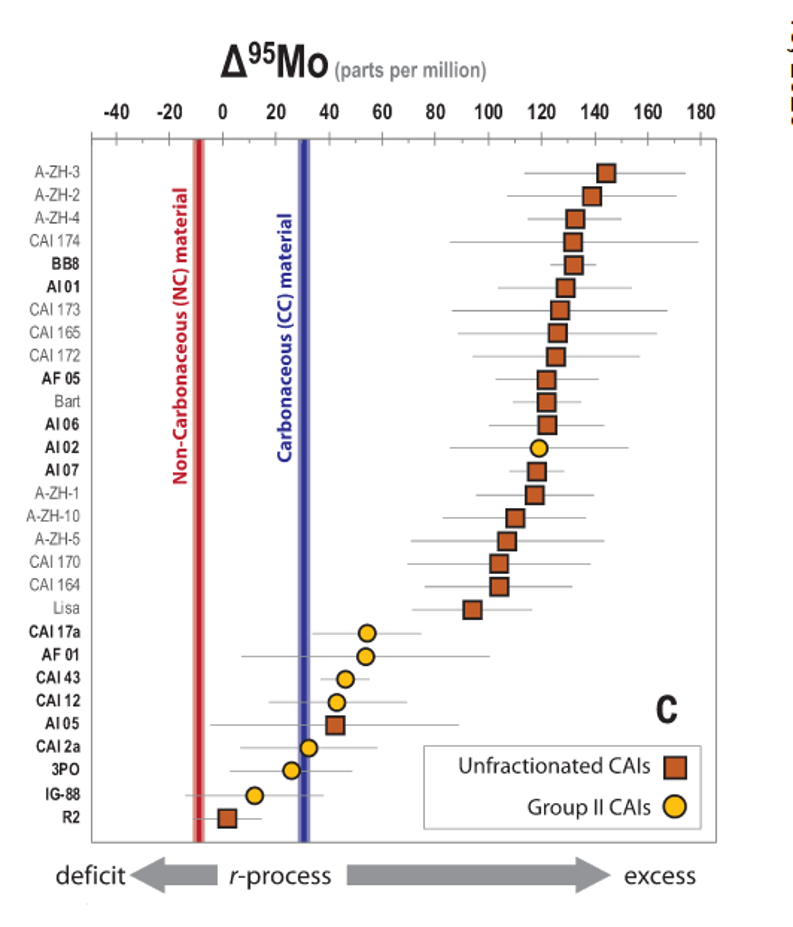}
    \caption{{Left}: a scheme of the accretion of material of changing isotopic composition in the inner part of the disk, as the disk is radially spreading, from \citep{nanne_origin_2019}. This established an isotopic gradient throughout the disk which is then turned into a dichotomy by the formation of early planetesimals at two distinct sites \citep{Morbidelli2022NatAs} and the establishment of a barrier against radial drift of dust, here symbolized by Jupiter (see Sect.~\ref{trapping}). {Right}: Isotopic composition in $\Delta^{95}$Mo of CAIs \citep{brennecka_astronomical_2020}, showing that the shift from the original isotopic composition towards the NC composition started to occur while condensation of calcium-aluminum inclusions was still on-going.}
    \label{nanne}
\end{figure*}

The isotopic characterization of CAIs and AOAs has revealed that the disk was initially enriched in neutron-rich r-process-derived nuclides\footnote{the r-process entails multiple neutron captures and beta decays in a very short time, and occurs typically during supernov\ae\ ou kilonov\ae\ explosions. They are distinct from the s-process, which entails slow neutron capture and beta decay over hundreds of years, typically in the envelopes of evolved stars.} (e.g. $^{48}$Ca, $^{50}$Ti, $^{54}$Cr, and Mo and Ni isotopes; Fig.~\ref{fig:anomalies}, \citealt{trinquier_widespread_2007, trinquier_origin_2009, burkhardt_planetary_2021, jansen_condensate_2024, torrano_common_2024}). Importantly, both CAIs and AOAs share similar isotopic signatures that indicate formation from the same isotopic reservoir despite having condensed at different temperatures (with $T_{\rm cond}^{\rm CAIs} > T_{\rm cond}^{\rm AOAs}$; \citealt{grossman_amoeboid_1976}). Another key constraint is that, for all elements displaying the NC-CC dichotomy, the CC meteorites always plot between the isotopic compositions of CAIs/AOAs and NC meteorites \citep{warren_stable-isotopic_2011, kleine_non-carbonaceouscarbonaceous_2020, burkhardt_planetary_2021}. Furthermore, the NC meteorites show correlated isotopic variations among the nucleosynthetic isotopic anomalies of various elements, regardless of their volatility and geochemical behavior. In fact, the observed correlation stands for lithophile (rock-loving; e.g. Ti, Cr) and siderophile (iron-loving; e.g. Ni, Mo) as well as refractory (e.g. Ti, Mo) and non-refractory (e.g. Cr, Ni, Zn) elements \citep{dauphas_calcium-48_2014, kruijer_age_2017, nanne_origin_2019, spitzer_isotopic_2020, steller_nucleosynthetic_2022}. Altogether, these observations indicate that the isotopic variations are unlikely to result from selective thermal processing in the disk, as different nucleosynthetic anomalies are hosted by different presolar carriers with variable thermal sensitivities. 

Considering all these constraints, the NC-CC dichotomy rather reflects the variable isotopic composition of the infalling material from the parental molecular cloud, which affected the inner and outer regions of the disk differently. In such a scenario \citep{nanne_origin_2019}, the cloud parcel at the origin of the protostar and of its surrounding initial compact disk is dominated by material whose isotopic composition is enriched in supernova-derived nuclides, as recorded in refractory inclusions. % These first condensates were then subsequently transported in the outer disk zone through 
The viscous spreading of the inner disk distributed this material up to large heliocentric distances. Meanwhile, the isotopic composition of latter infalling material changed towards the NC composition. This resulted in the dilution and mixing of the initial isotopic signature of the disk. The new material had to fall predominantly in the inner disk, so as to establish a radial isotopic gradient in the disk, dominated by NC material in the inner part and a mixture of NC and original material (traced by CAIs and AOAs) in the outer part (Fig.~\ref{nanne}, left panel). The contemporaneous formation of the first generation of planetesimals (parent bodies of iron meteorites) at two distinct sites recorded this gradient as a dichotomy \citep{Morbidelli2022NatAs}. This early process thus set the isotopic dichotomy, which would be kept throughout the accretion of later planetesimals thanks to the presence of a physical barrier separating the NC and CC reservoirs, that will be discussed in Sect.~\ref{trapping}. As an end result, all planetesimals formed in the inner Solar system are composed primarily by late infall material with NC composition, while those formed in the outer Solar system are made of a mixture of early and late infalling material and thus have an isotopic composition intermediate between CAIs/AOAs and NC (Fig.~\ref{fig:anomalies}). 

These considerations show that not only the early material had to fall close to the central star in order to promote an initial, vigorous radial spreading of the inner disk (if needed to advect CAIs and AOAs - see Sect.~\ref{comparing}), but also that the subsequent infalling material continued to feed the disk predominantly in its inner part, although not entirely (see Sect.~\ref{volatiles}), until the envelope was removed.

The change in isotopic composition of the infalling material must have started relatively early, that is, while the condensation of calcium-aluminum rich minerals was still ongoing. The evidence comes from the data illustrated in the right panel of Fig.~\ref{nanne}, which shows that most CAIs have an isotopic anomaly in Mo with $80 < \Delta^{95}$Mo$ < 150$ (see \citealt{kleine_non-carbonaceouscarbonaceous_2020} for a definition of $\Delta^{95}$Mo), but some have a composition deviating towards the NC value of $\Delta^{95}$Mo$ \sim -10$. Moreover, aluminum-rich chondrules in NC meteorites, which presumably incorporated an aluminum rich condensate, show no systematic isotopic deviation towards the CAI composition relative to Al-normal chondrules \citep{ebert_ti_2018}, suggesting that the incorporated condensate also had a NC composition. To put this information in context, consider that the condensation of calcium-aluminum rich minerals can occur only as long as the radial motion of the gas is positive across the calcium-aluminum condensation line (i.e. moving from the hotter to the cooler side), as discussed in Sect.~\ref{1D}. In the \cite{pignatale_making_2018} and \cite{pignatale_fingerprints_2019} models, this stops at $50$~Kyr, whereas in the \cite{Marschall2023} model it can continue till $\sim 200$~Kyr. In any case, in both these models the change in isotopic composition from the original to the NC material should have happened very early in time to occur while condensation of calcium-aluminum rich minerals was still ongoing. As an aside, this helps in understanding why all CAIs with large isotopic anomaly have been moved to the outer disk: in a disk spreading model, these would have formed first and therefore would have been advected the farthest, whereas the latest refractory-rich condensates, more likely to be left behind in the inner disk, would have no anomaly relative to the NC composition. 

We stress that even CAIs with the largest isotopic anomalies might not record the "real" solar isotopic composition \textit{stricto sensu} as they are not quite as $^{16}$O-enriched as the Sun (i.e. with $\Delta^{17}$O values of CAIs and the Sun being of -23 ‰ and -29 ‰, respectively, \citealt{mckeegan_oxygen_2011}, see Fig.\ref{3-isotopes}). This suggests that CAIs could have experienced some mixing with isotopically heavier oxygen from other Solar system reservoirs \citep{mckeegan_oxygen_2011} or underwent later gas-melt interactions \citep{aleon_closed_2018}. In addition, it is not known if CAIs represent the solar values for $^{50}$Ti, $^{54}$Cr and other refractory elements because no data are available for the Sun. From the model described in this section we expect that, because the Sun accreted some NC material during the final part of its growth, the solar value should plot somewhere in-between CAIs and NC in a $^{50}$Ti–$^{54}$Cr diagram (Fig.~\ref{fig:anomalies} and Fig.\ref{3-isotopes}). In any case, the fact that CAIs are often referred to as solar comes from the fact that their mineralogies can be at first order reproduced by condensation processes from a gas of solar compositions. Isotopically speaking, they likely do not represent the exact solar values but still represent our best proxy of the early composition of the disk.

\subsection{Streamers within the envelope and the origin of the isotopic anomaly}

The change in isotopic composition in the material infalling from the envelope to the disk at different times may appear surprising, but the reader should remember that the isotopic anomaly differences are very subtle, at a level of one part per 10,000 for some elements, or even less. Without the extreme resolution of modern mass spectrometers, the envelope feeding the proto-Sun and its disk would have appeared extremely uniform. 

Nevertheless, it is not unreasonable that different parts of the envelope had different histories in the interstellar medium, receiving different contaminations from evolved stars and supernovae. Observations of the gas at millimeter wavelengths \citep{Pineda2020, Flores2023} show that envelopes are not smooth spheres as often envisioned in models, but are clumpy, and the accretion towards the protostar does not occur as an ordered collapse of concentric shells, but rather via streamers and filaments within the envelope (not to be confused with late streamers during the Class II phase, discussed in Sect.~\ref{streamers}). These streams of material are also sometimes seen in the dust thermal emission \citep{Cacciapuoti2024b}, suggesting that the dust feeding the disk may originate from different envelope reservoirs, which may each have their own history in terms of irradiation, chemical evolution, and dust evolution.
In this framework, it is not so surprising that materials accreted onto the disk at different times had slightly different properties, as recorded by the meteorite dichotomy. 

%\am{To read on $^{16}$O isotopic depletion (''The relationships observed in Lumley indicate that the parent body incorporated material at the micrometre-scale from discrete diverse isotopic reservoirs, some of which are represented by inner Solar System material but others which must have formed in the outer Solar System. ''): https://ui.adsabs.harvard.edu/abs/2014GeCoA.142..115S/abstract}

\subsection{Isotopic anomalies of volatile elements reveal the infall of material also directly in the outer part of the disk}
\label{volatiles}

As discussed in ~\ref{sec:dichtomy}, the discovery of the nucleosynthetic isotopic dichotomy among meteorites has revolutionized our understanding of the structure of the disk and its evolution over time. In particular, the fact that the NC-CC dichotomy exists for elements of different volatility and thermal sensitivity provides strong constraints on its origin. This supports models where the change in the isotopic composition of the infalling material from the parental molecular cloud modified the initial solar composition of the disk \citep{nanne_origin_2019, burkhardt_elemental_2019, jansen_condensate_2024}. As a consequence, the outer Solar system sampled by CC meteorites has an isotopic composition closer to the solar value due to the presence of early-condensed refractory inclusions \citep{burkhardt_elemental_2019, kleine_non-carbonaceouscarbonaceous_2020} and, possibly, other less-refractory rich materials with the same isotopic composition as CAIs \citep{nanne_origin_2019, yap_nc-cc_2023}.

\begin{figure*}
    \centering
    \includegraphics[width=0.95\linewidth]{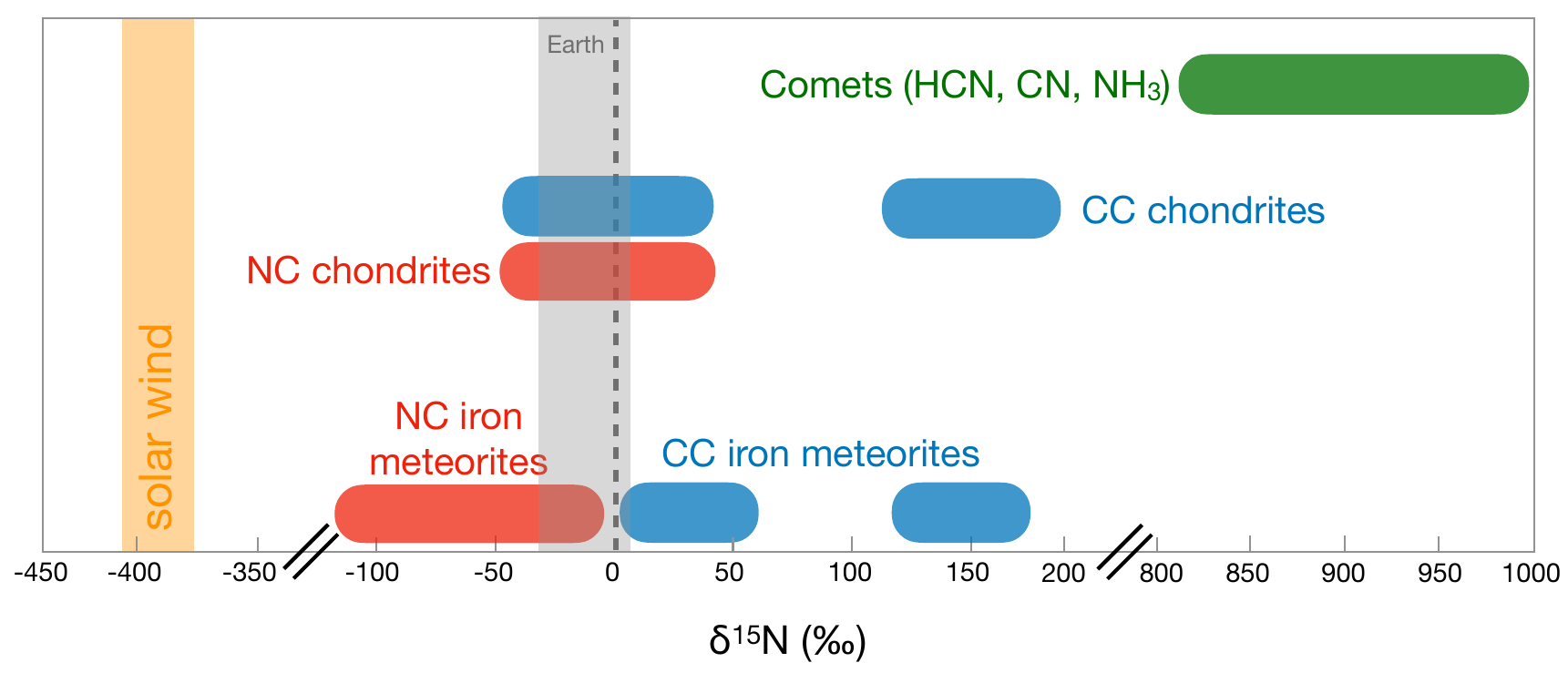}
    \caption{Nitrogen isotopic compositions of NC and CC meteorites. 
The bottom row  concerns iron meteorites, the middle row chondrites and the upper row comets, so that the vertical axis can be grossly interpreted as a temporal sequence, from bottom to top.  
Although both NC and CC meteorites are $^{15}$N-rich relative to nebular gas, the former are closer to the solar value estimated from the Genesis samples \citep{marty_15n-poor_2011}. Modified from \citep{grewal_very_2021}}
    \label{fig:volatiles}
\end{figure*}

Surprisingly, the picture painted by the extremely volatile elements H and N is the opposite of that inferred from more refractory elements. Early-formed iron meteorites thus reveal an opposite isotopic dichotomy, because those belonging to the NC group have compositions slightly closer to the solar value than their CC counterparts \citep{grewal_very_2021} 
(Fig.~\ref{fig:volatiles}).

\begin{figure*}
    \centering
    \includegraphics[width=0.95\linewidth]{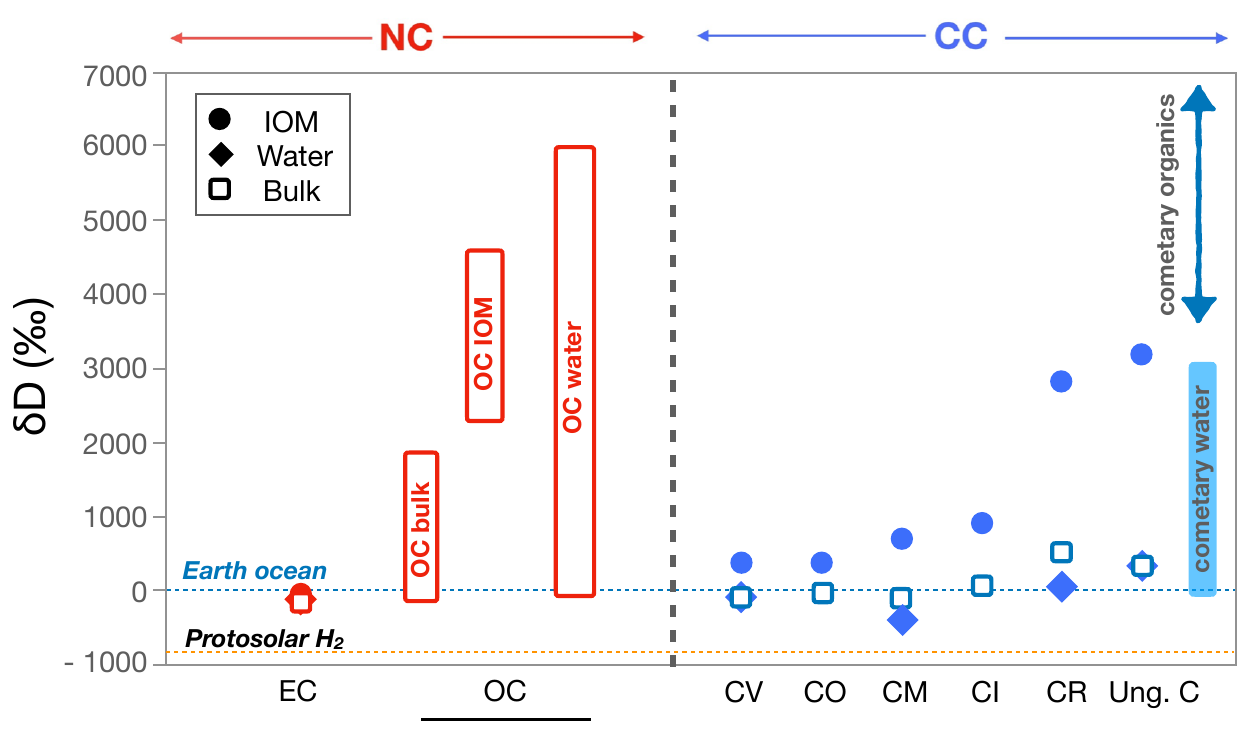}
    \caption{Bulk, water, and insoluble organic matter hydrogen isotopic compositions of non-carbonaceous  and carbonaceous chondrites. Modified from \citep{piani_origin_2021, grant_identification_2024}}
    \label{fig:iron}
\end{figure*}

In addition, later-formed chondrites also display this opposite dichotomy, with NC chondrites having bulk isotopic composition slightly closer to the solar values than CCs \citep{vacher_hydrogen_2020}, at the exception of few rare ordinary chondrites\footnote{Ordinary chondrites (OC) are the most prominent class of NC chondrites \citealt{jones_petrographic_2012}} showing large deuterium (D) enrichments \citep{grant_bulk_2023} (Fig.~\ref{fig:iron}). The same features are also observed in the H-N-bearing insoluble organic matter (IOM) with CC displaying slightly larger D-$^{15}$N-enrichments compared to most of NC chondrites \citep{alexander_deuterium_2010} (Fig.~\ref{fig:iron}). Moreover, comets exhibit much larger D/H and/or $^{15}$N/$^{14}$N ratios, up to a factor 2-3 larger than those observed in any meteorite \citep{altwegg_67p/churyumov-gerasimenko_2014, marty_origins_2016}. In summary, both NC iron and chondritic meteorites show H- and N-isotopic compositions different than, but slightly closer to, the solar values than their CC counterparts, while comets show isotopic compositions much farther away from the solar references. It should be noted, however, than IOM and H-bearing phyllosilicates in the least altered NC-related ordinary chondrites show extreme D-enrichments \citep{alexander_deuterium_2010, piani_dual_2018, piani_origin_2021, piani_hydrogen_2018, marrocchi_hydrogen_2023}.

Combining the information based on nucleosynthetic anomalies and isotopic anomalies of H and N gives the following picture:
\begin{itemize}
\item For nucleosynthetic anomalies (at least for elements with condensation temperatures larger or equal to that of Zn, which is approximately 800~K, \citealt{lodders_solar_2003}), CCs are closer to near-solar isotopic composition as recorded by CAI than NCs
\item Volatile H and N, NCs are closer to the solar isotopic composition than CC.
\end{itemize}	

These characteristics bear fundamental constraints on the disk evolution and its connection with the parental molecular cloud. Although irradiation can play a role in generating $^{15}$N enrichment \citep{Chakraborty2014}, no irradiation experiments have succeeded to produce the large D-enrichments observed in the IOM of CCs. while leaving NC meteorites less enriched \citep{robert_hydrogen_2017}. Altogether, this suggests that molecular cloud material could also have played a role in generating the isotopic characteristics of volatile elements in meteorites. 

Whatever the actual origin of the isotopic anomaly of volatile elements, we propose below a tentative scenario that can reconcile the opposite isotopic dichotomy observed for volatile elements compared to more refractory elements.

The model is mostly based on the intrinsic different behavior between refractory and moderately volatile elements on the one hand, and H and N on the other. Contrary to the former, the latter are carried by water-ice and/or organic grains that could experience sublimation and isotopic re-equilibration with the solar gas of the disk. In this scenario, infalling D-rich water-ice grains from the molecular cloud could undergo sublimation depending of the background temperature of the disk at the place and time of injection. If the temperature is hot enough, water-ice grains will thus sublimate and the resulting vapor will experience isotopic exchange with the D-poor protosolar H$_{2}$ via the following equation: 
\begin{equation}
%HDO + H$_{2}$ <=> H$_{2}$O + HD, 
\text{HDO} + \text{H}_2 \leftrightarrow \text{H}_2\text{O} + \text{HD},
\end{equation}
whose efficiency increases with increasing the temperature \citep{richet_review_1977}. The H-isotopic composition of sublimated water (expressed as the D/H ratio) will decrease toward solar values thanks to efficient isotopic exchange with the D-poor protosolar H$_{2}$ \citep{piani_origin_2021}. Conversely, the injection in colder zones of the disk will inhibit the sublimation and decrease the efficiency of the equilibration reaction. Because later ionization-driven chemical reactions during the disk’s evolution do not efficiently enrich water in deuterium \citep{cleeves_ancient_2014, robert_hydrogen_2017}, this implies that the range of water D/H ratios observed among chondrites likely corresponds to a mixture between (i) water that experienced sublimation, high-temperature isotopic equilibration, and subsequent recondensation during the evolution of the disk and (ii) inherited D-enriched interstellar ices unaffected by disk processes \citep{jacquet_water_2013, yang_dh_2013}. The inheritance of molecular cloud material is more pronounced when considering organic grains because, compared to water, they show higher and more variable D/H \citep{busemann_interstellar_2006} that cannot be reproduced by disk-ionization models \citep{cleeves_exploring_2016}.

Considering the potential sublimation and isotopic re-equilibration processes upon injection of both water-ice grains and IOM from the molecular cloud, three regions can be defined in the disk where: (i) refractory organic and water-ice grains are sublimated, (ii) only refractory organic grains are stable whereas water is still present only as vapor, and  (iii) both organic and water-ice grains are stable. The spatial distribution of these regions evolves with time as the disk cools. Because the outer Solar system disk cools more rapidly than the inner disk \citep{yang_dh_2013}, the organic and water-ice grains injected in the outer disk do not undergo sublimation and keep their D-$^{15}$N-enrichments inherited from cold reactions in the molecular cloud \citep{ceccarelli_we_2022}. Conversely, the hotter inner disk will still be the site of efficient sublimation, resulting in more solar values due to isotopic re-equilibration with the D-poor protosolar H$_{2}$ and $^{15}$N-poor protosolar N$_2$. These temperature-dependent processes thus naturally produce: (i) an inner disk with H-N-isotopic compositions close to the solar values and (ii) a less solar, D-$^{15}$N-enriched, outer disk. {We note, however, that no object of the inner Solar system, neither the Earth nor any NC chondrites or achondrites, shows solar D/H or $^{15}$N/$^{14}$N values \citep{piani_earths_2020, grewal_very_2021, grant_bulk_2023}, thus suggesting that the mixture between (i) and (ii) occurred early and ubiquitously. It has been recently proposed that the D-rich phases in NC chondrites could corresponds to amorphous silicates, which are much more refractory than icy grains and therefore could deliver unprocessed molecular cloud material directly into the inner disk even when the latter was still warm \citep{grant_identification_2024}. Moreover, punctual large D-enrichments in the inner disk (as observed in some NC chondrites, \citealt{grant_bulk_2023}), could be due to individual grains drifting from the outer disk or fallen from the envelope once the temperature in the inner disk had decreased sufficiently.}

Combined with the change in the isotopic compositions of the infalling refractory and moderately volatile materials described in the previous section, the scenario described above will produce a double isotopic dichotomy with the (i) inner disk being dominated by less-solar refractory elements, resulting from the change in the isotopic composition of the infalling material, and more-solar volatile elements due to re-equilibration with the gas of the disk and (ii) outer disk showing the opposite signature, with more-solar refractory elements being inherited from early condensed  CAIs and AOAs, and less-solar volatile elements, due to unprocessed grains inherited from the molecular cloud. In particular, the cold reservoir at the origin of comets could have {sampled a larger abundance of} unprocessed grains, resulting in the largest D/H and $^{15}$N/$^{14}$N ratios {among Solar system objects}  \citep{ceccarelli_we_2022}, whereas NC and CC meteorite parent bodies would contain {a smaller fraction of} unprocessed volatiles from the envelope. From the solar and interstellar D/H and $^{15}$N/$^{14}$N ratios, we can infer that 13\% of water and 5\% of nitrogen in the parent bodies of meteorites is inherited from unprocessed grains, while for comets these fractions increase to 28\% and 15\%.  

This model implies that not all material fed the disk close to the star, but some must have accreted directly onto the outer disk. This is in contrast with the \cite{pignatale_making_2018} model (and even more with the \cite{Marschall2023} model), where the infall of material occurs inwards of the water snowline (see Fig.~\ref{fig:condensations_fronts}), meaning that all ice should have sublimated and equilibrated with the solar H, in contrast with observations. 

{On the other hand, the fact that the fractions of water and nitrogen inherited from unprocessed grains are well below unity for both carbonaceous asteroids and comets implies that a large fraction of the solid volatiles of the outer disk (perhaps a majority) did condense from a hot gas. This is consistent with the radial spreading of the disk. In the outflow transport model, gas is not expected to fall back onto the disk, but we cannot exclude that the temperature in the outflow decays enough to condense volatile grains, which could then decouple from the gas and reach the outer disk. Thus, as in Sect.~\ref{comparing}, it is difficult to discriminate with confidence the two transport scenarios, even though radial spreading seems more promising. }

%. This is in line with the r-depleted Xe isotopic compositions measured in the 67P/C-P comet \citep{marty_xenon_2017}. We thus predict that comets should be dominated by late infall material and thus should belong the NC clan despite having been formed in the outer solar system. If correct, this would lead to a NC-CC-NC structure of the disk as a function of the heliocentric distance.

\subsection{Sample return from comets as a key constraint on the disk's radial expansion vs. direct accretion of envelope material at large distances}
\label{comets}

We have seen in the previous sections that the infall of material from the envelope to the disk has to occur predominantly, but not exclusively, near the star. This can be understood, at least qualitatively, from modern numerical models (see Sect.~\ref{MHD}) if the initial disk of solar composition (perhaps the inner disk in rapid radial expansion around the second Larson core) accretes from an envelope 
(possibly the remnant of the first Larson core and its outer disk).

Recent numerical simulations that resolved regions close to the protostar show that some infalling material from the collapsing envelope slides over the disk surface and is deposited in the inner regions \citep{Lee2021,Bhandare2024, Ahmad2024}. Unless hindered by an outflow, jet, or disk winds, this could continue as the disk evolves during Class 0 and Class I phases, continuously feeding the disk in its inner part and promoting a positive radial motion of gas on the midplane, as suggested by the isotopic evidence (Fig. ~\ref{nanne}). Nevertheless, part of the infalling material can find its way towards the midplane of the disk also at wider radii, thus explaining the existence of un-sublimated volatiles, captured by forming planetesimals in larger proportions at larger heliocentric distances, as discussed in Sect.~\ref{volatiles}. 

\begin{figure*}[t!]
    \centering
    \includegraphics[width=\linewidth]{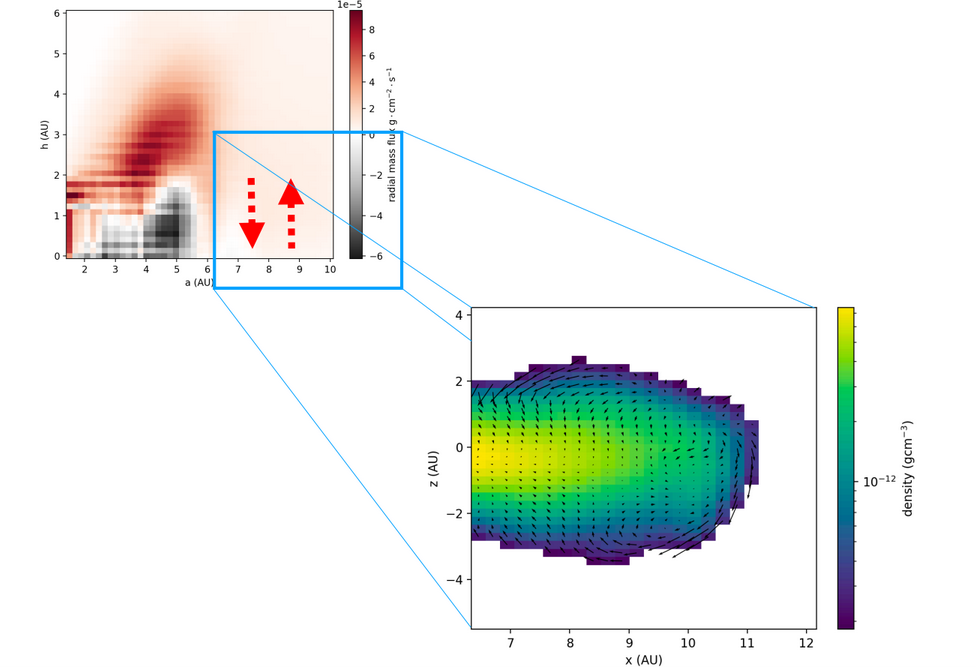}
    \caption{The dynamics of gas as the envelope accretes onto the disk. It is obtained from the simulations described in \cite{Lovascio2024}, after averaging over the elliptic motion of the gas streamlines around the star. The main plot shows the net radial flux, which is directed towards the star at the surface of the disk (red color) and outwards near the midplane (gray color), whereas there is no net radial component beyond $\sim 6$~au, where instead the motion of the gas is vertical. The latter is illustrated by black arrows in the inlet, where the color depicts the density of gas.}
    \label{Francesco}
\end{figure*}

Figure~\ref{Francesco} shows the motion of the gas from simulations presented in \citet{Lovascio2024}. In the inner part of the disk, that is, at $r< 5.5$~au, the infalling gas flows inwards at the surface of the disk and outwards on the midplane. Beyond this distance, however, the gas has a more complex vertical dynamics, as illustrated by the arrows in the inlet of the figure, and feeds directly the outer disk. This could deliver un-sublimated volatiles to the outer disk. Moreover, if we link this late infalling material to the NC isotopic composition, this scheme would predict a disk that is NC not only in its inner part, but also eventually in its outermost part, beyond the outer boundary of the initial disk. Thus, the CC composition could be found only in a central region, %where the early accreted gas was pushed by the radial spreading of the inner disk associated to the formation of the second Larson core (note: there is no such {inner} disk in the simulation presented in Fig.~\ref{Francesco})
the least contaminated part by the material infalling from the envelope. Assuming that the parent bodies of the CC meteorites formed in the giant planet region, it is unlikely that the outer transition from CC to NC occurred as close as 5-6 au, as suggested by Fig.~\ref{Francesco}, but it could have happened farther out. Do any planetesimals sample this transition?     

{Although the isotopic compositions of light elements (i.e., H, C, N, O) of cometary samples and interplanetary dust particles have been extensively documented (e.g., \citealt{floss_identification_2006, ogliore_comet_2023},  no data is available on the isotopic compositions of the refractory elements of comets (and their nucleosynthetic anomalies). Thus, we cannot yet locate the comets in the framework of the isotopic dichotomy of the Solar system, which would be of fundamental importance given that comets presumably formed farther out in the disk than the CC planetesimals, parent bodies of the carbonaceous chondrites.} To do so, a sample of a comet's nucleus would be required. Because there is no credible evidence that any of the meteorites in the current collection come from comets, obtaining a piece of a cometary nucleus requires a sample return mission. Mission concepts for cometary sample return are under study. We stress that, from the isotopic point of view, a cryogenic sample return is unnecessary, which hopefully can simplify the mission concept, reduce its cost and promote its feasibility. If comets turn out to be of CC composition, this will show that the outer disk up to $\sim 35$~au (where comets now composing the scattered disk and the Oort cloud - the sources of short and long period comets respectively - come from: \citealt{Nesvorny2018}) was mostly composed of early infalling gas, close to solar composition. If instead comets turn out to have a composition closer to NC, this will indicate a more limited radial extent of the initial disk and that the disk in the $\sim 15$--$35$~au region  was built primarily from the material infalling later from the envelope. 

%In absence of data from a comet sample return, we remark that the isotopic composition of Xe in comet 67P/C-G differs from the solar isotopic composition by a deficit in r-produced isotopes \citep{marty_xenon_2017}. We remind that the deficit in r-produced isotopes of refractory elements is what distinguishes NC from CC composition. Thus, the comet isotopic composition of Xe may suggest a general depletion of comets in {\it all} r-process-produced isotopes, which would make comets NC in nature. If true, this would be a revolutionary discovery, a mile-stone in the reconstruction of the dynamical formation and evolution of the protosolar disk.   

\section{After the main accretion (Class II stage)}

We now address the evolution of planetary disks after that the infall of material onto the disk from the natal envelope has ceased. In astrophysics, this is called the Class II phase of young stellar objects (Fig.~\ref{fig:seqevol}). This phase is the longest among the three evolutionary stages described in the introduction and therefore a larger number of disks are observable; moreover, due to the removal of the envelope, the disks are better exposed to our observation capabilities in visible light, infrared, and in mm-wavelength. Observations at visible or near-IR {wavelengths} detect stellar light scattered by small grains floating at the surface of the disk. Observations at mm-wavelength are sensitive to the radiation re-emitted by dust approximately millimeter in size, which is in general well settled on the midplane of the disk \citep{Villenave}. The advent of the Acatama Large Millimeter/submillimeter Array (ALMA) is providing an unprecedented level of angular resolution and sensitivity at (sub)mm wavelengths, leading to a real revolution in our understanding of protoplanetary disks. ALMA can also detect the gas emission lines (in particular the CO line) which trace the dynamics of the gas and probe the outer radius of the gaseous component of a disk (e.g. \citealt{Dutrey1998}). We use these observations of dust and gas below to discuss how protoplanetary disks evolve in isolation, that is, after they stop accreting a significant amount of new material from the interstellar medium (with possible the exception of episodic accretion through streamers, discussed in Sect.~\ref{streamers}). 

The objects of the Solar system also provide many pieces of information on the properties and evolution of the protosolar disk in the Class II phase. The distribution of small objects, what remains of the original planetesimal population, provides a lower bound to our dust-disk radius. Cosmochemistry, by comparing chemical and isotopic properties of objects formed at different times, gives us information on the circulation of dust over the age of the disk, putting strict constraints on dust radial drift and mixing. Moreover, the analysis of the dominant component of chondrites, that is, the chondrules, reveal a complex physical evolution of the dust, subject to aggregation, transient heating events and recycling. 

Once again, the astronomical observations of disks and a thorough analysis of the Solar system objects are very complementary and, if integrated, can provide quite a clear view, although certainly still incomplete in the details, of the evolution of protoplanetary disks.

\subsection{Dust radial drift: evidence from the extrasolar disks and the Solar system}

Observations show that disks come in a variety of sizes, from a few au in radius to a thousand of au. Interestingly, whatever the size, the outer radius of the gaseous component of a disk is generally larger by a factor 1.5-3.5 than that of the dust component, probed by (sub)mm continuum emission \citep{NajitaBergin2018, Ansdell2018}. Part of this difference can come from opacity effects, because the gas lines are more optically thick than the continuum emission at similar wavelengths \citep{Hughes2008, Facchini2017}. However, recent observations of well-resolved disks show that the dust distribution decreases too sharply with radius relative to the gas distribution {for opacity effects to be a sufficient explanation}, suggesting the existence of a real edge of the dust disk inward of the gas disk edge \citep{Andrews2012, deGregorio-Monsalvo2013, Andrews2016, Cleeves2016}.  This difference in disk's radii is usually considered to be evidence for grain growth and radial drift. 

In fact, grains are partially coupled to the gas through the equation $d\vec{v}/dt=-\Omega/\tau_f (\vec{v}-\vec{u})$, where $\vec{v}$ and $\vec{u}$ are the velocity vectors of the dust and gas respectively, $\Omega$ is the local orbital frequency and $\tau_f$ is an dimensional coupling parameter known as the {\it Stokes number} which is, in the most common regime, linearly proportional to the size of the dust grains \citep{LJ2012}.  The radial motion of the dust is: 
\begin{equation}
v_r=-2{{\tau_f}\over{\tau_f^2+1}}\eta v_K + {{1}\over{\tau_f^2+1}} u_r
\label{vrad}
\end{equation} 
where $v_K$ is the Keplerian orbital speed on a circular orbit and $\eta$ is the fraction of this speed that is deficient in the orbital motion of the gas. Indeed, for disks whose density declines with increasing distance from the star, the pressure gradient exerts a force directed away from the star that subtracts to the stellar gravity. Thus, for the gas component, the balance between stellar net attraction and centrifugal force is achieved with an orbital speed slightly smaller than Keplerian. {In viscous disks, } the typical value of $\eta$ is expected to be $\sim 3-5\times 10^{-3}$ \citep{bitsch_structure_2015}. {However, in disks whose evolution is dominated by magnetized winds of increasing strength approaching the central star, the surface density gradient can be reversed and $\eta$ can be much smaller or even change of sign \citep{Ogihara2018}. It is important to realize that we don't have observational measurements of the value of $\eta$ in the inner disk, whereas in the outer disk it can be determined by looking at the Doppler shift of the gas emission lines \citep{Teague2019}. However, these measurements are available only for Class II disks, where $\eta$ exhibits sign oscillations consistent with the appearance of dust rings, as discussed in the next section.} 

{If $\eta>0$ (disk in sub-Keplerian rotation)}, grains feel a headwind as they orbit around the star. By friction, they loose energy and angular momentum, which gives them a negative radial speed (first term on the right-hand-side of Eq.~\ref{vrad}). At the same time, grains are entrained in the gas radial flow at speed $u_r$ (second term on the right-hand-side of Eq.~\ref{vrad}). During a rapid radial expansion of the disk in the early stage, grains can be transported outwards, particular if their Stokes number is small. But, {after the expansion phase,} $u_r$ becomes negative {(the disk becomes an accretion disk delivering gas to the central star: \cite{lynden-bell_evolution_1974})}, so the dust motion is inwards, whatever the value of $\tau_f$, {because both terms in Eq. (\ref{vrad}) have the same sign (provided $\eta>0$)}. Thus, the fact that the radius of the dust component of protoplanetary disk is systematically smaller than the radius of the gas component is not a surprise; it is actually a prediction of any simple disk model.

The sizes of the dust and gas component of the protosolar disk are, of course, uncertain. The Kuiper belt of planetesimals of the Solar system extends, with its dynamically cold component, to 45~au. All the objects that are currently situated beyond this distance have eccentric and inclined orbits that suggest that they have been scattered by the planets (see \citealt{MorbyNesvorny}, for a review). Thus, there is evidence that planetesimals formed in the Solar system up to 45~au and no evidence that any formed beyond this threshold. It is thus tempting to conclude that the dust component of the Solar system was (or shrunk to) 45~au at the time of formation of Kuiper belt objects. A size of 45~au in dust is consistent with the median size of protoplanetary disks ($\sim 60$~au; \citealt{Tobin2020}).  We have no direct information of the size of the disk of gas around the Sun. \citet{Kretke} estimated that it was not larger than $\sim 100$~au, otherwise it would have exerted a gravitational torque on the planetesimals scattered by Jupiter during its growth, capturing many on inclined orbit decoupled from the giant planets orbits; these objects are not observed.  If we take $100$~au as a rough indication of the gas disk size, the ratio between the gas radius and the dust radius of the protosolar disk would have been $\sim 2.2$, again quite typical of protoplanetary disks \citep{Ansdell2018}.

The Solar system shows two aspects that can be interpreted by models as evidence that grains did drift towards the central star during at least part of the disk's lifetime: the rapid formation of planetesimals and the enrichment in heavy elements of the atmosphere of Jupiter and of the giant planets in general. We discuss these lines of evidence below. Many in the community would argue that the formation of the solid cores of the giant planets, which requires the process of pebble accretion, is also evidence for dust radial drift. However, it has been shown that the cores of the giant planets could have grown from a ring of radially trapped dust \citep{Morby2020, JiangOrmel2023, Velasco2024}. So, we think that the formation of the cores of the giant planets cannot be considered a strong proof for the radial drift of dust and therefore don't discuss this further.   

It is now commonly accepted that planetesimal accretion occurs via the formation of self-gravitating clumps of dust due to some hydrodynamical \citep{YoudinGoodman2005, Johansen2007, HartlepCuzzi2020} or gravitational instability during settling on the disk midplane \citep{GW1973, Youdin2011}. Triggering these instabilities, however, requires an enhanced dust-to-gas ratio with respect to that expected from the condensation of a solar gas, in particular if any small amount of turbulent diffusion is present in the disk \citep{LiYoudin2021, LimSimon2023}. The first planetesimals of the Solar system, that is, the parent bodies of iron meteorites, formed very quickly, namely within a few $10^5$ years. At that early time, there should have been a full load of gas in the disk, because processes such as magnetized or photoevaporative winds should not have had the time to remove a substantial fraction of the gas. Thus, the only way to achieve the enhanced dust-to-gas ratio needed for planetesimal formation is dust pile-up during radial drift \citep{Drazkowska2016}. The pile-up is caused by changes in the dust radial drift-speed due to changes in the gas disk properties or dust size \citep{Drazkowska2016, IdaGuillot2016} or the evaporation and recondensation of a volatile specie as dust drifts through the corresponding sublimation front \citep{SL1988, Schoonenberg2017}. Thus, early planetesimal formation can be considered evidence for grain drift in the protosolar disk.  \citet{Morbidelli2022NatAs} and \citet{Marschall2023} showed with 1D models that dust pile-up during radial drift is facilitated when the disk is still in radial expansion, so this evidence may concern the disk when it is still in Class I rather than Class II. 
\begin{figure*}[t!]
    \centering
    \includegraphics[width=\linewidth]{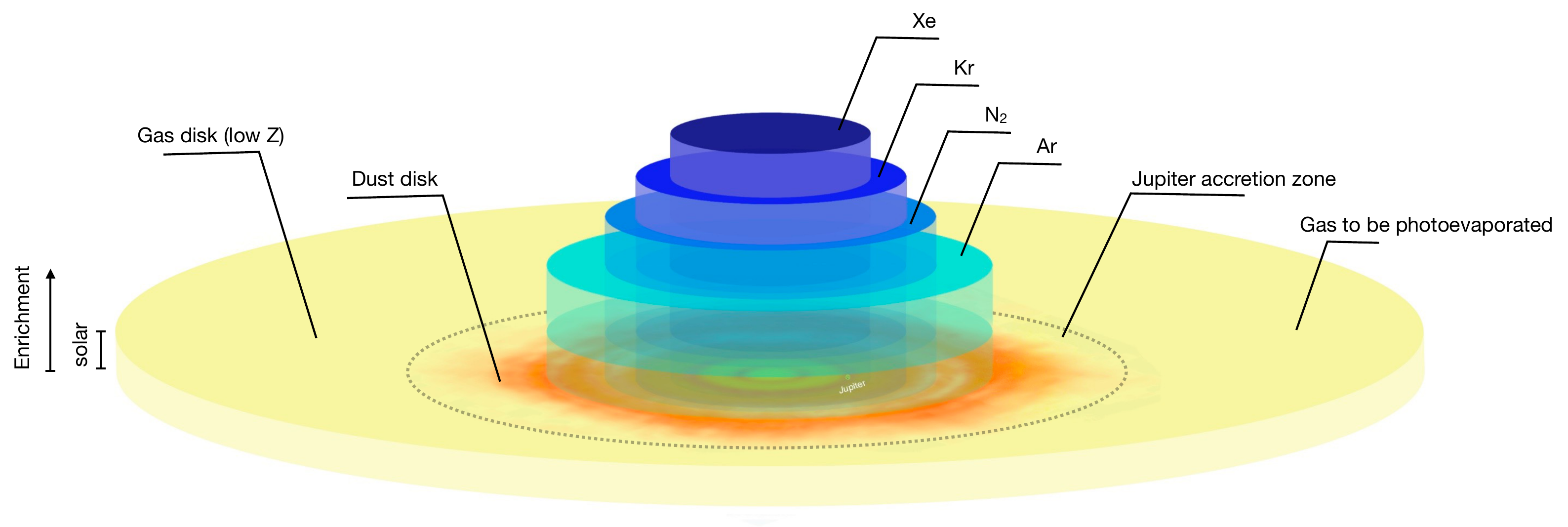}
    \caption{{Schematic view of radial extent and enrichment of the protosolar disk. The yellow disk indicates the extent of the hydrogen and helium gas disk which has, by definition, an enrichment of unity. The smaller dust disk is shown in red. The inner cylinders indicate possible enrichments in Ar, N$_2$, Kr and Xe, respectively, following the inner drift of pebbles and the release of these volatiles at their respective evaporation fronts. In this simple picture \citep{GuillotHueso2006,BitschMah2023}, the local enrichment is highest for the species with the highest condensation temperature. The capture of these volatiles in amorphous ice and their release at the location of the amorphous to crystalline ice transition \citep{MongaDesch2015, Mousis2019} has also been proposed. In this case the enrichment of the gas in all the considered volatiles would start at this location and would probably be more uniform. In the figure, the dashed circle indicates the zone over which Jupiter would accrete gas. Outside of this zone, the gas should be lost by photoevaporation or disk truncation, otherwise it would eventually flow into Jupiter's region and its accretion would dilute the atmospheric enrichemnt.}}
    \label{protoplanetary}
\end{figure*}

The enrichment in heavy volatile elements (relative to H and He) of the atmosphere of Jupiter is intriguing because it seems to be the same (about a factor of three) for all elements, with the possible exception of water, and the isotopic properties of N and Xe are solar \citep{Mahaffy2000}. No objects in the Solar system are known to have both N and Xe with isotopic ratios close to solar. Xe has an isotopic composition close to solar in chondrites, but radically different in comet 67P/C-G\footnote{This comet cannot be an isolated case. The Xe in the Earth's atmosphere is inherited by a 20-80 mixture of Xe with 67P-like isotopic properties and Xe of chondritic/solar properties \citep{marty_xenon_2017}, implying that most comets must have the same Xe isotopic composition of comet 67P/C-G.} \citep{marty_xenon_2017}, suggesting that the outermost part of the disk was non-solar in Xe isotopic composition. This  excludes the possibility that Jupiter acquired its heavy volatile element atmospheric enrichment via the accretion of planetesimals or dust, or the erosion of its solid core, even if Jupiter had formed very far from the Sun as sometimes proposed \citep{Owen+1999, OW2019, bosman+2019}. 

A possible explanation for this enrichment has been proposed by \cite{GuillotHueso2006} and revisited recently by \cite{BitschMah2023} in a general context. In this model, dust sublimation -- as it passes through the various volatile sublimation fronts during its radial drift -- can enrich the gas in volatile species at Jupiter's location. The accretion of gas by Jupiter would then result in a volatile-enriched atmosphere. 

This scenario requires several ingredients: First, the outer disk must cold, $\sim$20 to 30\,K, so that argon, the most volatile element, can condense onto grains and drift inward. {Second, the gas must have inward radial motion so that the enrichment in the evaporated specie at the corresponding evaporation front is eventually advected to Jupiter's location. Third}, most of {the gas of the outer disk, which has low metallicity due to volatile condensation and grain drift,} must be lost by photoevaporation or other mass loss processes {before that it can be advected to Jupiter's orbit}. Lastly, in order for Jupiter to have similar enrichments for species with condensation temperatures as different as that of Ar (25\,K or less) and Xe (45\,K), it had to accrete from a zone larger than that of the different sublimation lines, as depicted in Fig.\ref{protoplanetary}.

While these three ingredients are plausible, the scenario still remains to be quantitatively validated: the early work by \citet{GuillotHueso2006} does not consider all the species separately and, while the new calculations by \citet{BitschMah2023} show that giant planets can be enriched globally, they do not include noble gases and do not study whether the enrichment is uniform in all species. %The contributions of solids and gas as well as the role of the dilute core in the final atmospheric enrichments hence remains to be quantified. Also, the release of volatiles may occur at their respective evaporation temperatures as depicted in Fig.~\ref{protoplanetary}, or together, for example at the amorphous to crystalline ice transition \citep{MongaDesch2015, Mousis2019}. 
On the other hand, this scenario can naturally explains why Jupiter was enriched in nitrogen while keeping a solar $^{15}$N/$^{14}$N ratio. In fact, the original N$_2$ ice, whose drift and evaporation induced the gas enrichment in N, is expected to have a solar isotopic ratio, while the $^{15}$N enrichment of NH$_3$ and HCN in solar system planetesimals \citep{furi_nitrogen_2015,howard+2023invertedZ} is due to photo-dissociation (see Sect.~\ref{sec-anomaly}). This scenario yields an interesting puzzle for water, whose abundance in Jupiter’s atmosphere remains debated. If, as expected, Jupiter’s envelope formed while the planet was beyond the snow line, it should not accrete significant amounts of water. This could then account for the low O/H abundance in Jupiter’s atmosphere inferred from CO disequilibrium chemistry \citep{Cavalie2024}. Alternatively however, upward mixing from the core \citep{Vazan2018,MullerHelled2020} should enrich the planet envelope in core material, including water. The efficiency of this upward mixing is not clear, but it could be sufficient to yield the over-solar enrichment in water that is favored by several other studies \citep{Bjoraker2018,Li2024}.

In conclusion, the enrichment of Jupiter's atmosphere in ultra-volatile species (more volatile than water) is indicative of grain drift and of strong photoevaporation of the external disk.

\subsection{Formation of dust-trapping structures}
\label{trapping}

\begin{figure*}[t!]
    \centering
    \includegraphics[width=0.95\linewidth]{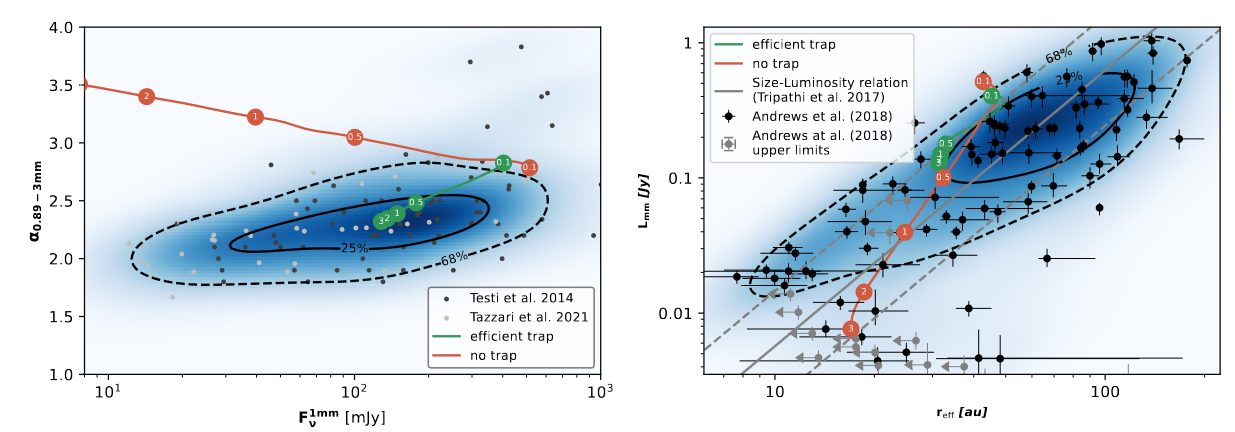}
    \caption{Left: Distribution of millimeter fluxes and spectral indices of protoplanetary disks. Right: their size-Luminosity correlation. In both panels, black and gray dots are the observed disks. Their combined kernel density estimate distribution is shown in color scale. Colored lines depict the expected evolutionary tracks in the cases whee dust traps exist (green) or not (orange). Numbers denote the evolution time in millions of years. From \cite{Birnstiel-ARAA}.}
    \label{fig:Til}
\end{figure*}

If the large ratio between the radii of the gas and dust components of protoplanetary disks is evidence for dust radial drift towards the star, an intriguing observation is the lack of any correlation between this ratio and the disk's age \citep{NajitaBergin2018}. This is in stark contrast with the predictions of simple models (see Fig.~\ref{fig:Til}, where dust radial drift makes this ratio larger and larger over time \citep{Appelgren2020,Birnstiel-ARAA}. This suggests that dust drift stops after sometime, {after that} the dust-disk has shrunk by approximately a factor of $\sim 2$.   

\begin{figure}[t!]
    \centering
    \includegraphics[width=\linewidth]{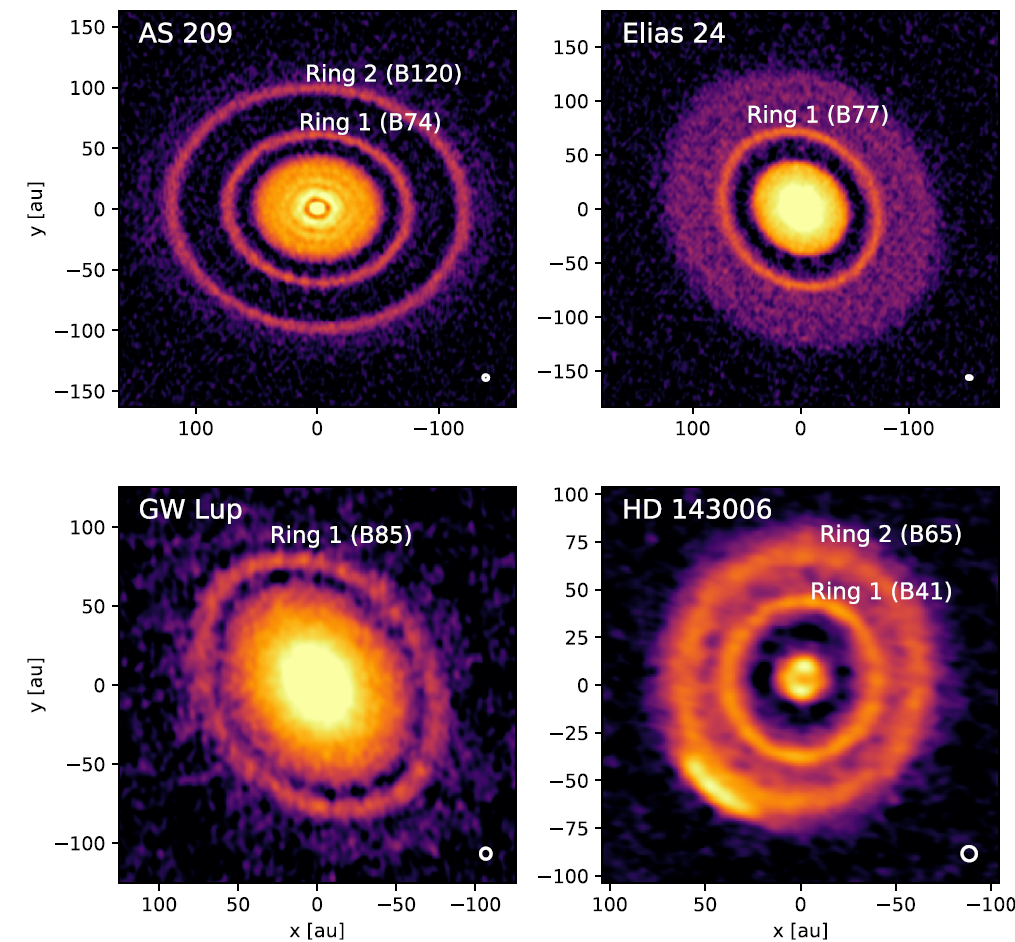}
    \caption{Four examples of ringed disks at mm-wavelength (from \citealt{Dullemond2018}). The white dot in each panel shows the resolution limit of the instrument.}
    \label{rings}
\end{figure}

The high-resolutions disk observations by ALMA \citep{Andrews2018} show that most of the disks display a multi-ringed structure (Fig.~\ref{rings}). The origin and nature of these rings is debated in the literature. They could be due to maxima in the radial distribution of disk pressure. In fact, inward of each maximum, the radial pressure gradient is positive, which makes the coefficient $\eta$ in Eq.~\ref{vrad} negative, as it is proportional to $-{\rm d}\log P/{\rm d}\log r$. Thus, dust with $\tau_f$ large enough that the first term on the right-hand side of Eq.~\ref{vrad} dominates, such as the mm-size dust observed by ALMA, has $v_r>0$ inward of the pressure maximum and $v_r<0$ outward of the pressure maximum. This implies that the dust tends to concentrate at the distance of the pressure maximum, that is, in a ring. The width of the ring is determined by the balance between turbulent diffusion, which tends to disperse any overdensity of dust, and drift towards the pressure maximum, as explained in Appendix~F of \citet{Dullemond2018}. Importantly, if the location of the pressure maximum does not change over time, there is no net drift towards the star of the dust trapped in a ring. 

Other explanations for the rings do not involve dust trapping, but just radial variations of the drift speed of dust which, in a steady state scenario, is inversely correlated to the dust surface density. These radial variations in $v_r$ could be due to (i) oscillations of the pressure gradient, but without the reversal of sign discussed above \citep{Pinilla2018}, possibly due to viscosity transitions induced by the dust itself \citep{DullemondPenzlin2018}, or (ii) condensation fronts which induce dust sintering, which in turn favors collisional fragmentation with a consequent reduction of the dust Stokes number $\tau_f$ \citep{Okuzumi2016}. Dust could also be concentrated by some bi-fluid instability such as the two-component secular gravitational instability \citep{Takahashi2014}. Of these scenarios, only dust trapping at pressure maxima can explain why the ratio between the gas and dust disk radii does not shrink with time. Recent work tracing gas kinematics confirms that rings are associated with pressure maxima \citep{Teague2019,Rosotti2020}. 

The origin of these pressure maxima is, in turn, debated. The most popular explanation is that they are due to the presence of planets opening gaps in the gas distribution, so that a pressure maximum is generated at the gap edges. In some cases, localized kinks in the gas velocity are indeed detected in between dust rings, strongly suggesting the existence of a planet \citep{Pinte2020, Pinte2023}. The planets should be quite massive (at least 20 Earth masses) to generate pressure maxima \citep{Bitsch2018}, otherwise they only generate oscillations of the pressure that modulate dust drift but do not trap dust. An alternative scenario is that pressure maxima emerge in low-viscosity disks as a result of the interaction with the magnetic field, when non-ideal MHD effects are taken into account and in particular in a regime dominated by ambipolar diffusion \citep{Bethune2017, Riols2019}. If the effective viscosity in a disk correlates with the infall of gas from the envelope onto the disk \citep{Kuznetsova2022}, then these pressure maxima may emerge in a natural way at the end of the Class I phase. 

The Solar system provides evidence for pressure maxima both related and unrelated to planets. Jupiter and Saturn achieved masses well above the gap-opening threshold \citep{Bitsch2018}, so they must have created dust-trapping pressure maxima. But Uranus and Neptune were always too small. Despite that, planetesimals formed in the Kuiper belt, up to $\sim 45$~au, as discussed above. The high porosity of the Kuiper belt objects (with densities well below that of 1g/cm$^3$ for objects smaller than 700~km in diameter - \citealt{Brown2013}; see also \citealt{Keane2022} for the specific case of the KBO (486958) Arrokoth) implies that these bodies never experienced a high temperature phase that could melt water or even more volatile molecules such as CO. This in turn implies that these objects formed "late", at least 3-4~Myr after CAIs, in order to avoid overheating by radioactive decay of $^{26}$Al. This means that either the outer disk was replenished of material at a late time (but we will discard this possibility in Section~\ref{streamers}), or dust must have remained trapped for so long at $\sim 45$~au, that is, a pressure maximum must have existed in the Kuiper belt. But, because Neptune is too small to generate one, this pressure maximum must have formed by a planet-unrelated process, possibly non-ideal MHD effects.      

How many pressure maxima formed in the protosolar disk? The answer is difficult, but is probably "several". One pressure maximum must have preserved the NC-CC isotopic dichotomy described in Sect.~\ref{sec:dichtomy}, because otherwise the dichotomy would have been erased by the radial drift of CC dust into the inner Solar system in about 0.5~My \citep{Morbidelli2022NatAs}. The formation of rings so early in the disk evolution is indeed observed in protoplanetary disks \citep{Segura-Cox2020}. We note that \cite{Liu2022} proposed that the frontier between NC and CC was initially at $\sim 30$~au, so that the inner disk could remain NC for millions of years despite of dust drift. This proposal, however, is problematic, because it would imply that the parent bodies of CC iron meteorites, which accreted in the first million year, should have formed in the Kuiper belt. However, we have no evidence for high density differentiated objects there. The pressure maximum preserving the Solar system isotopic dichotomy is usually attributed to the formation of Jupiter \citep{kruijer_age_2017}. However, the fact that it should have appeared within 1/2 Myr seems incompatible with the formation of such a massive planet \citep{Brasser2020}. If this is the case, this pressure maximum should have appeared spontaneously (i.e. without the action of a planet). Such a pressure maximum would have inevitably played a role for the accumulation of dust, the formation of icy planetesimals and ultimately of Jupiter itself, which then would have reinforced the pressure maximum by opening a gap in the gas distribution, thus enduring the dichotomy over the eons.  There is the empirical evidence that only dust smaller than 200 microns managed to pass through such a maximum, with a negligible contribution in the mass balance of inner Solar system objects \citep{Haugbolle2019}.

However, the dichotomy-preserving barrier and the Kuiper belt ring should not have been the sole pressure maxima in the protosolar disk. One evidence for this is that the objects belonging to the CC (resp. NC) class are well distinct from each other. This is true also for chondrites, despite the fact that they formed 2-4~Myr after CAIs, which implies that their constituent materials were not free to mix even on long timescales, unlike what should have happened if all CC objects (resp. NC) had formed in a unique ring. Moreover, as described below in Sect.~\ref{chondrules}, chondrules (the main constituent of chondrites) are distinct from one chondrite group to another, even within the NC or CC isotopic groups, and there is little evidence of any radial transport of chondrules in the timespan lasting from their formation up to  their incorporation in the chondritic parent body.

Strong evidence for the existence of sub-rings could be provided by the so-called chondrule-matrix complementarity  in carbonaceous chondrites \citep{hezel_chemical_2010}, the existence of which, however, is not universally accepted in the meteoritic community (e.g. \citealt{zanda_chondritic_2018}, \citealt{patzer_testing_2021}). We first stress the importance in this concept of the CI chondrites, a rare group of chondrites whose overall chemical composition closely resembles the elemental composition of the Sun \citep{lodders_solar_2003}.

{The concept of CC chondrule-matrix complementarity  relies on the observation that chondrules and matrix have different chemical ratios, respectively higher and lower than the solar value recorded in CI chondrites, but the bulk ratio in the chondrite, which is basically set by the mass-weighted average between the values of its chondrules and matrix, is close to the CI value. Although a secondary origin induced by exchange during alteration on the chondrite parent bodies have been proposed (\citealt{zanda_chondritic_2018} \citealt{van_kooten_unifying_2019}), it should be noted that the chondrule-matrix complementarity is also observed in chondrites with minimal thermal and aqueous alteration \citep{hezel_what_2018}. This seems hardly coincidence and indicates that chondrules and matrix are likely chemically connected. To set the scene, it is also important to stress that the matrix contain organic molecules and presolar grains that could not have resisted to high-temperature chondrule-forming events, implying that unprocessed material must also be present in the matrix of carbonaceous chondrites.  The complementarity has been reported for Mg/Si, Fe/Mg, Ti/Al and Hf/W ratios but does not stand for Al/Na, Al/Si or Ca/Si  \citep{zanda_chondritic_2018, hezel_what_2018}. Generally speaking, element pairs with comparable condensation temperatures often have CI chondritic ratios in bulk carbonaceous chondrites, although the individual elements may have different chemical behaviors. The non-complementarity of pair of elements with different condensation temperatures such as Al/Na, Al/Si or Ca/Si might come from fractionation in pre-accretionary environments such as the addition or loss of the more refractory component or incomplete condensation of more volatile element. In the following,  we focus on the Mg/Si ratio as it is the most commonly used in the literature.} The existence of a complementarity is reinforced by the finding of an isotopic complementarity in $^{183}$W between chondrules and matrix in the CV meteorite Allende, whose average is -- again -- the chondritic value \citep{Budde2016}.

In the most detailed scenario of chondrule-matrix complementarity \citep{hezel_what_2018}, it is thus proposed that: (i) chondrule precursors with supra-solar Mg/Si ratios (e.g. AOAs, \citealt{ruzicka_amoeboid_2012}) underwent exchanges of varying duration with an ambient gas rich in SiO (thus lowering and making  Mg/Si ratios variable but sill supra-solar, e.g. \citealt{tissandier_gas-melt_2002}, \citealt{marrocchi_sulfur_2013}) and (ii) CC matrices represent a mix of condensed material induced by the chondrules heating mechanism and leftover precursors material not affected by the chondrule-formation process \citep{hezel_what_2018}.

For what concerns our discussion on grain drift, chondrule-matrix complementarity  would imply that the matrix grains and the chondrules remained in the same region of the disk {for sometime}, until they have been incorporated into a planetesimal. {This time could be of several $10^5$~yrs according to the spread in chondrule ages belonging to the same meteorite \citep{pape_time_2019}.} Although, strictly speaking, it is enough that chondrules and matrix grain {aggregates} had the same Stokes number and therefore drifted through the disk at the same rate, chondrule-matrix complementarity more realistically  suggests -- once again -- confinement of dust particles in a narrow ring, {within which some particles were thermally processed and converted into chondrules}. 

The detractors of the complementarity \citep{zanda_chondritic_2018, patzer_testing_2021} argue that the chondritic average need to include also refractory inclusions and metal blebs in the mass balance, and stress that the chondrule-matrix complementarity breaks, as we said above, for elements of different condensation temperature (Al/Si, Ca/Si). Alternatives also exists where the CC matrices are all of CI-composition instead being complementary to chondrules \citep{hellmann_origin_2020, marrocchi_iron_2023}. Thus, we consider complementarity as a {\it potential} strong constraint on the existence of dust-trapping sub-rings in the CC reservoir.  

It remains nevertheless remarkable that within the range of isotopic properties proper of the NC or CC groups, there are pairs of achondrites-chondrites with almost identical properties: notable examples are Aubrites and enstatite chondrites  in the NC group \citep{Keil89} and Tafassites and CR chondrites in the CC group \citep{Ma}. If not coincidental, this is remarkable because achondrites and chondrites formed at least one million years apart. This implies that the constituent material did not disperse or mix with other materials of the same broad NC or CC reservoir, which can be understood only postulating the existence of sub-rings trapping particles.  
\begin{figure*}[t!]
    \centering
    \includegraphics[width=0.95\linewidth]{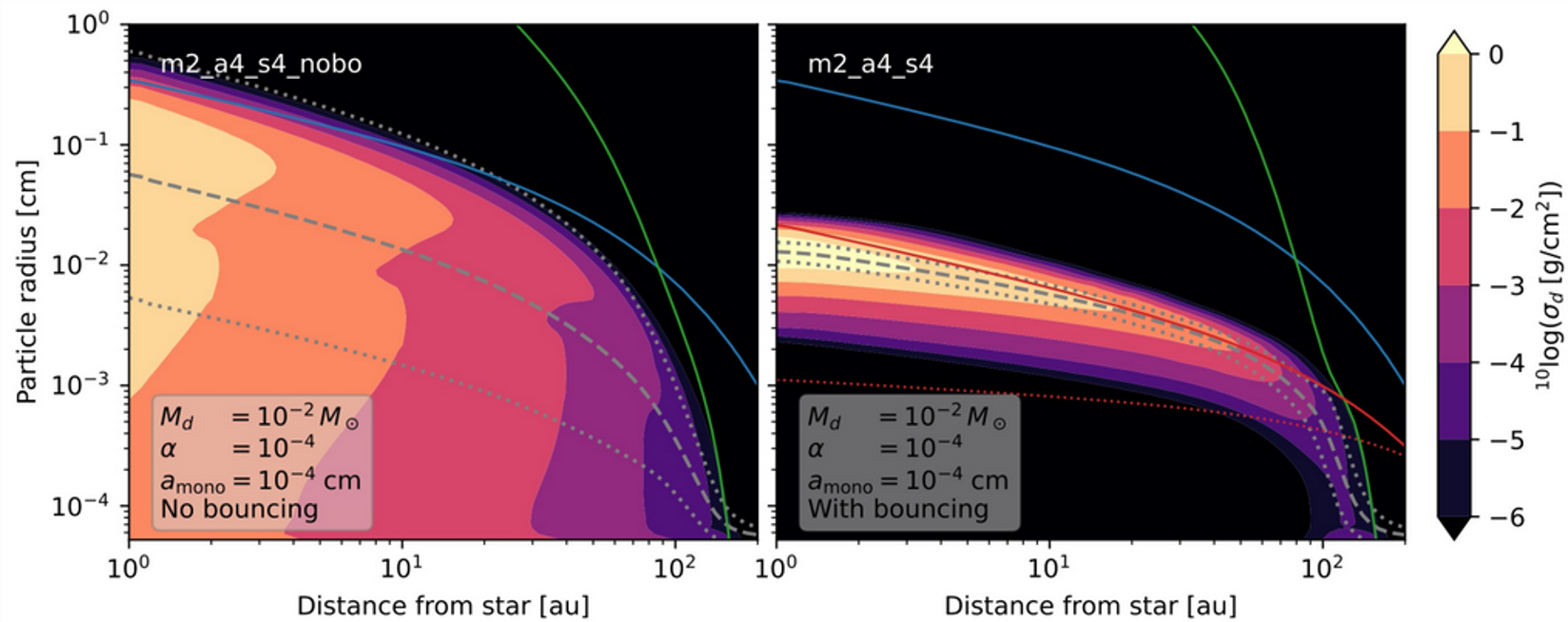}
    \caption{Dust size distribution for a fiducial model without the bouncing barrier  (left) and with bouncing barrier 
(right). The blue line is the fragmentation barrier, the green line is the radial drift barrier, and the red line is the bouncing barrier, all computed assuming turbulence to be the dominant driver of relative velocities. The dashed gray line is the mean grain size and the dotted lines are plus or minus one standard deviation in log space. Form \cite{DD2024}. }
    \label{DD24}
\end{figure*}

The existence of multiple pressure maxima is a necessary, but not sufficient condition for the preservation of material and to avoid mixing with neighboring material. The reason is that trapping in pressure maxima is size dependent. Pressure maxima trap only particles with Stokes number $\tau_f$ large enough that the first term in the right-hand side of Eq.~\ref{vrad} dominates. The smaller particles follow the radial motion of the gas ($u_r$) which has no reason to be null at pressure maxima and therefore are not trapped. If dust particles underwent a collisional cycle, that is, growing until collisional fragmentation \citep{Birnstiel2011}, there would be a constant production of small particles which would leave the parent ring and would then coagulate and be trapped in other rings of the disk, causing a significant mixing of material in the end. The preservation of strict isotopic properties for millions of years is evidence that this does not occur (or is irrelevant in terms of mass balance), suggesting that particle growth is stopped by the {\it bouncing barrier}, rather than the {\it fragmentation barrier}. The bouncing barrier does not generate a large amount of small particle debris and produces a very narrow, almost mono-disperse, size distribution (\citealt{Zsom2010, Jacquet2014, DD2024}; see Fig.~\ref{DD24}), more consistent with strict confinement requirements.    

Taking all these indications for dust trapping together, we can conclude that the large dust fluxes of hundreds of Earth masses over the lifetime of the disk used in several models forming planets by pebble accretion are not realistic and can be excluded for the specific case of the Solar system. The growth of giant planet cores by the accretion of dust is still possible, but only feeding from the dust confined in a ring \citep{Morby2020, Velasco2024}. The confinement of dust also rules out the sometimes-proposed idea that Jupiter formed in the very distant part of the disk (i.e. beyond $\sim 30$~au) and migrated to its current position, because the ring structure would not survive the migration of a gap-opening planet. 

\subsection{Reprocessing of dust as revealed by chondrules}
\label{chondrules}

Chondrules are magmatic submillimeter to millimeter-scale spheroids (Fig.~\ref{fig:chondrites}) that represent the most abundant high-temperature dust in NC and C chondrites, thus implying that both the inner and outer Solar system were affected by the chondrule formation process(es). A key point here is to assess whether both NC and CC chondrules accreted where they formed or experienced disk-wide transport across and between the NC and CC reservoirs. Both models have been proposed and are highly debated \citep[e.g.][]{schrader_outward_2020, schneider_early_2020, williams_chondrules_2020}, with the latter including mostly outward transport of NC chondrules into the CC reservoirs \citep{schrader_outward_2020, williams_chondrules_2020, van_kooten_hybrid_2021, schrader_prolonged_2022}, although inward-drift of CC chondrules was also advocated in the pebble-accretion scenario \citep{johansen_pebble_2021}. In addition to the potential transport of chondrules in the disk, the timing of the onset of chondrule formation is also the subject of an intense debate \citep{bollard_early_2017, fukuda_temporal_2022, piralla_unified_2023}. Nevertheless, all chronometers point to formation processes spanning several million years \citep{pape_time_2019, villeneuve_homogeneous_2009,siron_new_2021, siron_high_2022, fukuda_temporal_2022, piralla_unified_2023} making thus chondrules invaluable witnesses of dust production and processing during the Class II stage of the disk, at a time where the NC/CC barrier was already established.

It should be noted that a large diversity of chondrules exist in both NCs and CCs \citep[see][for a review]{connolly_chondrules:_2016} . They are broadly divided into two varieties, porphyritic (i.e. with large crystals) and non-porphyritic. As the former is overabundant in both NCs and CCs, we will focus on this specific type. Porphyritic chondrules are sub-classified into type I and type II chondrules. This distinction is based on the valence state of iron \citep{scott_chondrules_1983} as indicated by the magnesium number (Mg\#) = 100 $\times$ (Molar [Mg/(Mg+Fe)]), with type I chondrules having a Mg\# $>$ 90 (Fe-poor) and type II chondrules having Mg\# $<$ 90 (Fe-rich). In other words, type I chondrules were formed under reduced conditions (i.e. at C/O ratios where iron is stable in metal form) while type II crystallized under oxidizing conditions (i.e. at C/O ratios where iron is under its oxide form FeO). Whereas type I chondrules are abundant in all types of chondrites, type II chondrules are predominately found in ordinary chondrites. They are nevertheless present in all types of carbonaceous chondrites, although as a minority component \citep{jones_petrographic_2012, pinto_deciphering_2024}.

The origin of chondrules is a long-standing debate and countless models have been proposed to describe their formation. This is due to the difficulty to (i) integrate the complex chemical, petrographic, and isotopic signatures of chondrules and (ii) to find a physical process capable of heating the chondrule precursors to magmatic temperatures ($>$ 1500 K) episodically over several million years, in both the inner and outer parts of the disk. This has led to the development of a plethora of models, either considering chondrule production in nebular environments (e.g. shock waves, lightning discharges \citep{desch_model_2002, DeschCuzzi2000, hubbard_making_2017, kaneko_cooling_2023} or planetary settings (e.g. solid debris of collisions, impact splashes \citep{johnson_impact_2015, libourel_evidence_2007, lichtenberg_impact_2018, sanders_origin_2012}. Models also exist where chondrules formed through the melting of nebular dust by the bow shock wave generated by eccentric large planetary embryos \citep{morris_chondrule_2012}. The problem of chondrule formation might appear unsolvable, but recent years have seen important progress by revealing new, overlooked petrographic characteristics and isotopic constraints. Here, we highlight recently reported constraints and discuss implications regarding chondrule formation conditions, their potential transport in the disk, and the plausible astrophysical process at their origin.

\begin{figure*}[t!]
    \centering
    \includegraphics[width=0.95\linewidth]{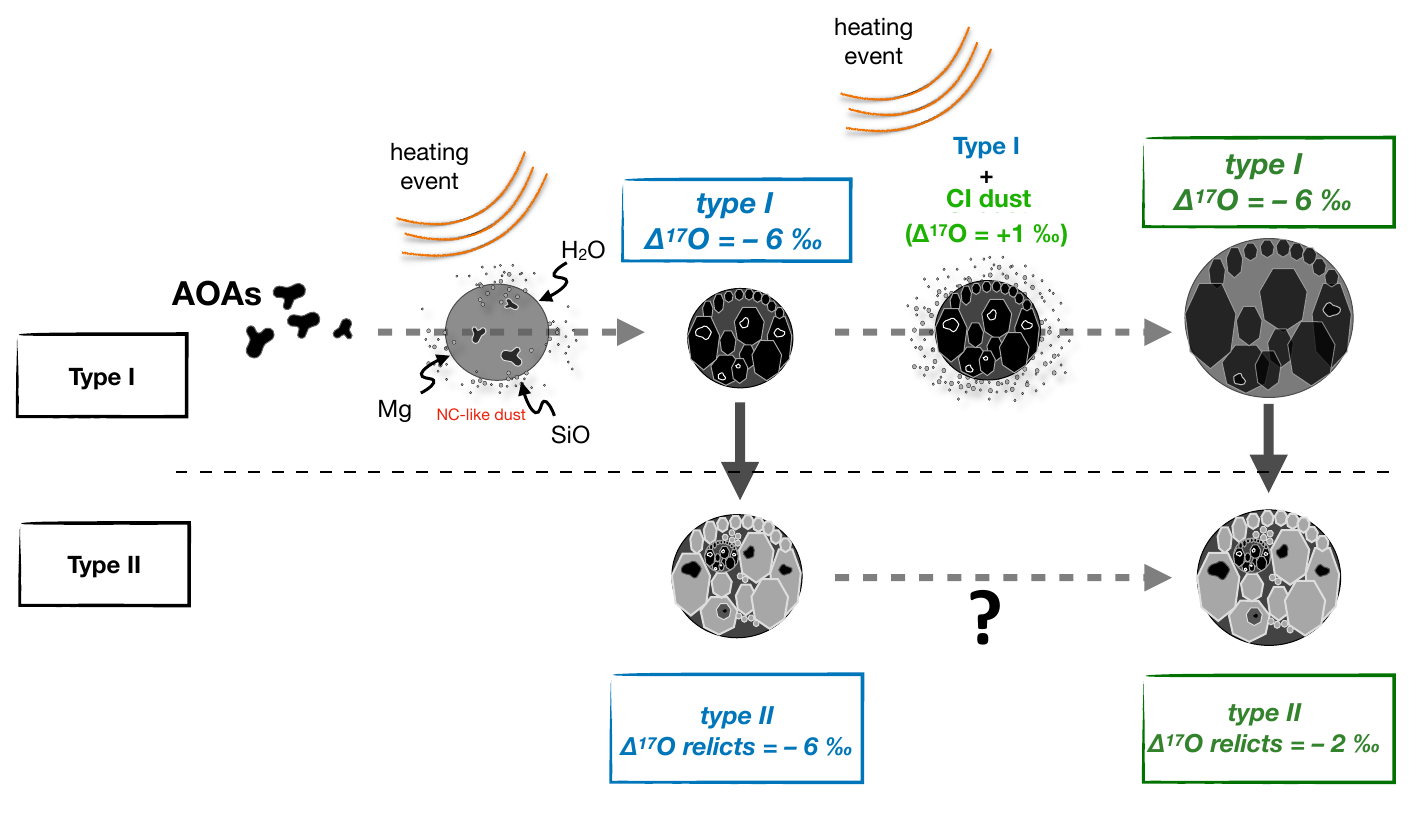}
    \caption{Schematic representation of formation of type I and II chondrules in CC. Previously condensed $^{16}$O-rich AOAs, mixed with other precursor material, experienced melting and gas–melt interactions that led to the formation of olivine-rich type I chondrules characterized by $^{50}$Ti-$^{54}$Cr anomalies and the presence of $^{16}$O-rich olivine grains with $\Delta^{17}$O$ = \sim-6$\textperthousand. In CR chondrites, these type I chondrules mix with solar-like dust and experienced another melting episode. This results in the formation of larger chondrules with $\Delta^{17}$O$ = \sim -2$\textperthousand. Both types of type I  chondrules were then recycled into type II chondrules with Mg-rich relic grains having $\Delta^{17}$O characteristics of the two generations of type I chondrules. This implies that the origin of chondrules must be extremely localized for producing type II chondrules without recycling all type I chondrules.}
    \label{fig:chondrules}
\end{figure*}
We first focus on CC chondrules as they have been the object of the most comprehensive studies. We deliberately exclude from this discussion the type I non-porphyritic chondrules in the CB/CH chondrites, for which there is a consensus on an origin by impact between planetesimals \citep{krot_young_2005}. We will discuss separately the formation of CO/CM/CV chondrules from that of CR chondrules, where CO, CM, CV and CR are all different classes of CC chondrites.
\begin{itemize}
\item Type I CO/CM/CV chondrules. Several key arguments support their formation by recycling of early-formed amoeboid olivine aggregate-like precursors. Type I CO/CM/CV chondrules have heterogeneous oxygen isotopic compositions and commonly display relic grains with chemical (Al-Ti-poor) and isotopic ($\Delta^{17}$O$ = \sim -15$\textperthousand) properties similar to those of olivine grains present within AOAs \citep{tenner_chondrules_2018, chaumard_oxygen_2018, rudraswami_oxygen_2011, marrocchi_oxygen_2018, marrocchi_formation_2019, schneider_early_2020, jacquet_origin_2021}. In addition, mechanically isolated type I chondrules have variable $^{50}$Ti and $^{54}$Cr nucleosynthetic anomalies \citep{schneider_early_2020}. The fact that chondrules from the same chondrites show oxygen mass-independent variations as well as variable $^{50}$Ti and $^{54}$Cr signatures (both among chondrules of the same meteorite and within any given chondrule) strongly weakens models where chondrules are solid or molten by-products of impact between differentiated planetesimals \citep{libourel_evidence_2007, faure_origin_2012, johnson_impact_2015, libourel_oxygen_2022, libourel_significance_2023}, as these models would predict a more homogeneous properties. Conversely, these isotopic variations suggest that type I CC chondrules were formed from a dust aggregate comprising a precursor grain isotopically similar to $^{50}$Ti-$^{54}$Cr-rich CAIs and AOAs \citep{ruzicka_amoeboid_2012, ruzicka_relict_2007, ruzicka_relict_2008, marrocchi_oxygen_2018, marrocchi_formation_2019, schneider_early_2020}, together with other grains of NC isotopic properties (Fig.~\ref{fig:chondrites})\citep{schneider_early_2020}. The presence of material with NC isotopic signature in CC chondrules is supported by the observation that, in bulk, individual CC chondrules have $^{50}$Ti and $^{54}$Cr signatures intermediate between the CAIs/AOAs values and NC chondrites values and is consistent with the model where the isotopic differences observed between the NC and CC reservoirs result from the change in the isotopic composition of late infalling material from the Solar system's parental envelope \citep{nanne_origin_2019}. For chondrule formation, the precursor dust aggregate must have experienced melting by some nebular process and protracted gas–melt interactions (Fig.~\ref{fig:chondrules}). 
\item Type I CR chondrules. Compared to type I CO/CM/CV chondrules, type I CR chondrules show more rarely the presence of  $^{16}$O-rich relic grains and display a smaller chondrule-to-chondrule variability in $\epsilon^{54}$Cr \citep{schrader_formation_2013, tenner_oxygen_2015, olsen_magnesium_2016, van_kooten_isotopic_2016, schneider_early_2020, marrocchi_isotopic_2022}. This is generally ascribed to a formation of type I CR chondrules from different precursor material than that of CO/CM/CV chondrules, either because they formed at larger heliocentric distance \citep{van_kooten_isotopic_2016, van_kooten_role_2020} and/or at later times \citep{budde_hf-w_2018, schrader_distribution_2017, tenner_extended_2019}. Recent petrographic and isotopic characterization of type I CR chondrules show that two populations of chondrules co-exist in CR chondrites: (i) small chondrules with textural features and oxygen isotopic compositions similar to type I CO/CM/CV (with $\Delta^{17}$O$ = \sim -6$\textperthousand) and (ii) large chondrules with more complex textures and higher $\Delta^{17}$O values around -2\textperthousand\ \citep{marrocchi_isotopic_2022}. Based on these observations and a mass balance calculations for Cr isotopes, it has been proposed that large type I CR chondrules derive from earlier-formed smaller CO/CM/CV chondrules which aggregated with unprocessed dust before undergoing a new chondrule-forming event (Fig.~\ref{fig:chondrules}; \cite{marrocchi_isotopic_2022, marrocchi_whom_2023}). This interpretation is supported by Al-Mg dating, which shows that large CR chondrules formed $\sim$ 1.5 Myr after smaller CR chondrules, which in turn formed 2 to 3 Myr after CAIs \citep{tenner_extended_2019}. Moreover, mass balance calculations based on O, Te, and $^{54}$Cr isotopic compositions indicate that the unprocessed dust that aggregated with small CO/CM/CV chondrules had isotopic composition similar to the matrix of CC meteorites and amounted to $\sim$ 70-80 wt\% of the full aggregate  \citep{bryson_constraints_2021, hellmann_origin_2023, marrocchi_isotopic_2022}.
\item Type II CC chondrules. Compared to type I, type II CC chondrules are characterized by FeO-rich silicates and more chondritic abundances of moderately volatile elements \citep{jones_petrology_1990}. This attests to their formation under oxidizing conditions that could be produced (i) in regions with enhanced dust/gas ratios \citep{schrader_formation_2013} or from precursors with low carbon contents \citep{connolly_carbon_1994}. {Alternatively, it has also been proposed that type II chondrules are derived from their type I counterparts (Fig.~\ref{fig:chondrules})  within impact-generated vapor plumes \citep{libourel_significance_2023} or through so-called nebular oxidation shocks, which correspond to drastic changes in oxygen fugacity compared to the reduced formation conditions of type I chondrules \citep{villeneuve_silicon_2020, villeneuve_relationships_2015, pinto_deciphering_2024}.}

\color{black} We now focus on the comparison between type I chondrules in NC (especially ordinary chondrites) and CC chondrites, and especially those from ordinary chondrites. NC chondrules display (i) more homogeneous $^{50}$Ti and $^{54}$Cr isotopic compositions, which are similar to the compositions of their host chondrites \citep{schneider_early_2020} and (ii) more homogeneous oxygen isotopic compositions \citep{kita_high_2010, piralla_conditions_2021}. Although the compositions of the analyzed CC and NC chondrules may overlap for either $^{50}$Ti, $^{54}$Cr or $\Delta^{17}$O \citep[e.g.][]{olsen_magnesium_2016}, in multi-isotope space, none of the CC chondrules plot in the compositional field of NC chondrites, and no NC chondrule plots within the field of CC chondrites \citep{schneider_early_2020}. These data thus reveal a fundamental isotopic difference between NC and CC chondrules, which is inconsistent with a disk-wide transport of chondrules across and between the NC and CC reservoirs. Conversely, this supports that NC and CC chondrules formed in spatially separate reservoirs and likely have been incorporated in chondritic planetesimals where they formed, unlike CAIs. This is also supported by the fact that each chondrite group is characterized by chondrules showing specific petrographic characteristics, suggesting that they sampled separated chondrule reservoirs \citep{jones_petrographic_2012}. To reconcile the local chondrule formation and accretion with the fact that refractory inclusions have undergone significant inside-out radial transport, it should be noted that chondrules formed later, during the Class II stage of the disk, when outward transport mechanisms were more limited or non-existent (both the radial expansion of the disk and the ballistic transport in the outflow) and turbulent diffusion also becomes very weak \citep{Villenave, Lesur-PPVII}. 

The higher abundance of chondrules in NC chondrites attests that the inner disk was the site of more efficient chondrule production processes. Compared to CC chondrules, the more homogeneous internal oxygen isotopic compositions of large NC chondrules ($>$ 300 $\muup$m) suggests they experienced enhanced recycling processes and interactions with the gas of their forming-regions \citep{piralla_conditions_2021}. However, rare $^{16}$O-rich olivine grains have been also spotted in type I OC porphyritic chondrules \citep{kita_high_2010, piralla_conditions_2021}. Interestingly, these relic olivine grains have chemical and oxygen isotopic compositions similar to the recently discovered population of small OC chondrules ($<$ 300 $\muup$m, \citealt{marrocchi_isotopic_2024}). This suggests that larger type I OC chondrules could derive from the small type I OC chondrule population, which would have formed previously, during the evolution of the inner disk \citep{marrocchi_isotopic_2024}. The small chondrules can themselves be derived from $^{16}$O-rich precursors akin to AOAs but whose condensation would have occurred after the isotopic composition of the inner disk had changed toward NC values, as recorded in rare OC CAIs \citep{ebert_ti_2018, nanne_origin_2019, brennecka_astronomical_2020}. How to keep this population of small chondrules in the inner disk {until being recycled in large chondrules}, without being flushed into the Sun is a key question. {They could (i) be accreted in pressure bumps that emerged in the inner disk, as discussed in Sect.~\ref{trapping} or (ii) be stored in planetesimals small enough ($<$ 10 km) to avoid melting or (iii) be accreted as part of an undifferentiated layer on top of earlier-formed differentiated bodies \citep{elkins-tanton_chondrites_2011}.}

%Interestingly, small OC chondrules have more heterogeneous oxygen isotopic compositions compared to their larger NC counterparts \citep{marrocchi_isotopic_2024}. This implies that they cannot derive from large differentiated planetesimals that would have homogeneous isotopic compositions, {\color{red}unless invoking that chondrites correspond to the unmelted crust accreted on top of differentiated parent bodies \citep{elkins-tanton_chondrites_2011}. Their storage in the inner disk can thus be the result of (i) the accretion on small ($<$ 10 km), undifferentiated, planetesimals, (ii) trapping in pressure bumps that emerged in the inner disk, as discussed in Sect.~\ref{trapping} or (iii) late accretion of early-formed differentiated bodies \citep{elkins-tanton_chondrites_2011}.}
 
Taken together, the NC and CC observations point toward complex episodes of dust reprocessing that transform early condensates to chondrules and chondrules to new bigger chondrules. This thus attests the complex history of dust evolution. In addition, as discussed above, petrographic and isotopic systematics do not support disk-wide transport of chondrules but point toward local formation of chondrules within their respective accretion reservoirs. A key observation stands in the fact that several generations of genetically-related chondrules (i.e. deriving from each other) co-exist in chondrites. In addition to supporting the nebular brand of chondrule-forming scenarios, this argues for repetitive and extremely localized heating events to transform certain chondrules without affecting others. In astrophysical terms, it requires the occurrences of repetitive and spatially-limited nebular heating events, efficient in both the inner and outer disk. This could be achieved under different conditions such as bow shocks \citep{morris_chondrule_2012}, lightning \citep{kaneko_cooling_2023}, local shock waves \citep{hood_nebular_2009}, or current sheets \citep{lebreuilly_dust_2023}. 

\end{itemize}

\subsection{Streamers and the late delivery of fresh material to disks}
\label{streamers}
Although the Class II phase starts by definition when the envelope surrounding the young star has been removed, the protoplanetary disk can still accrete material at various rates as the star moves through the natal giant molecular cloud. The first description of this process was provided in a seminal work by \citet{HL1941} and \citet{BH1944} and still goes under the name of BHL-accretion. Recent observations have caught this process in action, imaging streamers as they fall towards the protoplanetary disk \citep{Pineda2020, Valdivia-Mena2022, Cacciapuoti2024b}.

These late episodes of accretion of material onto the disk may have several important consequences: they can provide fresh material, allowing the late formation of planetesimals and planets \citep{Throop2008}; because the angular momentum vector of the new infalling material has no reason to be aligned with the angular momentum vector of the disk, they can form misaligned disks, which are also observed in quite large numbers \citep{Kuffmeier2021}; they can create stresses in the disk, effectively enhancing the disk's effective viscosity \citep{Kuznetsova2022}. 

Because the Solar system presents many planetesimals that formed millions of years after CAI condensation and exhibit a heterogeneous isotopic distribution (the dichotomy discussed in Sect.~\ref{sec:dichtomy}) it is tempting to invoke a late streamer to explain these properties. However, a careful look at cosmochemical constraints drives to the exact opposite conclusion: no streamer delivered any substantial amount of material during the Class II phase. These constraints are
\begin{itemize}
\item there is no systematic difference in isotopic properties between achondrites (samples of early-formed planetesimals) and chondrites (samples of late-formed planetesimals) in each of the NC and CC classes, indicating that no new material was delivered in between the two planetesimal formation periods. 
\item the switch from solar to NC material during the formation of the disk occurred very early, when formation of refractory condensates was still ongoing (see right panel of Fig.~\ref{nanne}).
\item the material incorporated in CC planetesimals from the outer part of the disk, is closer to the solar composition than the material incorporated in the NC planetesimals from the inner part of the disk. This is exactly the opposite of what would be expected if exotic material had been delivered by a streamer which, given its typical large angular momentum with respect to the star, feeds predominantly the outer disk.
\item the eight planets and the cold Kuiper belt are all within a few degrees of a unique plane, orthogonal to the full angular momentum of the system. This excludes any misalignment of the protosolar disk up to at least 45~au.
\end{itemize}

The absence of late streamers in the history of the protosolar disk is probably linked to its enrichment in $^{26}$Al via the astrophysical environment in which the Sun formed \citep{Desch-SSR}. The injection of a large amount of $^{26}$Al without the injection of a significant amount of $^{60}$Fe is evidence for the formation in proximity of an evolved Wolf-Rayet star \citep{GounelleMeynet2012, Dwarkadas2017}. The winds emitted from these stars form around them bubbles of emptiness in the molecular cloud embedding the star-forming region. Protostars and their disks that enter these expanding bubbles, in addition to being seeded with radioactive elements launched by the Wolf-Rayet star, find themselves isolated from the rest of the molecular cloud. In these conditions, it is unlikely that they can accrete a streamer of material at a late time \citep{Dwarkadas2017, Desch-SSR}.

\section{Summary and Conclusions}

In this paper we have summarized the consensual view that the authors achieved on the history of the protosolar disk, after a week of intense discussions in the peaceful atmosphere of the Treilles foundation in southern France. Our approach was mostly guided by cosmochemistry, which provides exquisite constraints on the composition of the disk as a function of time and location. Numerical simulations and astronomical observations gave us an interpretative framework in which the cosmochemistry data could be put in context to form a coherent and, we believe, a credible picture of the protosolar disk origin and evolution. 
\begin{figure*}[h!]
    \centering
    \includegraphics[width=0.6\textwidth]{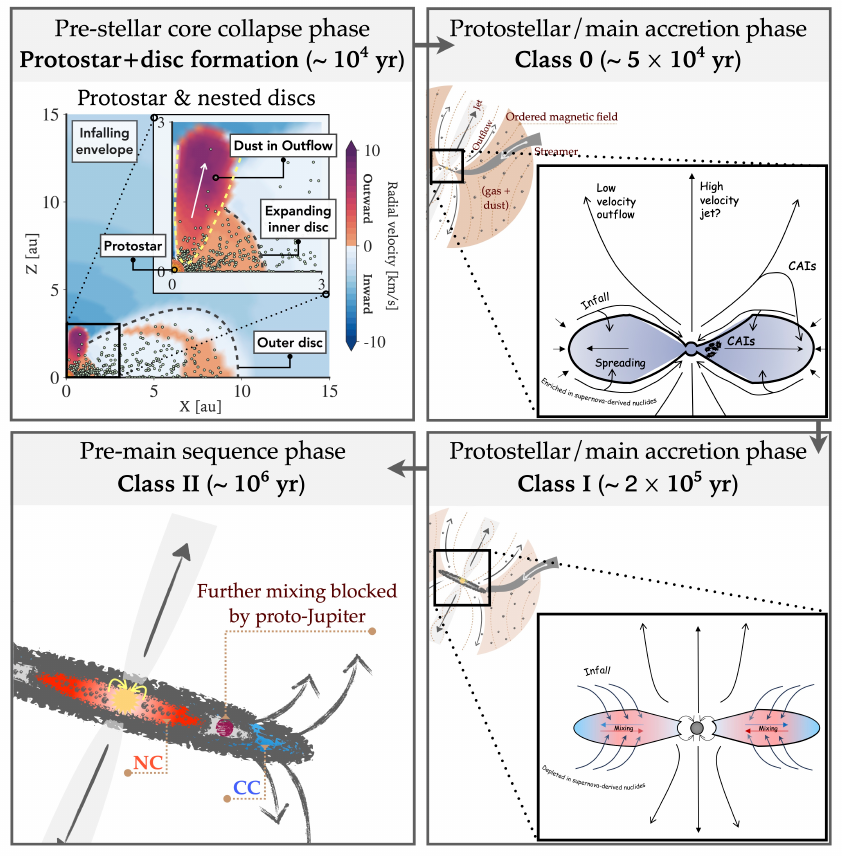}
    \caption{Schematic representation summarizing each key phase in the birth process of the Solar system, with $t=0$ marking the onset of collapse of the parent dense molecular cloud core. Top left panel is adapted from \citet{Bhandare2024}.}
    \label{fig:recap}
\end{figure*}

With the help of Fig.~\ref{fig:recap} we summarize here the main steps of the formation and evolution of our protoplanetary disk that emerged from this analysis. 

The history started when the first Larson core collapsed to form the second core, that is, the protostar (top left panel of Fig.~\ref{fig:recap}). In this process, the disk was born in the vicinity of the protostar and vigorously spread outward, transporting the refractory material that condensed from the hot gas (CAIs and AOAs) and all other dust grains up to distances of at least $\sim 10$~au. Some inner disk grains (e.g. CAIs) could also be transported to the outer disk ballistically via the stellar outflow (top panels of Fig.~\ref{fig:recap}). 

The inflow of material from the envelope continued to feed the disk predominantly in its inner part, promoting the continuous spreading of the disk, but not exclusively, which allowed volatile elements to be incorporated in the outer disk without being sublimated and equilibrated with the gas at high temperature. Early in this process, the gas changed of isotopic composition, evolving from the initial solar composition to another characterized by a deficit in nucleosynthetic isotopes produced in the r-process, which characterizes the  NC composition.  This produced a disk that was dominated by NC material in the inner part and, in the outer part, a mixture of NC and solar material (the latter being carried by CAIs, AOAs and other possibly less refractory grains with similar isotopic composition). Such a mixture characterizes the CC composition \citep{schneider_early_2020}. This is illustrated in the left panels of Fig.~\ref{nanne}, which can be considered as sub-panels of the top right and bottom right panels of Fig.~\ref{fig:recap}. If material was accreted from the envelope also beyond the spreading disk, then the outermost part of the resulting disk should be again of NC composition. Comets may sample that outermost part, but only the isotopic analysis of refractory elements, possible via a comet sample return mission, could tell. 

As time proceeded, the infall of material on the disk waned and became more uniform, which allowed the disk to evolve as an accretion disk. In this phase the dust started to drift inwards, shrinking the radius of the dust component to $\sim 45$~au, probably about 1/2 of the width of the gas component. The end of the infall marked the end of the Class I phase and the drop in viscosity of the disk, so that MHD structures emerged, producing a series of pressure maxima in the disk. The dust got trapped in rings associated to these pressure maxima and stopped drifting towards the Sun (bottom left panel in Fig.~\ref{fig:recap}). Planetesimals could thus form over millions of years without changing of isotopic properties. Nevertheless, the disk was not quiet: heating events periodically processed the dust turning dust aggregates around refractory seeds into chondrules, and new aggregates around chondrules into new bigger chondrules etc. The origin of these heating events remains unclear.

There was no late accretion of material onto the disk via streamers. The disk disappeared in 5~Myr, as indicated by paleomagnetic data in meteorites \citep{Weiss2021}. At the time of the disappearance of the disk, all planetesimals and the giant planets were already formed. Only terrestrial planet formation continued via collisions, for several tens of Myr, possibly 100~Myr. A phase of dynamical instability of the giant planets sometimes during this period dispersed the trans-Neptunian planetesimals \citep{Nesvorny2018} and the asteroid belt \citep{Deienno2018}, imprinting the final structure of the Solar system.

%%%%%%%%%%%%%%%%%%%%%%%%%%%%%%%%%%%%%%%%%%%%%%%%%%%%%%%%%%%%%%%%%%%%%%%%%%%%%%
%%%%%%%%%%%%%%%%%%%%%%%%%%%%%%%%%%%%%%%%%%%%%%%%%%%%%%%%%%%%%%%%%%%%%%%%%%%%%%
%%%%%%%%%%%%%%%%%%%%%%%%%%%%%%%%%%%%%%%%%%%%%%%%%%%%%%%%%%%%%%%%%%%%%%%%%%%%%%
%%%%%%%%%%%%%%%%%%%%%%%%%%%%%%%%%%%%%%%%%%%%%%%%%%%%%%%%%%%%%%%%%%%%%%%%%%%%%%
%%%%%%%%%%%%%%%%%%%%%%%%%%%%%%%%%%%%%%%%%%%%%%%%%%%%%%%%%%%%%%%%%%%%%%%%%%%%%%
%%%%%%%%%%%%%%%%%%%%%%%%%%%%%%%%%%%%%%%%%%%%%%%%%%%%%%%%%%%%%%%%%%%%%%%%%%%%%%

%%%%%%%%%%%%%%%%%%%%%%%%%%%%%%%%%%%%%%%%%%%%%%%%%%%%%%%%%%%%%%%%%%%%%%%%%%%%%%
%%%%%%%%%%%%%%%%%%%%%%%%%%%%%%%%%%%%%%%%%%%%%%%%%%%%%%%%%%%%%%%%%%%%%%%%%%%%%%
%%%%%%%%%%%%%%%%%%%%%%%%%%%%%%%%%%%%%%%%%%%%%%%%%%%%%%%%%%%%%%%%%%%%%%%%%%%%%%
%%%%%%%%%%%%%%%%%%%%%%%%%%%%%%%%%%%%%%%%%%%%%%%%%%%%%%%%%%%%%%%%%%%%%%%%%%%%%%
%%%%%%%%%%%%%%%%%%%%%%%%%%%%%%%%%%%%%%%%%%%%%%%%%%%%%%%%%%%%%%%%%%%%%%%%%%%%%%
%%%%%%%%%%%%%%%%%%%%%%%%%%%%%%%%%%%%%%%%%%%%%%%%%%%%%%%%%%%%%%%%%%%%%%%%%%%%%%

\begin{acknowledgements}
The group of authors is extremely thankful to Fondation de Treilles for sponsoring and hosting the meeting that made possible the present and discuss the various data, models and lines of argument that converged into the view of the protosolar disk evolution presented in this paper. \\ 
S.Charnoz, B.Commer\,con. and A.Morbidelli also acknowledge the funding from programme ANR-20-CE49-0006 (ANR DISKBUILD).\\
A.Morbidelli acknowledge support the funding from the European Research Council (ERC) under the European Union’s Horizon 2020 research and innovation programme (Grant agreement No. 101019380 - HolyEarth).\\
A. Maury acknowledges support the funding from the European Research Council (ERC) under the European Union’s Horizon 2020 research and innovation programme (Grant agreement No. 101098309 - PEBBLES)\\
A. Bhandare acknowledges funding by the Deutsche Forschungsgemeinschaft (DFG, German Research Foundation) under Germany's Excellence Strategy - EXC-2094 - 390783311. \\
All authors are grateful to Fred Ciesla for his constructive comments during the review process. 

\end{acknowledgements}

\bibliography{references,Ma-bibliotheque}
\bibliographystyle{aasjournal}



\end{document}